\documentclass[longauth]{aa}

\usepackage[utf8]{inputenc}
\usepackage{txfonts}
\usepackage{textcomp}
\usepackage{graphicx}
\usepackage{float}
\usepackage{tabularx}
\usepackage{booktabs}
\usepackage[table]{xcolor}
\usepackage{colortbl}
\usepackage{amsmath,amssymb}
\usepackage{siunitx}
\usepackage{xspace}
\usepackage{scalerel}
\usepackage{fontawesome}
\usepackage{multirow}
\usepackage{verbatim}
\usepackage[ruled,vlined]{algorithm2e}
\usepackage{natbib}
\usepackage[pdfencoding=auto,psdextra]{hyperref}
\hypersetup{
    colorlinks=true,
    linkcolor=blue,
    citecolor=blue,
    urlcolor=blue
}
\usepackage[nameinlink,capitalise]{cleveref}
\usepackage[switch,modulo]{lineno}
\usepackage{glossaries}
\glsdisablehyper
\usepackage{orcidlink}
\newcommand{\orcid}[1]{\orcidlink{#1}}
\usepackage{lastpage}
\makeatletter
\renewcommand*\aa@pageof{, page \thepage{} of \pageref*{LastPage}}
\makeatother
\usepackage{euclid}
\usepackage{ISTL_macros}
\usepackage[normalem]{ulem}

\newcommand{\kpnu}{EP$\nu$\xspace}
\newcommand{\dnnu}{\ensuremath{\Delta N_\mathrm{eff}}\xspace}
\newcommand{\pcb}{P_\mathrm{cb}}
\newcommand{\pmm}{P_\mathrm{m}}
\newcommand{\threetimestwo}{3\texttimes2pt\xspace}

\newcommand{\cloe}{\texttt{CLOE}\xspace}

\definecolor{crisp}{HTML}{C7DDF2}
\definecolor{crispier}{HTML}{E8F1FA}
\definecolor{gray}{gray}{0.9}

\begin{document}
\graphicspath{ {./Figures/} }

\title{\Euclid preparation.}
\subtitle{XCVIII. Cosmology Likelihood for Observables in Euclid (CLOE). \\ 5: Extensions beyond the standard modelling of theoretical probes and systematic effects}    

\author{Euclid Collaboration: L.~W.~K.~Goh\orcid{0000-0002-0104-8132}\thanks{\email{lgoh@roe.ac.uk}}\inst{\ref{aff1}}
\and A.~Nouri-Zonoz\orcid{0009-0006-6164-8670}\inst{\ref{aff2}}
\and S.~Pamuk\orcid{0009-0004-0852-8624}\inst{\ref{aff3}}
\and M.~Ballardini\orcid{0000-0003-4481-3559}\inst{\ref{aff4},\ref{aff5},\ref{aff6}}
\and B.~Bose\orcid{0000-0003-1965-8614}\inst{\ref{aff7}}
\and G.~Ca\~nas-Herrera\orcid{0000-0003-2796-2149}\inst{\ref{aff8},\ref{aff9},\ref{aff10}}
\and S.~Casas\orcid{0000-0002-4751-5138}\inst{\ref{aff11}}
\and G.~Franco-Abell\'an\orcid{0000-0001-9742-5408}\inst{\ref{aff12}}
\and S.~Ili\'c\orcid{0000-0003-4285-9086}\inst{\ref{aff13},\ref{aff14}}
\and F.~Keil\orcid{0000-0002-8108-1679}\inst{\ref{aff14}}
\and M.~Kunz\orcid{0000-0002-3052-7394}\inst{\ref{aff2}}
\and A.~M.~C.~Le~Brun\orcid{0000-0002-0936-4594}\inst{\ref{aff15}}
\and F.~Lepori\orcid{0009-0000-5061-7138}\inst{\ref{aff16}}
\and M.~Martinelli\orcid{0000-0002-6943-7732}\inst{\ref{aff17},\ref{aff18}}
\and Z.~Sakr\orcid{0000-0002-4823-3757}\inst{\ref{aff19},\ref{aff14},\ref{aff20}}
\and F.~Sorrenti\orcid{0000-0001-7141-9659}\inst{\ref{aff2}}
\and E.~M.~Teixeira\orcid{0000-0001-7417-0780}\inst{\ref{aff21}}
\and I.~Tutusaus\orcid{0000-0002-3199-0399}\inst{\ref{aff14}}
\and L.~Blot\orcid{0000-0002-9622-7167}\inst{\ref{aff22},\ref{aff15}}
\and M.~Bonici\orcid{0000-0002-8430-126X}\inst{\ref{aff23},\ref{aff24}}
\and C.~Bonvin\orcid{0000-0002-5318-4064}\inst{\ref{aff2}}
\and S.~Camera\orcid{0000-0003-3399-3574}\inst{\ref{aff25},\ref{aff26},\ref{aff27}}
\and V.~F.~Cardone\inst{\ref{aff17},\ref{aff18}}
\and P.~Carrilho\orcid{0000-0003-1339-0194}\inst{\ref{aff7}}
\and S.~Di~Domizio\orcid{0000-0003-2863-5895}\inst{\ref{aff28},\ref{aff29}}
\and R.~Durrer\orcid{0000-0001-9833-2086}\inst{\ref{aff2}}
\and S.~Farrens\orcid{0000-0002-9594-9387}\inst{\ref{aff1}}
\and S.~Gouyou~Beauchamps\inst{\ref{aff30},\ref{aff31}}
\and S.~Joudaki\orcid{0000-0001-8820-673X}\inst{\ref{aff32},\ref{aff33}}
\and C.~Moretti\orcid{0000-0003-3314-8936}\inst{\ref{aff34},\ref{aff35},\ref{aff36},\ref{aff37},\ref{aff38}}
\and A.~Pezzotta\orcid{0000-0003-0726-2268}\inst{\ref{aff39},\ref{aff40}}
\and A.~G.~S\'anchez\orcid{0000-0003-1198-831X}\inst{\ref{aff40}}
\and D.~Sciotti\orcid{0009-0008-4519-2620}\inst{\ref{aff17},\ref{aff18}}
\and K.~Tanidis\orcid{0000-0001-9843-5130}\inst{\ref{aff41}}
\and A.~Amara\inst{\ref{aff42}}
\and S.~Andreon\orcid{0000-0002-2041-8784}\inst{\ref{aff43}}
\and N.~Auricchio\orcid{0000-0003-4444-8651}\inst{\ref{aff6}}
\and C.~Baccigalupi\orcid{0000-0002-8211-1630}\inst{\ref{aff37},\ref{aff36},\ref{aff38},\ref{aff34}}
\and D.~Bagot\inst{\ref{aff44}}
\and M.~Baldi\orcid{0000-0003-4145-1943}\inst{\ref{aff45},\ref{aff6},\ref{aff46}}
\and S.~Bardelli\orcid{0000-0002-8900-0298}\inst{\ref{aff6}}
\and P.~Battaglia\orcid{0000-0002-7337-5909}\inst{\ref{aff6}}
\and A.~Biviano\orcid{0000-0002-0857-0732}\inst{\ref{aff36},\ref{aff37}}
\and E.~Branchini\orcid{0000-0002-0808-6908}\inst{\ref{aff28},\ref{aff29},\ref{aff43}}
\and M.~Brescia\orcid{0000-0001-9506-5680}\inst{\ref{aff47},\ref{aff48}}
\and V.~Capobianco\orcid{0000-0002-3309-7692}\inst{\ref{aff27}}
\and C.~Carbone\orcid{0000-0003-0125-3563}\inst{\ref{aff24}}
\and J.~Carretero\orcid{0000-0002-3130-0204}\inst{\ref{aff32},\ref{aff49}}
\and M.~Castellano\orcid{0000-0001-9875-8263}\inst{\ref{aff17}}
\and G.~Castignani\orcid{0000-0001-6831-0687}\inst{\ref{aff6}}
\and S.~Cavuoti\orcid{0000-0002-3787-4196}\inst{\ref{aff48},\ref{aff50}}
\and K.~C.~Chambers\orcid{0000-0001-6965-7789}\inst{\ref{aff51}}
\and A.~Cimatti\inst{\ref{aff52}}
\and C.~Colodro-Conde\inst{\ref{aff53}}
\and G.~Congedo\orcid{0000-0003-2508-0046}\inst{\ref{aff7}}
\and C.~J.~Conselice\orcid{0000-0003-1949-7638}\inst{\ref{aff54}}
\and L.~Conversi\orcid{0000-0002-6710-8476}\inst{\ref{aff55},\ref{aff56}}
\and Y.~Copin\orcid{0000-0002-5317-7518}\inst{\ref{aff57}}
\and F.~Courbin\orcid{0000-0003-0758-6510}\inst{\ref{aff58},\ref{aff59}}
\and H.~M.~Courtois\orcid{0000-0003-0509-1776}\inst{\ref{aff60}}
\and M.~Cropper\orcid{0000-0003-4571-9468}\inst{\ref{aff61}}
\and A.~Da~Silva\orcid{0000-0002-6385-1609}\inst{\ref{aff62},\ref{aff63}}
\and H.~Degaudenzi\orcid{0000-0002-5887-6799}\inst{\ref{aff64}}
\and S.~de~la~Torre\inst{\ref{aff65}}
\and G.~De~Lucia\orcid{0000-0002-6220-9104}\inst{\ref{aff36}}
\and H.~Dole\orcid{0000-0002-9767-3839}\inst{\ref{aff66}}
\and M.~Douspis\orcid{0000-0003-4203-3954}\inst{\ref{aff66}}
\and F.~Dubath\orcid{0000-0002-6533-2810}\inst{\ref{aff64}}
\and X.~Dupac\inst{\ref{aff56}}
\and S.~Escoffier\orcid{0000-0002-2847-7498}\inst{\ref{aff67}}
\and M.~Farina\orcid{0000-0002-3089-7846}\inst{\ref{aff68}}
\and F.~Faustini\orcid{0000-0001-6274-5145}\inst{\ref{aff17},\ref{aff69}}
\and S.~Ferriol\inst{\ref{aff57}}
\and F.~Finelli\orcid{0000-0002-6694-3269}\inst{\ref{aff6},\ref{aff70}}
\and P.~Fosalba\orcid{0000-0002-1510-5214}\inst{\ref{aff30},\ref{aff31}}
\and S.~Fotopoulou\orcid{0000-0002-9686-254X}\inst{\ref{aff71}}
\and M.~Frailis\orcid{0000-0002-7400-2135}\inst{\ref{aff36}}
\and E.~Franceschi\orcid{0000-0002-0585-6591}\inst{\ref{aff6}}
\and M.~Fumana\orcid{0000-0001-6787-5950}\inst{\ref{aff24}}
\and S.~Galeotta\orcid{0000-0002-3748-5115}\inst{\ref{aff36}}
\and B.~Gillis\orcid{0000-0002-4478-1270}\inst{\ref{aff7}}
\and C.~Giocoli\orcid{0000-0002-9590-7961}\inst{\ref{aff6},\ref{aff46}}
\and J.~Gracia-Carpio\inst{\ref{aff40}}
\and A.~Grazian\orcid{0000-0002-5688-0663}\inst{\ref{aff72}}
\and F.~Grupp\inst{\ref{aff40},\ref{aff73}}
\and L.~Guzzo\orcid{0000-0001-8264-5192}\inst{\ref{aff74},\ref{aff43},\ref{aff75}}
\and H.~Hoekstra\orcid{0000-0002-0641-3231}\inst{\ref{aff10}}
\and W.~Holmes\inst{\ref{aff76}}
\and F.~Hormuth\inst{\ref{aff77}}
\and A.~Hornstrup\orcid{0000-0002-3363-0936}\inst{\ref{aff78},\ref{aff79}}
\and K.~Jahnke\orcid{0000-0003-3804-2137}\inst{\ref{aff80}}
\and M.~Jhabvala\inst{\ref{aff81}}
\and B.~Joachimi\orcid{0000-0001-7494-1303}\inst{\ref{aff82}}
\and E.~Keih\"anen\orcid{0000-0003-1804-7715}\inst{\ref{aff83}}
\and S.~Kermiche\orcid{0000-0002-0302-5735}\inst{\ref{aff67}}
\and A.~Kiessling\orcid{0000-0002-2590-1273}\inst{\ref{aff76}}
\and M.~Kilbinger\orcid{0000-0001-9513-7138}\inst{\ref{aff1}}
\and B.~Kubik\orcid{0009-0006-5823-4880}\inst{\ref{aff57}}
\and M.~K\"ummel\orcid{0000-0003-2791-2117}\inst{\ref{aff73}}
\and H.~Kurki-Suonio\orcid{0000-0002-4618-3063}\inst{\ref{aff84},\ref{aff85}}
\and O.~Lahav\orcid{0000-0002-1134-9035}\inst{\ref{aff82}}
\and S.~Ligori\orcid{0000-0003-4172-4606}\inst{\ref{aff27}}
\and P.~B.~Lilje\orcid{0000-0003-4324-7794}\inst{\ref{aff86}}
\and V.~Lindholm\orcid{0000-0003-2317-5471}\inst{\ref{aff84},\ref{aff85}}
\and I.~Lloro\orcid{0000-0001-5966-1434}\inst{\ref{aff87}}
\and G.~Mainetti\orcid{0000-0003-2384-2377}\inst{\ref{aff88}}
\and D.~Maino\inst{\ref{aff74},\ref{aff24},\ref{aff75}}
\and E.~Maiorano\orcid{0000-0003-2593-4355}\inst{\ref{aff6}}
\and O.~Mansutti\orcid{0000-0001-5758-4658}\inst{\ref{aff36}}
\and O.~Marggraf\orcid{0000-0001-7242-3852}\inst{\ref{aff89}}
\and K.~Markovic\orcid{0000-0001-6764-073X}\inst{\ref{aff76}}
\and N.~Martinet\orcid{0000-0003-2786-7790}\inst{\ref{aff65}}
\and F.~Marulli\orcid{0000-0002-8850-0303}\inst{\ref{aff90},\ref{aff6},\ref{aff46}}
\and R.~Massey\orcid{0000-0002-6085-3780}\inst{\ref{aff91}}
\and E.~Medinaceli\orcid{0000-0002-4040-7783}\inst{\ref{aff6}}
\and S.~Mei\orcid{0000-0002-2849-559X}\inst{\ref{aff92},\ref{aff93}}
\and Y.~Mellier\inst{\ref{aff94},\ref{aff95}}
\and M.~Meneghetti\orcid{0000-0003-1225-7084}\inst{\ref{aff6},\ref{aff46}}
\and E.~Merlin\orcid{0000-0001-6870-8900}\inst{\ref{aff17}}
\and G.~Meylan\inst{\ref{aff96}}
\and A.~Mora\orcid{0000-0002-1922-8529}\inst{\ref{aff97}}
\and M.~Moresco\orcid{0000-0002-7616-7136}\inst{\ref{aff90},\ref{aff6}}
\and L.~Moscardini\orcid{0000-0002-3473-6716}\inst{\ref{aff90},\ref{aff6},\ref{aff46}}
\and C.~Neissner\orcid{0000-0001-8524-4968}\inst{\ref{aff98},\ref{aff49}}
\and S.-M.~Niemi\orcid{0009-0005-0247-0086}\inst{\ref{aff8}}
\and C.~Padilla\orcid{0000-0001-7951-0166}\inst{\ref{aff98}}
\and S.~Paltani\orcid{0000-0002-8108-9179}\inst{\ref{aff64}}
\and F.~Pasian\orcid{0000-0002-4869-3227}\inst{\ref{aff36}}
\and K.~Pedersen\inst{\ref{aff99}}
\and W.~J.~Percival\orcid{0000-0002-0644-5727}\inst{\ref{aff23},\ref{aff100},\ref{aff101}}
\and V.~Pettorino\inst{\ref{aff8}}
\and S.~Pires\orcid{0000-0002-0249-2104}\inst{\ref{aff1}}
\and G.~Polenta\orcid{0000-0003-4067-9196}\inst{\ref{aff69}}
\and M.~Poncet\inst{\ref{aff44}}
\and L.~A.~Popa\inst{\ref{aff102}}
\and F.~Raison\orcid{0000-0002-7819-6918}\inst{\ref{aff40}}
\and R.~Rebolo\orcid{0000-0003-3767-7085}\inst{\ref{aff53},\ref{aff103},\ref{aff104}}
\and A.~Renzi\orcid{0000-0001-9856-1970}\inst{\ref{aff105},\ref{aff106}}
\and J.~Rhodes\orcid{0000-0002-4485-8549}\inst{\ref{aff76}}
\and G.~Riccio\inst{\ref{aff48}}
\and E.~Romelli\orcid{0000-0003-3069-9222}\inst{\ref{aff36}}
\and M.~Roncarelli\orcid{0000-0001-9587-7822}\inst{\ref{aff6}}
\and R.~Saglia\orcid{0000-0003-0378-7032}\inst{\ref{aff73},\ref{aff40}}
\and D.~Sapone\orcid{0000-0001-7089-4503}\inst{\ref{aff107}}
\and B.~Sartoris\orcid{0000-0003-1337-5269}\inst{\ref{aff73},\ref{aff36}}
\and J.~A.~Schewtschenko\orcid{0000-0002-4913-6393}\inst{\ref{aff7}}
\and T.~Schrabback\orcid{0000-0002-6987-7834}\inst{\ref{aff108}}
\and A.~Secroun\orcid{0000-0003-0505-3710}\inst{\ref{aff67}}
\and E.~Sefusatti\orcid{0000-0003-0473-1567}\inst{\ref{aff36},\ref{aff37},\ref{aff38}}
\and G.~Seidel\orcid{0000-0003-2907-353X}\inst{\ref{aff80}}
\and M.~Seiffert\orcid{0000-0002-7536-9393}\inst{\ref{aff76}}
\and P.~Simon\inst{\ref{aff89}}
\and C.~Sirignano\orcid{0000-0002-0995-7146}\inst{\ref{aff105},\ref{aff106}}
\and G.~Sirri\orcid{0000-0003-2626-2853}\inst{\ref{aff46}}
\and A.~Spurio~Mancini\orcid{0000-0001-5698-0990}\inst{\ref{aff109}}
\and L.~Stanco\orcid{0000-0002-9706-5104}\inst{\ref{aff106}}
\and J.~Steinwagner\orcid{0000-0001-7443-1047}\inst{\ref{aff40}}
\and P.~Tallada-Cresp\'{i}\orcid{0000-0002-1336-8328}\inst{\ref{aff32},\ref{aff49}}
\and A.~N.~Taylor\inst{\ref{aff7}}
\and I.~Tereno\orcid{0000-0002-4537-6218}\inst{\ref{aff62},\ref{aff110}}
\and S.~Toft\orcid{0000-0003-3631-7176}\inst{\ref{aff111},\ref{aff112}}
\and R.~Toledo-Moreo\orcid{0000-0002-2997-4859}\inst{\ref{aff113}}
\and F.~Torradeflot\orcid{0000-0003-1160-1517}\inst{\ref{aff49},\ref{aff32}}
\and A.~Tsyganov\inst{\ref{aff114}}
\and J.~Valiviita\orcid{0000-0001-6225-3693}\inst{\ref{aff84},\ref{aff85}}
\and T.~Vassallo\orcid{0000-0001-6512-6358}\inst{\ref{aff73},\ref{aff36}}
\and G.~Verdoes~Kleijn\orcid{0000-0001-5803-2580}\inst{\ref{aff115}}
\and A.~Veropalumbo\orcid{0000-0003-2387-1194}\inst{\ref{aff43},\ref{aff29},\ref{aff28}}
\and Y.~Wang\orcid{0000-0002-4749-2984}\inst{\ref{aff116}}
\and J.~Weller\orcid{0000-0002-8282-2010}\inst{\ref{aff73},\ref{aff40}}
\and G.~Zamorani\orcid{0000-0002-2318-301X}\inst{\ref{aff6}}
\and E.~Zucca\orcid{0000-0002-5845-8132}\inst{\ref{aff6}}
\and M.~Bolzonella\orcid{0000-0003-3278-4607}\inst{\ref{aff6}}
\and E.~Bozzo\orcid{0000-0002-8201-1525}\inst{\ref{aff64}}
\and C.~Burigana\orcid{0000-0002-3005-5796}\inst{\ref{aff117},\ref{aff70}}
\and R.~Cabanac\orcid{0000-0001-6679-2600}\inst{\ref{aff14}}
\and M.~Calabrese\orcid{0000-0002-2637-2422}\inst{\ref{aff118},\ref{aff24}}
\and A.~Cappi\inst{\ref{aff6},\ref{aff119}}
\and D.~Di~Ferdinando\inst{\ref{aff46}}
\and J.~A.~Escartin~Vigo\inst{\ref{aff40}}
\and L.~Gabarra\orcid{0000-0002-8486-8856}\inst{\ref{aff41}}
\and W.~G.~Hartley\inst{\ref{aff64}}
\and J.~Mart\'{i}n-Fleitas\orcid{0000-0002-8594-569X}\inst{\ref{aff120}}
\and M.~Maturi\orcid{0000-0002-3517-2422}\inst{\ref{aff19},\ref{aff121}}
\and N.~Mauri\orcid{0000-0001-8196-1548}\inst{\ref{aff52},\ref{aff46}}
\and R.~B.~Metcalf\orcid{0000-0003-3167-2574}\inst{\ref{aff90},\ref{aff6}}
\and M.~P\"ontinen\orcid{0000-0001-5442-2530}\inst{\ref{aff84}}
\and C.~Porciani\orcid{0000-0002-7797-2508}\inst{\ref{aff89}}
\and I.~Risso\orcid{0000-0003-2525-7761}\inst{\ref{aff122}}
\and V.~Scottez\inst{\ref{aff94},\ref{aff123}}
\and M.~Sereno\orcid{0000-0003-0302-0325}\inst{\ref{aff6},\ref{aff46}}
\and M.~Tenti\orcid{0000-0002-4254-5901}\inst{\ref{aff46}}
\and M.~Viel\orcid{0000-0002-2642-5707}\inst{\ref{aff37},\ref{aff36},\ref{aff34},\ref{aff38},\ref{aff35}}
\and M.~Wiesmann\orcid{0009-0000-8199-5860}\inst{\ref{aff86}}
\and Y.~Akrami\orcid{0000-0002-2407-7956}\inst{\ref{aff124},\ref{aff125}}
\and I.~T.~Andika\orcid{0000-0001-6102-9526}\inst{\ref{aff126},\ref{aff127}}
\and S.~Anselmi\orcid{0000-0002-3579-9583}\inst{\ref{aff106},\ref{aff105},\ref{aff128}}
\and M.~Archidiacono\orcid{0000-0003-4952-9012}\inst{\ref{aff74},\ref{aff75}}
\and F.~Atrio-Barandela\orcid{0000-0002-2130-2513}\inst{\ref{aff129}}
\and A.~Balaguera-Antolinez\orcid{0000-0001-5028-3035}\inst{\ref{aff53},\ref{aff130}}
\and D.~Bertacca\orcid{0000-0002-2490-7139}\inst{\ref{aff105},\ref{aff72},\ref{aff106}}
\and M.~Bethermin\orcid{0000-0002-3915-2015}\inst{\ref{aff131}}
\and A.~Blanchard\orcid{0000-0001-8555-9003}\inst{\ref{aff14}}
\and H.~B\"ohringer\orcid{0000-0001-8241-4204}\inst{\ref{aff40},\ref{aff132},\ref{aff133}}
\and S.~Borgani\orcid{0000-0001-6151-6439}\inst{\ref{aff134},\ref{aff37},\ref{aff36},\ref{aff38},\ref{aff35}}
\and M.~L.~Brown\orcid{0000-0002-0370-8077}\inst{\ref{aff54}}
\and S.~Bruton\orcid{0000-0002-6503-5218}\inst{\ref{aff135}}
\and A.~Calabro\orcid{0000-0003-2536-1614}\inst{\ref{aff17}}
\and B.~Camacho~Quevedo\orcid{0000-0002-8789-4232}\inst{\ref{aff37},\ref{aff34},\ref{aff36},\ref{aff30},\ref{aff31}}
\and F.~Caro\inst{\ref{aff17}}
\and C.~S.~Carvalho\inst{\ref{aff110}}
\and T.~Castro\orcid{0000-0002-6292-3228}\inst{\ref{aff36},\ref{aff38},\ref{aff37},\ref{aff35}}
\and F.~Cogato\orcid{0000-0003-4632-6113}\inst{\ref{aff90},\ref{aff6}}
\and S.~Conseil\orcid{0000-0002-3657-4191}\inst{\ref{aff57}}
\and S.~Contarini\orcid{0000-0002-9843-723X}\inst{\ref{aff40}}
\and A.~R.~Cooray\orcid{0000-0002-3892-0190}\inst{\ref{aff136}}
\and O.~Cucciati\orcid{0000-0002-9336-7551}\inst{\ref{aff6}}
\and S.~Davini\orcid{0000-0003-3269-1718}\inst{\ref{aff29}}
\and F.~De~Paolis\orcid{0000-0001-6460-7563}\inst{\ref{aff137},\ref{aff138},\ref{aff139}}
\and G.~Desprez\orcid{0000-0001-8325-1742}\inst{\ref{aff115}}
\and A.~D\'iaz-S\'anchez\orcid{0000-0003-0748-4768}\inst{\ref{aff140}}
\and J.~J.~Diaz\orcid{0000-0003-2101-1078}\inst{\ref{aff53}}
\and J.~M.~Diego\orcid{0000-0001-9065-3926}\inst{\ref{aff3}}
\and P.~Dimauro\orcid{0000-0001-7399-2854}\inst{\ref{aff17},\ref{aff141}}
\and A.~Enia\orcid{0000-0002-0200-2857}\inst{\ref{aff45},\ref{aff6}}
\and Y.~Fang\inst{\ref{aff73}}
\and A.~G.~Ferrari\orcid{0009-0005-5266-4110}\inst{\ref{aff46}}
\and P.~G.~Ferreira\orcid{0000-0002-3021-2851}\inst{\ref{aff41}}
\and A.~Finoguenov\orcid{0000-0002-4606-5403}\inst{\ref{aff84}}
\and A.~Franco\orcid{0000-0002-4761-366X}\inst{\ref{aff138},\ref{aff137},\ref{aff139}}
\and K.~Ganga\orcid{0000-0001-8159-8208}\inst{\ref{aff92}}
\and J.~Garc\'ia-Bellido\orcid{0000-0002-9370-8360}\inst{\ref{aff124}}
\and T.~Gasparetto\orcid{0000-0002-7913-4866}\inst{\ref{aff36}}
\and E.~Gaztanaga\orcid{0000-0001-9632-0815}\inst{\ref{aff31},\ref{aff30},\ref{aff33}}
\and F.~Giacomini\orcid{0000-0002-3129-2814}\inst{\ref{aff46}}
\and F.~Gianotti\orcid{0000-0003-4666-119X}\inst{\ref{aff6}}
\and G.~Gozaliasl\orcid{0000-0002-0236-919X}\inst{\ref{aff142},\ref{aff84}}
\and A.~Gruppuso\orcid{0000-0001-9272-5292}\inst{\ref{aff6},\ref{aff46}}
\and M.~Guidi\orcid{0000-0001-9408-1101}\inst{\ref{aff45},\ref{aff6}}
\and C.~M.~Gutierrez\orcid{0000-0001-7854-783X}\inst{\ref{aff143}}
\and H.~Hildebrandt\orcid{0000-0002-9814-3338}\inst{\ref{aff144}}
\and J.~Hjorth\orcid{0000-0002-4571-2306}\inst{\ref{aff99}}
\and J.~J.~E.~Kajava\orcid{0000-0002-3010-8333}\inst{\ref{aff145},\ref{aff146}}
\and Y.~Kang\orcid{0009-0000-8588-7250}\inst{\ref{aff64}}
\and V.~Kansal\orcid{0000-0002-4008-6078}\inst{\ref{aff147},\ref{aff148}}
\and D.~Karagiannis\orcid{0000-0002-4927-0816}\inst{\ref{aff4},\ref{aff149}}
\and K.~Kiiveri\inst{\ref{aff83}}
\and C.~C.~Kirkpatrick\inst{\ref{aff83}}
\and S.~Kruk\orcid{0000-0001-8010-8879}\inst{\ref{aff56}}
\and F.~Lacasa\orcid{0000-0002-7268-3440}\inst{\ref{aff150},\ref{aff66}}
\and M.~Lattanzi\orcid{0000-0003-1059-2532}\inst{\ref{aff5}}
\and V.~Le~Brun\orcid{0000-0002-5027-1939}\inst{\ref{aff65}}
\and L.~Legrand\orcid{0000-0003-0610-5252}\inst{\ref{aff151},\ref{aff152}}
\and M.~Lembo\orcid{0000-0002-5271-5070}\inst{\ref{aff95},\ref{aff5}}
\and G.~Leroy\orcid{0009-0004-2523-4425}\inst{\ref{aff153},\ref{aff91}}
\and J.~Lesgourgues\orcid{0000-0001-7627-353X}\inst{\ref{aff11}}
\and L.~Leuzzi\orcid{0009-0006-4479-7017}\inst{\ref{aff90},\ref{aff6}}
\and T.~I.~Liaudat\orcid{0000-0002-9104-314X}\inst{\ref{aff154}}
\and S.~J.~Liu\orcid{0000-0001-7680-2139}\inst{\ref{aff68}}
\and A.~Loureiro\orcid{0000-0002-4371-0876}\inst{\ref{aff155},\ref{aff156}}
\and J.~Macias-Perez\orcid{0000-0002-5385-2763}\inst{\ref{aff157}}
\and G.~Maggio\orcid{0000-0003-4020-4836}\inst{\ref{aff36}}
\and M.~Magliocchetti\orcid{0000-0001-9158-4838}\inst{\ref{aff68}}
\and F.~Mannucci\orcid{0000-0002-4803-2381}\inst{\ref{aff158}}
\and R.~Maoli\orcid{0000-0002-6065-3025}\inst{\ref{aff159},\ref{aff17}}
\and C.~J.~A.~P.~Martins\orcid{0000-0002-4886-9261}\inst{\ref{aff160},\ref{aff161}}
\and L.~Maurin\orcid{0000-0002-8406-0857}\inst{\ref{aff66}}
\and M.~Miluzio\inst{\ref{aff56},\ref{aff162}}
\and P.~Monaco\orcid{0000-0003-2083-7564}\inst{\ref{aff134},\ref{aff36},\ref{aff38},\ref{aff37}}
\and G.~Morgante\inst{\ref{aff6}}
\and S.~Nadathur\orcid{0000-0001-9070-3102}\inst{\ref{aff33}}
\and K.~Naidoo\orcid{0000-0002-9182-1802}\inst{\ref{aff33}}
\and A.~Navarro-Alsina\orcid{0000-0002-3173-2592}\inst{\ref{aff89}}
\and S.~Nesseris\orcid{0000-0002-0567-0324}\inst{\ref{aff124}}
\and L.~Pagano\orcid{0000-0003-1820-5998}\inst{\ref{aff4},\ref{aff5}}
\and F.~Passalacqua\orcid{0000-0002-8606-4093}\inst{\ref{aff105},\ref{aff106}}
\and K.~Paterson\orcid{0000-0001-8340-3486}\inst{\ref{aff80}}
\and L.~Patrizii\inst{\ref{aff46}}
\and D.~Potter\orcid{0000-0002-0757-5195}\inst{\ref{aff16}}
\and A.~Pourtsidou\orcid{0000-0001-9110-5550}\inst{\ref{aff7},\ref{aff163}}
\and S.~Quai\orcid{0000-0002-0449-8163}\inst{\ref{aff90},\ref{aff6}}
\and M.~Radovich\orcid{0000-0002-3585-866X}\inst{\ref{aff72}}
\and P.-F.~Rocci\inst{\ref{aff66}}
\and S.~Sacquegna\orcid{0000-0002-8433-6630}\inst{\ref{aff137},\ref{aff138},\ref{aff139}}
\and M.~Sahl\'en\orcid{0000-0003-0973-4804}\inst{\ref{aff164}}
\and D.~B.~Sanders\orcid{0000-0002-1233-9998}\inst{\ref{aff51}}
\and E.~Sarpa\orcid{0000-0002-1256-655X}\inst{\ref{aff34},\ref{aff35},\ref{aff38}}
\and J.~Schaye\orcid{0000-0002-0668-5560}\inst{\ref{aff10}}
\and A.~Schneider\orcid{0000-0001-7055-8104}\inst{\ref{aff16}}
\and M.~Schultheis\inst{\ref{aff119}}
\and E.~Sellentin\inst{\ref{aff165},\ref{aff10}}
\and C.~Tao\orcid{0000-0001-7961-8177}\inst{\ref{aff67}}
\and G.~Testera\inst{\ref{aff29}}
\and R.~Teyssier\orcid{0000-0001-7689-0933}\inst{\ref{aff166}}
\and S.~Tosi\orcid{0000-0002-7275-9193}\inst{\ref{aff28},\ref{aff29},\ref{aff43}}
\and A.~Troja\orcid{0000-0003-0239-4595}\inst{\ref{aff105},\ref{aff106}}
\and M.~Tucci\inst{\ref{aff64}}
\and C.~Valieri\inst{\ref{aff46}}
\and A.~Venhola\orcid{0000-0001-6071-4564}\inst{\ref{aff167}}
\and D.~Vergani\orcid{0000-0003-0898-2216}\inst{\ref{aff6}}
\and F.~Vernizzi\orcid{0000-0003-3426-2802}\inst{\ref{aff168}}
\and G.~Verza\orcid{0000-0002-1886-8348}\inst{\ref{aff169}}
\and N.~A.~Walton\orcid{0000-0003-3983-8778}\inst{\ref{aff170}}}
                                                                                   
\institute{Universit\'e Paris-Saclay, Universit\'e Paris Cit\'e, CEA, CNRS, AIM, 91191, Gif-sur-Yvette, France\label{aff1}
\and
Universit\'e de Gen\`eve, D\'epartement de Physique Th\'eorique and Centre for Astroparticle Physics, 24 quai Ernest-Ansermet, CH-1211 Gen\`eve 4, Switzerland\label{aff2}
\and
Instituto de F\'isica de Cantabria, Edificio Juan Jord\'a, Avenida de los Castros, 39005 Santander, Spain\label{aff3}
\and
Dipartimento di Fisica e Scienze della Terra, Universit\`a degli Studi di Ferrara, Via Giuseppe Saragat 1, 44122 Ferrara, Italy\label{aff4}
\and
Istituto Nazionale di Fisica Nucleare, Sezione di Ferrara, Via Giuseppe Saragat 1, 44122 Ferrara, Italy\label{aff5}
\and
INAF-Osservatorio di Astrofisica e Scienza dello Spazio di Bologna, Via Piero Gobetti 93/3, 40129 Bologna, Italy\label{aff6}
\and
Institute for Astronomy, University of Edinburgh, Royal Observatory, Blackford Hill, Edinburgh EH9 3HJ, UK\label{aff7}
\and
European Space Agency/ESTEC, Keplerlaan 1, 2201 AZ Noordwijk, The Netherlands\label{aff8}
\and
Institute Lorentz, Leiden University, Niels Bohrweg 2, 2333 CA Leiden, The Netherlands\label{aff9}
\and
Leiden Observatory, Leiden University, Einsteinweg 55, 2333 CC Leiden, The Netherlands\label{aff10}
\and
Institute for Theoretical Particle Physics and Cosmology (TTK), RWTH Aachen University, 52056 Aachen, Germany\label{aff11}
\and
GRAPPA Institute, Institute for Theoretical Physics Amsterdam, University of Amsterdam, Science Park 904, 1098 XH Amsterdam, The Netherlands\label{aff12}
\and
Universit\'e Paris-Saclay, CNRS/IN2P3, IJCLab, 91405 Orsay, France\label{aff13}
\and
Institut de Recherche en Astrophysique et Plan\'etologie (IRAP), Universit\'e de Toulouse, CNRS, UPS, CNES, 14 Av. Edouard Belin, 31400 Toulouse, France\label{aff14}
\and
Laboratoire d'etude de l'Univers et des phenomenes eXtremes, Observatoire de Paris, Universit\'e PSL, Sorbonne Universit\'e, CNRS, 92190 Meudon, France\label{aff15}
\and
Department of Astrophysics, University of Zurich, Winterthurerstrasse 190, 8057 Zurich, Switzerland\label{aff16}
\and
INAF-Osservatorio Astronomico di Roma, Via Frascati 33, 00078 Monteporzio Catone, Italy\label{aff17}
\and
INFN-Sezione di Roma, Piazzale Aldo Moro, 2 - c/o Dipartimento di Fisica, Edificio G. Marconi, 00185 Roma, Italy\label{aff18}
\and
Institut f\"ur Theoretische Physik, University of Heidelberg, Philosophenweg 16, 69120 Heidelberg, Germany\label{aff19}
\and
Universit\'e St Joseph; Faculty of Sciences, Beirut, Lebanon\label{aff20}
\and
Laboratoire univers et particules de Montpellier, Universit\'e de Montpellier, CNRS, 34090 Montpellier, France\label{aff21}
\and
Center for Data-Driven Discovery, Kavli IPMU (WPI), UTIAS, The University of Tokyo, Kashiwa, Chiba 277-8583, Japan\label{aff22}
\and
Waterloo Centre for Astrophysics, University of Waterloo, Waterloo, Ontario N2L 3G1, Canada\label{aff23}
\and
INAF-IASF Milano, Via Alfonso Corti 12, 20133 Milano, Italy\label{aff24}
\and
Dipartimento di Fisica, Universit\`a degli Studi di Torino, Via P. Giuria 1, 10125 Torino, Italy\label{aff25}
\and
INFN-Sezione di Torino, Via P. Giuria 1, 10125 Torino, Italy\label{aff26}
\and
INAF-Osservatorio Astrofisico di Torino, Via Osservatorio 20, 10025 Pino Torinese (TO), Italy\label{aff27}
\and
Dipartimento di Fisica, Universit\`a di Genova, Via Dodecaneso 33, 16146, Genova, Italy\label{aff28}
\and
INFN-Sezione di Genova, Via Dodecaneso 33, 16146, Genova, Italy\label{aff29}
\and
Institut d'Estudis Espacials de Catalunya (IEEC),  Edifici RDIT, Campus UPC, 08860 Castelldefels, Barcelona, Spain\label{aff30}
\and
Institute of Space Sciences (ICE, CSIC), Campus UAB, Carrer de Can Magrans, s/n, 08193 Barcelona, Spain\label{aff31}
\and
Centro de Investigaciones Energ\'eticas, Medioambientales y Tecnol\'ogicas (CIEMAT), Avenida Complutense 40, 28040 Madrid, Spain\label{aff32}
\and
Institute of Cosmology and Gravitation, University of Portsmouth, Portsmouth PO1 3FX, UK\label{aff33}
\and
SISSA, International School for Advanced Studies, Via Bonomea 265, 34136 Trieste TS, Italy\label{aff34}
\and
ICSC - Centro Nazionale di Ricerca in High Performance Computing, Big Data e Quantum Computing, Via Magnanelli 2, Bologna, Italy\label{aff35}
\and
INAF-Osservatorio Astronomico di Trieste, Via G. B. Tiepolo 11, 34143 Trieste, Italy\label{aff36}
\and
IFPU, Institute for Fundamental Physics of the Universe, via Beirut 2, 34151 Trieste, Italy\label{aff37}
\and
INFN, Sezione di Trieste, Via Valerio 2, 34127 Trieste TS, Italy\label{aff38}
\and
INAF - Osservatorio Astronomico di Brera, via Emilio Bianchi 46, 23807 Merate, Italy\label{aff39}
\and
Max Planck Institute for Extraterrestrial Physics, Giessenbachstr. 1, 85748 Garching, Germany\label{aff40}
\and
Department of Physics, Oxford University, Keble Road, Oxford OX1 3RH, UK\label{aff41}
\and
School of Mathematics and Physics, University of Surrey, Guildford, Surrey, GU2 7XH, UK\label{aff42}
\and
INAF-Osservatorio Astronomico di Brera, Via Brera 28, 20122 Milano, Italy\label{aff43}
\and
Centre National d'Etudes Spatiales -- Centre spatial de Toulouse, 18 avenue Edouard Belin, 31401 Toulouse Cedex 9, France\label{aff44}
\and
Dipartimento di Fisica e Astronomia, Universit\`a di Bologna, Via Gobetti 93/2, 40129 Bologna, Italy\label{aff45}
\and
INFN-Sezione di Bologna, Viale Berti Pichat 6/2, 40127 Bologna, Italy\label{aff46}
\and
Department of Physics "E. Pancini", University Federico II, Via Cinthia 6, 80126, Napoli, Italy\label{aff47}
\and
INAF-Osservatorio Astronomico di Capodimonte, Via Moiariello 16, 80131 Napoli, Italy\label{aff48}
\and
Port d'Informaci\'{o} Cient\'{i}fica, Campus UAB, C. Albareda s/n, 08193 Bellaterra (Barcelona), Spain\label{aff49}
\and
INFN section of Naples, Via Cinthia 6, 80126, Napoli, Italy\label{aff50}
\and
Institute for Astronomy, University of Hawaii, 2680 Woodlawn Drive, Honolulu, HI 96822, USA\label{aff51}
\and
Dipartimento di Fisica e Astronomia "Augusto Righi" - Alma Mater Studiorum Universit\`a di Bologna, Viale Berti Pichat 6/2, 40127 Bologna, Italy\label{aff52}
\and
Instituto de Astrof\'{\i}sica de Canarias, V\'{\i}a L\'actea, 38205 La Laguna, Tenerife, Spain\label{aff53}
\and
Jodrell Bank Centre for Astrophysics, Department of Physics and Astronomy, University of Manchester, Oxford Road, Manchester M13 9PL, UK\label{aff54}
\and
European Space Agency/ESRIN, Largo Galileo Galilei 1, 00044 Frascati, Roma, Italy\label{aff55}
\and
ESAC/ESA, Camino Bajo del Castillo, s/n., Urb. Villafranca del Castillo, 28692 Villanueva de la Ca\~nada, Madrid, Spain\label{aff56}
\and
Universit\'e Claude Bernard Lyon 1, CNRS/IN2P3, IP2I Lyon, UMR 5822, Villeurbanne, F-69100, France\label{aff57}
\and
Institut de Ci\`{e}ncies del Cosmos (ICCUB), Universitat de Barcelona (IEEC-UB), Mart\'{i} i Franqu\`{e}s 1, 08028 Barcelona, Spain\label{aff58}
\and
Instituci\'o Catalana de Recerca i Estudis Avan\c{c}ats (ICREA), Passeig de Llu\'{\i}s Companys 23, 08010 Barcelona, Spain\label{aff59}
\and
UCB Lyon 1, CNRS/IN2P3, IUF, IP2I Lyon, 4 rue Enrico Fermi, 69622 Villeurbanne, France\label{aff60}
\and
Mullard Space Science Laboratory, University College London, Holmbury St Mary, Dorking, Surrey RH5 6NT, UK\label{aff61}
\and
Departamento de F\'isica, Faculdade de Ci\^encias, Universidade de Lisboa, Edif\'icio C8, Campo Grande, PT1749-016 Lisboa, Portugal\label{aff62}
\and
Instituto de Astrof\'isica e Ci\^encias do Espa\c{c}o, Faculdade de Ci\^encias, Universidade de Lisboa, Campo Grande, 1749-016 Lisboa, Portugal\label{aff63}
\and
Department of Astronomy, University of Geneva, ch. d'Ecogia 16, 1290 Versoix, Switzerland\label{aff64}
\and
Aix-Marseille Universit\'e, CNRS, CNES, LAM, Marseille, France\label{aff65}
\and
Universit\'e Paris-Saclay, CNRS, Institut d'astrophysique spatiale, 91405, Orsay, France\label{aff66}
\and
Aix-Marseille Universit\'e, CNRS/IN2P3, CPPM, Marseille, France\label{aff67}
\and
INAF-Istituto di Astrofisica e Planetologia Spaziali, via del Fosso del Cavaliere, 100, 00100 Roma, Italy\label{aff68}
\and
Space Science Data Center, Italian Space Agency, via del Politecnico snc, 00133 Roma, Italy\label{aff69}
\and
INFN-Bologna, Via Irnerio 46, 40126 Bologna, Italy\label{aff70}
\and
School of Physics, HH Wills Physics Laboratory, University of Bristol, Tyndall Avenue, Bristol, BS8 1TL, UK\label{aff71}
\and
INAF-Osservatorio Astronomico di Padova, Via dell'Osservatorio 5, 35122 Padova, Italy\label{aff72}
\and
Universit\"ats-Sternwarte M\"unchen, Fakult\"at f\"ur Physik, Ludwig-Maximilians-Universit\"at M\"unchen, Scheinerstrasse 1, 81679 M\"unchen, Germany\label{aff73}
\and
Dipartimento di Fisica "Aldo Pontremoli", Universit\`a degli Studi di Milano, Via Celoria 16, 20133 Milano, Italy\label{aff74}
\and
INFN-Sezione di Milano, Via Celoria 16, 20133 Milano, Italy\label{aff75}
\and
Jet Propulsion Laboratory, California Institute of Technology, 4800 Oak Grove Drive, Pasadena, CA, 91109, USA\label{aff76}
\and
Felix Hormuth Engineering, Goethestr. 17, 69181 Leimen, Germany\label{aff77}
\and
Technical University of Denmark, Elektrovej 327, 2800 Kgs. Lyngby, Denmark\label{aff78}
\and
Cosmic Dawn Center (DAWN), Denmark\label{aff79}
\and
Max-Planck-Institut f\"ur Astronomie, K\"onigstuhl 17, 69117 Heidelberg, Germany\label{aff80}
\and
NASA Goddard Space Flight Center, Greenbelt, MD 20771, USA\label{aff81}
\and
Department of Physics and Astronomy, University College London, Gower Street, London WC1E 6BT, UK\label{aff82}
\and
Department of Physics and Helsinki Institute of Physics, Gustaf H\"allstr\"omin katu 2, 00014 University of Helsinki, Finland\label{aff83}
\and
Department of Physics, P.O. Box 64, 00014 University of Helsinki, Finland\label{aff84}
\and
Helsinki Institute of Physics, Gustaf H{\"a}llstr{\"o}min katu 2, University of Helsinki, Helsinki, Finland\label{aff85}
\and
Institute of Theoretical Astrophysics, University of Oslo, P.O. Box 1029 Blindern, 0315 Oslo, Norway\label{aff86}
\and
SKA Observatory, Jodrell Bank, Lower Withington, Macclesfield, Cheshire SK11 9FT, UK\label{aff87}
\and
Centre de Calcul de l'IN2P3/CNRS, 21 avenue Pierre de Coubertin 69627 Villeurbanne Cedex, France\label{aff88}
\and
Universit\"at Bonn, Argelander-Institut f\"ur Astronomie, Auf dem H\"ugel 71, 53121 Bonn, Germany\label{aff89}
\and
Dipartimento di Fisica e Astronomia "Augusto Righi" - Alma Mater Studiorum Universit\`a di Bologna, via Piero Gobetti 93/2, 40129 Bologna, Italy\label{aff90}
\and
Department of Physics, Institute for Computational Cosmology, Durham University, South Road, Durham, DH1 3LE, UK\label{aff91}
\and
Universit\'e Paris Cit\'e, CNRS, Astroparticule et Cosmologie, 75013 Paris, France\label{aff92}
\and
CNRS-UCB International Research Laboratory, Centre Pierre Bin\'etruy, IRL2007, CPB-IN2P3, Berkeley, USA\label{aff93}
\and
Institut d'Astrophysique de Paris, 98bis Boulevard Arago, 75014, Paris, France\label{aff94}
\and
Institut d'Astrophysique de Paris, UMR 7095, CNRS, and Sorbonne Universit\'e, 98 bis boulevard Arago, 75014 Paris, France\label{aff95}
\and
Institute of Physics, Laboratory of Astrophysics, Ecole Polytechnique F\'ed\'erale de Lausanne (EPFL), Observatoire de Sauverny, 1290 Versoix, Switzerland\label{aff96}
\and
Telespazio UK S.L. for European Space Agency (ESA), Camino bajo del Castillo, s/n, Urbanizacion Villafranca del Castillo, Villanueva de la Ca\~nada, 28692 Madrid, Spain\label{aff97}
\and
Institut de F\'{i}sica d'Altes Energies (IFAE), The Barcelona Institute of Science and Technology, Campus UAB, 08193 Bellaterra (Barcelona), Spain\label{aff98}
\and
DARK, Niels Bohr Institute, University of Copenhagen, Jagtvej 155, 2200 Copenhagen, Denmark\label{aff99}
\and
Department of Physics and Astronomy, University of Waterloo, Waterloo, Ontario N2L 3G1, Canada\label{aff100}
\and
Perimeter Institute for Theoretical Physics, Waterloo, Ontario N2L 2Y5, Canada\label{aff101}
\and
Institute of Space Science, Str. Atomistilor, nr. 409 M\u{a}gurele, Ilfov, 077125, Romania\label{aff102}
\and
Consejo Superior de Investigaciones Cientificas, Calle Serrano 117, 28006 Madrid, Spain\label{aff103}
\and
Universidad de La Laguna, Departamento de Astrof\'{\i}sica, 38206 La Laguna, Tenerife, Spain\label{aff104}
\and
Dipartimento di Fisica e Astronomia "G. Galilei", Universit\`a di Padova, Via Marzolo 8, 35131 Padova, Italy\label{aff105}
\and
INFN-Padova, Via Marzolo 8, 35131 Padova, Italy\label{aff106}
\and
Departamento de F\'isica, FCFM, Universidad de Chile, Blanco Encalada 2008, Santiago, Chile\label{aff107}
\and
Universit\"at Innsbruck, Institut f\"ur Astro- und Teilchenphysik, Technikerstr. 25/8, 6020 Innsbruck, Austria\label{aff108}
\and
Department of Physics, Royal Holloway, University of London, TW20 0EX, UK\label{aff109}
\and
Instituto de Astrof\'isica e Ci\^encias do Espa\c{c}o, Faculdade de Ci\^encias, Universidade de Lisboa, Tapada da Ajuda, 1349-018 Lisboa, Portugal\label{aff110}
\and
Cosmic Dawn Center (DAWN)\label{aff111}
\and
Niels Bohr Institute, University of Copenhagen, Jagtvej 128, 2200 Copenhagen, Denmark\label{aff112}
\and
Universidad Polit\'ecnica de Cartagena, Departamento de Electr\'onica y Tecnolog\'ia de Computadoras,  Plaza del Hospital 1, 30202 Cartagena, Spain\label{aff113}
\and
Centre for Information Technology, University of Groningen, P.O. Box 11044, 9700 CA Groningen, The Netherlands\label{aff114}
\and
Kapteyn Astronomical Institute, University of Groningen, PO Box 800, 9700 AV Groningen, The Netherlands\label{aff115}
\and
Infrared Processing and Analysis Center, California Institute of Technology, Pasadena, CA 91125, USA\label{aff116}
\and
INAF, Istituto di Radioastronomia, Via Piero Gobetti 101, 40129 Bologna, Italy\label{aff117}
\and
Astronomical Observatory of the Autonomous Region of the Aosta Valley (OAVdA), Loc. Lignan 39, I-11020, Nus (Aosta Valley), Italy\label{aff118}
\and
Universit\'e C\^{o}te d'Azur, Observatoire de la C\^{o}te d'Azur, CNRS, Laboratoire Lagrange, Bd de l'Observatoire, CS 34229, 06304 Nice cedex 4, France\label{aff119}
\and
Aurora Technology for European Space Agency (ESA), Camino bajo del Castillo, s/n, Urbanizacion Villafranca del Castillo, Villanueva de la Ca\~nada, 28692 Madrid, Spain\label{aff120}
\and
Zentrum f\"ur Astronomie, Universit\"at Heidelberg, Philosophenweg 12, 69120 Heidelberg, Germany\label{aff121}
\and
INAF-Osservatorio Astronomico di Brera, Via Brera 28, 20122 Milano, Italy, and INFN-Sezione di Genova, Via Dodecaneso 33, 16146, Genova, Italy\label{aff122}
\and
ICL, Junia, Universit\'e Catholique de Lille, LITL, 59000 Lille, France\label{aff123}
\and
Instituto de F\'isica Te\'orica UAM-CSIC, Campus de Cantoblanco, 28049 Madrid, Spain\label{aff124}
\and
CERCA/ISO, Department of Physics, Case Western Reserve University, 10900 Euclid Avenue, Cleveland, OH 44106, USA\label{aff125}
\and
Technical University of Munich, TUM School of Natural Sciences, Physics Department, James-Franck-Str.~1, 85748 Garching, Germany\label{aff126}
\and
Max-Planck-Institut f\"ur Astrophysik, Karl-Schwarzschild-Str.~1, 85748 Garching, Germany\label{aff127}
\and
Laboratoire Univers et Th\'eorie, Observatoire de Paris, Universit\'e PSL, Universit\'e Paris Cit\'e, CNRS, 92190 Meudon, France\label{aff128}
\and
Departamento de F{\'\i}sica Fundamental. Universidad de Salamanca. Plaza de la Merced s/n. 37008 Salamanca, Spain\label{aff129}
\and
Instituto de Astrof\'isica de Canarias (IAC); Departamento de Astrof\'isica, Universidad de La Laguna (ULL), 38200, La Laguna, Tenerife, Spain\label{aff130}
\and
Universit\'e de Strasbourg, CNRS, Observatoire astronomique de Strasbourg, UMR 7550, 67000 Strasbourg, France\label{aff131}
\and
Ludwig-Maximilians-University, Schellingstrasse 4, 80799 Munich, Germany\label{aff132}
\and
Max-Planck-Institut f\"ur Physik, Boltzmannstr. 8, 85748 Garching, Germany\label{aff133}
\and
Dipartimento di Fisica - Sezione di Astronomia, Universit\`a di Trieste, Via Tiepolo 11, 34131 Trieste, Italy\label{aff134}
\and
California Institute of Technology, 1200 E California Blvd, Pasadena, CA 91125, USA\label{aff135}
\and
Department of Physics \& Astronomy, University of California Irvine, Irvine CA 92697, USA\label{aff136}
\and
Department of Mathematics and Physics E. De Giorgi, University of Salento, Via per Arnesano, CP-I93, 73100, Lecce, Italy\label{aff137}
\and
INFN, Sezione di Lecce, Via per Arnesano, CP-193, 73100, Lecce, Italy\label{aff138}
\and
INAF-Sezione di Lecce, c/o Dipartimento Matematica e Fisica, Via per Arnesano, 73100, Lecce, Italy\label{aff139}
\and
Departamento F\'isica Aplicada, Universidad Polit\'ecnica de Cartagena, Campus Muralla del Mar, 30202 Cartagena, Murcia, Spain\label{aff140}
\and
Observatorio Nacional, Rua General Jose Cristino, 77-Bairro Imperial de Sao Cristovao, Rio de Janeiro, 20921-400, Brazil\label{aff141}
\and
Department of Computer Science, Aalto University, PO Box 15400, Espoo, FI-00 076, Finland\label{aff142}
\and
Instituto de Astrof\'\i sica de Canarias, c/ Via Lactea s/n, La Laguna 38200, Spain. Departamento de Astrof\'\i sica de la Universidad de La Laguna, Avda. Francisco Sanchez, La Laguna, 38200, Spain\label{aff143}
\and
Ruhr University Bochum, Faculty of Physics and Astronomy, Astronomical Institute (AIRUB), German Centre for Cosmological Lensing (GCCL), 44780 Bochum, Germany\label{aff144}
\and
Department of Physics and Astronomy, Vesilinnantie 5, 20014 University of Turku, Finland\label{aff145}
\and
Serco for European Space Agency (ESA), Camino bajo del Castillo, s/n, Urbanizacion Villafranca del Castillo, Villanueva de la Ca\~nada, 28692 Madrid, Spain\label{aff146}
\and
ARC Centre of Excellence for Dark Matter Particle Physics, Melbourne, Australia\label{aff147}
\and
Centre for Astrophysics \& Supercomputing, Swinburne University of Technology,  Hawthorn, Victoria 3122, Australia\label{aff148}
\and
Department of Physics and Astronomy, University of the Western Cape, Bellville, Cape Town, 7535, South Africa\label{aff149}
\and
Universit\'e Libre de Bruxelles (ULB), Service de Physique Th\'eorique CP225, Boulevard du Triophe, 1050 Bruxelles, Belgium\label{aff150}
\and
DAMTP, Centre for Mathematical Sciences, Wilberforce Road, Cambridge CB3 0WA, UK\label{aff151}
\and
Kavli Institute for Cosmology Cambridge, Madingley Road, Cambridge, CB3 0HA, UK\label{aff152}
\and
Department of Physics, Centre for Extragalactic Astronomy, Durham University, South Road, Durham, DH1 3LE, UK\label{aff153}
\and
IRFU, CEA, Universit\'e Paris-Saclay 91191 Gif-sur-Yvette Cedex, France\label{aff154}
\and
Oskar Klein Centre for Cosmoparticle Physics, Department of Physics, Stockholm University, Stockholm, SE-106 91, Sweden\label{aff155}
\and
Astrophysics Group, Blackett Laboratory, Imperial College London, London SW7 2AZ, UK\label{aff156}
\and
Univ. Grenoble Alpes, CNRS, Grenoble INP, LPSC-IN2P3, 53, Avenue des Martyrs, 38000, Grenoble, France\label{aff157}
\and
INAF-Osservatorio Astrofisico di Arcetri, Largo E. Fermi 5, 50125, Firenze, Italy\label{aff158}
\and
Dipartimento di Fisica, Sapienza Universit\`a di Roma, Piazzale Aldo Moro 2, 00185 Roma, Italy\label{aff159}
\and
Centro de Astrof\'{\i}sica da Universidade do Porto, Rua das Estrelas, 4150-762 Porto, Portugal\label{aff160}
\and
Instituto de Astrof\'isica e Ci\^encias do Espa\c{c}o, Universidade do Porto, CAUP, Rua das Estrelas, PT4150-762 Porto, Portugal\label{aff161}
\and
HE Space for European Space Agency (ESA), Camino bajo del Castillo, s/n, Urbanizacion Villafranca del Castillo, Villanueva de la Ca\~nada, 28692 Madrid, Spain\label{aff162}
\and
Higgs Centre for Theoretical Physics, School of Physics and Astronomy, The University of Edinburgh, Edinburgh EH9 3FD, UK\label{aff163}
\and
Theoretical astrophysics, Department of Physics and Astronomy, Uppsala University, Box 516, 751 37 Uppsala, Sweden\label{aff164}
\and
Mathematical Institute, University of Leiden, Einsteinweg 55, 2333 CA Leiden, The Netherlands\label{aff165}
\and
Department of Astrophysical Sciences, Peyton Hall, Princeton University, Princeton, NJ 08544, USA\label{aff166}
\and
Space physics and astronomy research unit, University of Oulu, Pentti Kaiteran katu 1, FI-90014 Oulu, Finland\label{aff167}
\and
Institut de Physique Th\'eorique, CEA, CNRS, Universit\'e Paris-Saclay 91191 Gif-sur-Yvette Cedex, France\label{aff168}
\and
Center for Computational Astrophysics, Flatiron Institute, 162 5th Avenue, 10010, New York, NY, USA\label{aff169}
\and
Institute of Astronomy, University of Cambridge, Madingley Road, Cambridge CB3 0HA, UK\label{aff170}}    
\date{\today}

\authorrunning{Euclid Collaboration: L. W. K. Goh et al.}

\titlerunning{CLOE 5: Extensions beyond the standard modelling and systematic effects}

  \abstract
{\Euclid is expected to establish new state-of-the-art constraints on extensions beyond the standard \LCDM cosmological model by measuring the positions and shapes of billions of galaxies. Specifically, its goal is to shed light on the nature of dark matter and dark energy. Achieving this requires developing and validating advanced statistical tools and theoretical prediction software capable of testing extensions of the \LCDM model. In this work, we describe how the \Euclid likelihood pipeline, Cosmology Likelihood for Observables in Euclid (\cloe), has been extended to accommodate alternative cosmological models and to refine the theoretical modelling of \Euclid primary probes. In particular, we describe the modifications made to \cloe to incorporate the magnification bias term into the spectroscopic two-point correlation function of galaxy clustering. Additionally, we explain the adaptations made to \cloe's implementation of \Euclid primary photometric probes to account for massive neutrinos and modified gravity extensions. Finally, we present the validation of these \cloe modifications through dedicated forecasts on synthetic \Euclid-like data by sampling the full posterior distribution and comparing with the results drawn from the literature. In conclusion, we have identified several functionalities with regard to beyond-\LCDM modelling that could be further improved within \cloe. We also outline potential research directions to enhance the pipeline efficiency and flexibility through novel inference and machine learning techniques.}


   \maketitle

\section{Introduction}
The next generation of cosmological large-scale structure (LSS) surveys, distinguished by their unprecedented precision and ability to probe high redshifts, will allow us to map vast regions of the sky and trace the Universe's evolution history with exceptional accuracy. This advancement will be driven by upcoming and ongoing missions such as  \Euclid satellite~\citep{EuclidSkyOverview},  \textit{Nancy Grace Roman} Space Telescope~\citep{Green:2012mj},  Vera C. Rubin Observatory's Legacy Survey of Space and Time \citep[LSST;][]{LSST}, and  Dark Energy Spectroscopic Instrument \citep[DESI;][]{DESI:2019jxc,DESI:2023dwi}. A central goal of these surveys, particularly for \Euclid, is to unravel the nature of dark matter and dark energy and to assess whether the simple cosmological constant ($\varLambda$) scenario survives as a viable explanation for the late-time accelerated expansion of the Universe~\citep[see][for a review on observational evidence]{Huterer:2017buf}. 

While the Lambda cold dark matter (\LCDM) model still stands as the most successful framework for explaining a wide range of cosmological observations, the fundamental nature of dark matter and the cosmological constant remain elusive. In addition, the increasing precision of these measurements has revealed systematic tensions between different data sets~\citep[see e.g.][for a review and references]{Abdalla:2022yfr, CosmoVerseNetwork:2025alb}. These challenges suggest that extensions to the baseline model may be required to fully capture the underlying phenomenology spanning the Universe's expansion history and the evolution of the LSS. 

In response to these issues, several alternative models have been proposed, ranging from models of modified gravity (MG), to new physics in the dark sector~ \citep[see e.g.][]{Tsujikawa:2013fta,Joyce:2016vqv,2016RPPh...79i6901W,CANTATA:2021asi,Khalife:2023qbu,2024RPPh...87c6901W}. Nevertheless, a comprehensive and overarching approach is needed to assess the viability and robustness of these models in light of current and future data.

These models typically introduce a new time-dependent scalar degree of freedom to general relativity (GR). This degree of freedom, which alters the background evolution in some cases, can also exhibit spatial fluctuations, thus affecting the LSS on both counts. Such fluctuations can arise either from a non-minimal coupling of the field to gravity~\citep[see e.g.][]{Amendola:2003wa} or from the field adopting a low characteristic speed of sound~\citep[see e.g.][]{Gleyzes:2014rba}. If this field also ends up coupled to the matter sector, it can mediate an additional `fifth force', and if it is then also coupled to baryons, it must be screened at small scales to evade stringent constraints from Solar System tests of gravity~\citep{Will:2005va}. This is typically achieved by including a screening mechanism that suppresses this force locally~\citep{Brax:2013ida}.

These extensions typically induce specific phenomenological effects on the observables of interest. These include scale-dependent modifications to the linear growth of structure, characteristic of $f(R)$-gravity theories~\citep{Carroll:2006jn, Hu:2007nk}, as well as scale-independent enhancements of the linear growth, as seen in the Dvali–Gabadadze–Porrati (DGP) braneworld model~\citep{Dvali:2000hr}. Scalar-tensor modifications to gravity generally fall within one of these two categories. A set of theoretically viable models can be found within the Horndeski class of theories~\citep{Horndeski:1974wa}, which has been extensively studied and constrained~\citep[see][for a review]{Koyama:2015vza}. This class includes both modifications to gravity, typically characterised by direct couplings to the gravitational sector and dark energy models. In more exotic models, dark energy can also be coupled to dark matter in various ways~\citep{Pourtsidou:2013nha}, which might not necessarily impact the \LCDM background expansion~\citep[see e.g.][]{Simpson:2010vh}. 

Phenomenological parametrisations are a good way of probing the vast space of theoretical alternatives, whether modifications to GR or $\varLambda$. Many such parametrisations have been developed to this end, including Taylor expansions to the dark energy equation of state~\citep{Chevallier:2000qy,Linder:2002et}, simple modifications to the linear relationship between density and gravity or the lensing potential via the Poisson equation ~\citep{Zhang:2007nk, Amendola:2007rr,Pogosian:2010tj,2016A&A...594A..14P} or its non-linear counterpart~\citep[see e.g.][]{SpurioMancini:2019rxy,Bose:2022vwi}.

From the large set of viable extensions to \LCDM, a specific subset will be chosen for testing with \Euclid. In particular, various modifications to gravity and the dark sector will be considered~\citep{EP-Adamek,EP-Racz}. A wide range of common phenomenology found in the most general scalar-tensor theories will be covered, including scale-dependent~\citep{Casas23a, EP-Koyama} and scale-independent~\citep{Frusciante23} modifications to the growth of structures. Alongside the selected models, model-independent parametrisations will also be considered \cite{Euclid:2025tpw}. In the dark sector, both evolving and interacting dark energy models will be considered, as well as exotic dark matter models \citep{EP-Lesgourgues}. In addition, extensions that do not change the dark sector or gravity will be tested, namely non-standard initial conditions \citep{Ballardini23, Andrews24,EP-Finelli}, departures from the cosmological principle and relativistic effects~\citep[see e.g.][]{Lepori-EP19, EP-Lesgourgues}. Similar models have been recognised as primary candidates for testing in other large galaxy surveys~\citep[see][for example for an assessment by the Vera Rubin Observatory]{Ishak:2019aay}.

A more significant challenge is to test and provide frameworks for probing these selected extensions, while ensuring that the combination of different model extensions is done in a self-consistent manner. This effort encompasses, for example, delivering validated and accurate non-linear models applicable to both of \Euclid's primary probes: galaxy clustering and weak lensing~\citep{EP-Bose, EP-Koyama}. It also involves testing standard approximations that may not hold under the precision of \Euclid's measurements. For example, this includes considerations such as omitting the magnification bias in predicting the primary observables~\citep{Lepori-EP19, EP-JelicCizmek}, or accounting for non-linear modified gravity effects in galaxy clustering~\citep{EP-Bose}. These issues are addressed in this paper. The protocols and models developed through this process will then be available for reliably analysing the forthcoming data.

To draw robust data-driven conclusions regarding the detection of new physics and potential model preferences over \LCDM, minimising any differences in the analysis methodology and tools is crucial. This underscores the importance of constructing a single, well-validated analysis pipeline capable of handling both standard \LCDM-based analyses and analyses of selected extended models using \Euclid data. The solution is provided by \Euclid's Cosmology Likelihood for Observables in Euclid (\cloe) software \citep[see][for details]{EP-CLOE2}. This software has been developed using a mirroring repository system that promotes collaborative efficiency and strengthens the robustness of cosmological inference from \Euclid data.

The primary probes of \Euclid have been computationally implemented in \cloe, providing a robust and reliable foundation for further explorations. These implementations serve as the starting points for the modifications carried out in this work, ensuring consistency with the data and methodologies established by the \Euclid mission. The modifications introduced in this study build upon these implementations, aiming to extend their applicability and enhance their capacity to explore models beyond the standard cosmological paradigm.

To assess the feasibility of extending \cloe to test models beyond \LCDM, a combination of theoretical modelling and validations against existing observational data is required. This can be tested through extensive simulations that incorporate such models and compare them to forecasts for \Euclid data. Moreover, it is crucial to identify the impact of these extensions on the cosmological parameters and to determine whether these models are expected to provide a statistically significant improvement over the \LCDM model.

This paper introduces three key user cases that strengthen \cloe's pipeline for testing models beyond the standard \LCDM\ model. The first case focuses on the impact of magnification bias in galaxy clustering spectroscopic (GCsp) data. Gravitational lensing effects are known to alter the observed galaxy counts, leading to a magnification bias. This effect must be accounted for in \Euclid's spectroscopic and photometric surveys to ensure an accurate treatment of the bias and avoid systematic errors. The second case consists of bypassing the Weyl potential, a key quantity in understanding modified gravity theories. The Weyl potential governs the lensing effect for distant galaxies and is typically where modified gravity signatures are directly manifested. Finally, the third case concerns the role of massive neutrinos in shaping the LSS. Massive neutrinos suppress structure formation at small scales due to their free-streaming behaviour. Therefore, the evolution of this contribution must be carefully modelled to capture its impact on the matter power spectrum and growth rates. By incorporating these three effects into the \cloe pipeline, we can improve the accuracy of the LSS data analysis, while ensuring that \cloe is well-equipped to examine alternative theories of gravity that modify both the lensing and the growth of structures.

This paper is organised as follows. Section ~\ref{sec:main_probes} introduces the recipes for the theoretical computation of \Euclid's main probes. In Sect.~\ref{sec:mag_bias} we discuss the impact of magnification bias on galaxy clustering and its relevance for \Euclid's spectroscopic and photometric surveys. We summarise the methodology for incorporating this effect within \cloe and the validation tests conducted. In Sect.~\ref{sec:by_gamma}, we describe the implementation of extensions to the standard model within \cloe, including the incorporation of modified gravity effects through a Boltzmann solver and adjustments to the lensing window function. For completeness, we also discuss the theoretical details and numerical predictions for these frameworks. Section ~\ref{sec:nus} investigates the effects of neutrino masses and their integration in the solver, emphasising their impact on cosmological observables and the modifications needed for accurate predictions. Finally, Sect.~\ref{sec:conc} presents a summary of the outcomes of this work and an outlook on future developments in the \cloe pipeline, particularly its role in advancing \Euclid's scientific objectives and strengthening the accuracy and efficiency of its predictions.

\section{\Euclid main probes}\label{sec:main_probes}

 Euclid Wide Survey (EWS) will image over one billion galaxies and measure their position in the sky, photometric redshift, and shape. The catalogue is then separated into 13 bins, within the range $0.2<z<2.5$, as discussed in \cite{EuclidSkyOverview}. This traces the galaxy density and the cosmic shear fields, from which we use their two-point correlation functions: two auto-correlation functions of each field and their cross-correlation function. This probe is referred to as the \threetimestwo probe. \Euclid will additionally conduct a spectroscopic survey, which measures the cosmological redshifts of galaxies from their spectra for the spectroscopic galaxy-clustering probe. These observations will cover the comparatively smaller redshift range of $0.9<z<1.8,$ with higher precision than their photometric counterparts and for over 25 million galaxies. 

In the following paragraphs, we briefly introduce the theoretical recipes for the \Euclid primary probes. For a more detailed overview, we refer to the work presented by \cite{EP-CLOE1}.

The photometric observables provide us with tomographic angular power spectra. If we have an observable \textit{A} in redshift bin, \textit{i}, and an observable \textit{B} in redshift bin, \textit{j}, the corresponding angular power spectrum is then
\begin{equation}
    C_{ij}^{AB}(\ell) = c\int_0^\infty {\rm d} z\;\frac{W_i^A(z)\,W_j^B(z)}{H(z)\,f_K^2[r(z)]}\,P_{AB}\left[\frac{\ell+1/2}{f_K[r(z)]},\,z\right]\;,
\end{equation}
where $P_{AB}$ is the power spectrum for the observable combination $AB$, $W^A_i(z)$ is the radial window function for observable $A$ in the \textit{i}th bin, and $c$ is the speed of light.
This assumes the Limber approximation \citep{kaiser_1992} which evaluates the power spectrum at 
\begin{equation}
    k_\ell(z) = (\ell+1/2)/f_K[r(z)]\;. 
\end{equation}
Here, $f_K[r(z)]$ is the comoving angular-diameter distance as a function of the comoving distance $r$, which depends on the parameter for spatial curvature $K$ in a Friedmann--Lemaître--Robertson--Walker (FLRW) Universe.
This is expressed as\begin{equation}
    f_K[r(z)]=
    \begin{cases}
    \frac{\sinh\,\left[\sqrt{-K}\,r(z)\right]}{\sqrt{-K}} & K <0\;, \\
    r(z) & K =0\;,  \\
    \frac{\sin\,\left[\sqrt{K}\,r(z)\right]}{\sqrt{K}} & K >0\;.  \\
    \end{cases}
\end{equation}
The Limber approximation is valid for $\ell \gtrsim 100$ depending on the redshift bin. We refer to \cite{Simon:2006gm} for the accuracy of the approximation and to \cite{EP-CLOE2} for the implementation in \cloe. The non-Limber calculation will be part of future \cloe development.

The spectroscopic observable is the galaxy-galaxy power spectrum, which traces the total matter power spectrum $P_{\rm m}(k,z)$ with a bias and redshift-space distortions~\citep[RSD;][]{Villa:2017yfg}. At the linear order, this can be expressed with the Kaiser effect \citep{Kaiser1987}:
\begin{equation}
    P_{\rm {gg}}^{\rm{spectro}}(k,\mu,z)=\left[b^{\rm{spectro}}_{\rm {gal}}(z)+f(z)\,\mu^2\right]^2 P_{\rm m}(k,z)\;,  
\end{equation}
where $b^{\rm{spectro}}_{\rm {gal}}(z)$ is the linear galaxy bias. This effect scales with the growth rate $f(z)$ and the square of $\mu$, which is the cosine of the angle between the line-of-sight and the wave vector, $\textbf{k}$, where $k \equiv |\textbf{k}|$.
To compute the non-linear corrections to this, we followed the effective field theory of LSS (EFTofLSS) formalism \citep[see e.g. Euclid Collaboration: Moretti et al.\ in prep. or][]{Carrasco_2012}.We then considered the Legendre multipoles of order $\ell$ obtained by integrating over the Legendre polynomials $L_\ell(\mu)$.
\begin{equation}
    P^{\rm gg}_\ell(k,z)=\frac{2\ell+1}{2}\int_{-1}^{1}{\rm d}\mu\;L_\ell(\mu)\,P_{\rm {gg}}^{\rm{spectro}}(k,\mu,z)\;.
\end{equation}
Next, we can pass from Fourier space to configuration space to compute the multipoles of the two-point correlation function (2PCF) using the spherical Bessel function of the first kind, $j_\ell$,
\begin{equation}
\label{eq:2pcf_gg}
    \xi^{\rm gg}_{\ell} (s,z)=\frac{\rm{i}^\ell}{2\pi^2}\int_0^\infty {\rm d} k\; k^2 \, P_\ell^{\rm gg}(k,z)\,j_\ell(k s)\;. 
\end{equation} 

\section{Magnification bias for spectroscopic galaxy clustering}\label{sec:mag_bias}

The clustering of galaxies on large scales is not only affected by the peculiar velocities of the observed objects, but also by gravitational lensing~\citep{Matsubara:2004fr, Bonvin:2011bg, Challinor:2011bk}. Lensing causes a transverse distortion of an observed volume of the sky: behind an overdense region, the measured solid angle appears stretched, causing the observed number density of galaxies to appear smaller than the physical one. Furthermore, lensing conserves surface brightness and, therefore, objects appear magnified. Since galaxy surveys can detect sources above a magnitude threshold, galaxies that are intrinsically too faint to be observed might end up being included in the \Euclid catalogue due to this effect. 

The lensing contribution to the galaxy counts is known as lensing magnification and is a survey-dependent effect. In the ideal case of a purely magnitude-limited sample, the amplitude of the lensing contribution depends on the slope of the luminosity function of the galaxy population at the faint end, called the local count slope. Cosmic magnification has been detected with the cross-correlation of high-redshift quasars and low-redshift lens galaxies~\citep{Scranton:2005ci}, a background galaxy sample at high redshift with foreground lens galaxies~\citep{Hildebrandt:2009ez}, and the cross-correlation of galaxy shapes with a foreground galaxy counts field~\citep{Liu:2021gbm}.
Furthermore, there is extensive literature showing that magnification has a significant impact 
on the analysis of current and future photometric galaxy surveys such as Dark Energy Survey \citep[DES;][]{DES:2022fqx}, LSST~\citep{LSSTDarkEnergyScience:2021bah}, and \Euclid~\citep{Lepori-EP19}.

The analysis of spectroscopic galaxy surveys, which have better redshift resolution, is expected to be less affected by lensing than in the photometric case. Improving the redshift resolution will not significantly boost the number of modes induced by lensing, but only the modes dominated by density fluctuations and RSD. Furthermore, clustering analyses generally do not include information from the cross-correlations of different redshift bins, where the cosmological information is dominated by magnification. 

A study on the impact of magnification in the \Euclid spectroscopic survey was carried out in \cite{EP-JelicCizmek}. They showed that magnification does not add cosmological information to the standard analysis, which includes density and RSD. However, neglecting this effect can systematically shift the best-fit estimation of cosmological parameters. The significance of these shifts is model-dependent. In \LCDM, they reported that when using a mock galaxy catalogue from the \textit{Euclid} flagship simulations \citep{EuclidSkyFlagship}, constraints on the cosmological parameters were shifted by $(0.5-0.7)\,\sigma$. In \wowaCDM, it was at the level of roughly $0.4\sigma$.

Furthermore, model-independent measurements of the growth rate, $f(z)$, are also affected by magnification: neglecting it would lead to biases up to $1\sigma$ in the farthest redshift bin, $ z \in [1.5, 1.8]$. Thus, this work has motivated the effort to include this effect in \cloe. The forecast presented in~\cite{EP-JelicCizmek} employs the multipoles of the 2PCF as their summary statistic. The correlation functions in configuration space, and their Fourier-space counterpart (i.e. the power spectrum) are expected to contain the same cosmological information. Consequently, lensing magnification should affect the Fourier-space analysis similarly. However, since gravitational lensing is an integrated effect along the past light cone and inherently non-local, estimating its impact in Fourier space becomes challenging. Performing a Fourier transform requires knowledge of the lensing signal along arbitrary trajectories, many of which are not part of the observer's past light cone. A consistent way to compute the magnification contribution to the Fourier space power spectrum is presented in~\cite{Castorina:2021xzs}. In this paper, however,  we focus on the analysis in configuration space, leaving the implementation in Fourier space for future work.

In the following subsections, we describe the recipe implemented in \cloe. We also detail the tests carried out to validate the implementation: both on the level of 2PCFs and posterior distribution constraints.

\subsection{Magnification contributions to the spectroscopic galaxy clustering two-point correlation function} \label{subsec:mag_bias_theory}
The contribution of magnification to the 2PCF multipoles was computed in~\cite{Tansella:2018sld} for the full sky. However, the flat-sky approximation is sufficiently accurate, while reducing computational cost substantially~\citep{Jelic-Cizmek:2020jsn}. For this reason, we implemented this effect using the flat-sky recipe in \cloe. 

In the flat-sky Limber approximation, the contribution of lensing magnification to the 2PCF can be split into two terms: the cross-correlation of magnification and density and the magnification-magnification auto-correlation. The cross-correlation between magnification and RSD vanishes under this approximation and the full-sky contribution is negligible, as discussed in~\citet{Jelic-Cizmek:2020jsn}; hence, we do not include it in our modelling.

In summary, we modelled the effect of magnification by adding the two aforementioned contributions to the redshift-space multipoles of the 2PCF $\xi^{\rm gg}_{{\rm obs},\ell}(s^{\rm fid};z)$, already implemented in \cloe,
\begin{equation}
\xi_{{\rm obs},\ell}(s^{\rm fid};z) = \xi^{\rm gg}_{{\rm obs},\ell}(s^{\rm fid};z) + 2\xi^{{\rm g}\mu}_{\ell} (s^{\rm fid};z) + \xi^{\mu\mu}_{\ell}(s^{\rm fid};z)\;, 
\label{eq:pl2xil}
\end{equation}
where $\xi^{\rm gg}_{{\rm obs},\ell}(s^{\rm fid};z)$ is the true galaxy density auto-correlation term. It is computed as presented in Eq.~\eqref{eq:2pcf_gg}. The latter two terms are the density-magnification and the magnification-magnification correlation functions, respectively. 

The magnification-magnification 2PCF $\xi^{\mu\mu}_{\ell}(s^{\rm fid};z)$ can be explicitly written as
\begin{align}
\xi^{\mu\mu}_{\ell}(s^{\rm fid};z) = \,&C_{\mu\mu}(\ell) \frac{9 \Omega^2_{{\rm m}, 0} H_0^4}{8\pi c^4} \left[2-5 s_{\rm{magn}}(z)\right]^2 r^3(z) \notag\\
&\times\int_0^1 {\rm d}x\; f_\ell(x, s^{\rm fid}, z)\;,
\label{eq:pl2xil-magn-magn}
\end{align}
where the coefficient $C(\ell)$ is defined as
\begin{equation}
C_{\mu\mu}(\ell) = (2\ell + 1)\, \frac{\ell!}{2^\ell \left[(\ell/2)!\right]^2}\;, 
\label{eq:coeff-ell}
\end{equation}
with $!$ being the factorial operator.
The redshift-dependent quantity $s_{\rm magn}(z)$ is known as the local count slope of the spectroscopic sample, where $b^{\rm spectro}_{\rm magn} \equiv 2-5s_{\rm{magn}}(z)$, analogous to the magnification contribution to photometric galaxy clustering. The integrand in Eq.~\eqref{eq:pl2xil-magn-magn} is given by
\begin{equation}
f_\ell(x, s^{\rm fid}) = x^2 (1-x)^2 \left[1+z(x r)\right]^2 K_\ell\left(x s^{\rm fid}\right)\;,  
\label{eq:pl2xil-magn-magn-int}
\end{equation}
with
\begin{equation}
K_\ell\left(xs^{\rm fid}\right) = \left(x s^{\rm fid}\right) \int_0^\infty {\rm d}k\;k^2\, P_{\rm m}\left[k_\ell(z), z(xr)\right]\, \frac{j_\ell(x\,k s^{\rm fid})}{x\,k s^{\rm fid}}\;,
\label{eq:pl2xil-kell}
\end{equation}
where $z(xr)$ is the redshift corresponding to the radial comoving distance $xr$. 

The cross-correlation 2PCF between density and magnification $\xi^{{\rm g}\mu}_{\ell} (s^{\rm fid};z)$ is computed as
\begin{align}
\xi^{{\rm g}\mu}_{\ell}(s^{\rm fid};z) & =  - C_{{\rm g}\mu}(\ell) \frac{3 \Omega_{{\rm m}, 0} H_0^2}{4\pi c^2} b^{\rm spectro}_{\rm gal}(z) \left[2-5 s_{\rm{magn}}(z)\right]\nonumber\\
&\times(1+z) \left(s^{\rm fid}\right)^2\label{eq:pl2xil-dens-magn}\\
&\times\sum^{\ell/2}_{n=0} \frac{(-1)^n}{2^n} \binom{\ell}{n}\,\binom{2\ell - 2 n}{\ell} \left(\frac{\ell}{2} - n\right)! \,I^{\ell/2 - n + 3/2}_{\ell/2 - n + 1/2}(s^{\rm fid};z)\;, \nonumber
\end{align}
with
\begin{equation}
C_{{\rm g}\mu}(\ell) = \frac{2\ell + 1}{2} \pi^{3/2} \frac{2^{3/2}}{2^{\ell/2}}\;,
\label{eq:pl2xil-gmu-coeff}
\end{equation}
and
\begin{equation}
I^n_\ell(s^{\rm fid}, z) = \frac{1}{2\pi^2} \int_0^\infty {\rm d} k\,k^2  P_{\rm m}[k_\ell(z),z] \frac{j_\ell(k s^{\rm fid})}{(k s^{\rm fid})^n}\;.
\label{eq:pl2xil-integ-half-int}
\end{equation}
Since the integrals in Eq.~\eqref{eq:pl2xil-dens-magn} involve integrals of the spherical Bessel function of half-integer orders, it is convenient for a numerical evaluation to write them in terms of the Bessel function of the first kind, $J_\ell$, using
\begin{equation}
j_\ell(x) = \sqrt{\frac{\pi}{2x}}\, J_{\ell + 1/2}(x)\;.
\end{equation}
Therefore, the three types of integrals that are relevant for the computation of Eq.~\eqref{eq:pl2xil-dens-magn} are 
\begin{align} 
I^{3/2}_{1/2}(s^{\rm fid}, z) &= \frac{1}{2\pi^2}  \sqrt{\frac{\pi}{2}} \int_0^\infty {\rm d}k\;k^2 \,  P_{\rm m}[k_\ell(z),z]\, \frac{J_1(k s^{\rm fid})}{(k s^{\rm fid})^2}\;, \label{eq:pl2xil-Iell-32} \\
I^{5/2}_{3/2}(s^{\rm fid}, z) &= \frac{1}{2\pi^2}  \sqrt{\frac{\pi}{2}} \int_0^\infty {\rm d}k\;k^2 \,  P_{\rm m}[k_\ell(z),z]\, \frac{J_2(k s^{\rm fid})}{(k s^{\rm fid})^3}\;,\label{eq:pl2xil-Iell-52}\\
I^{7/2}_{5/2}(s^{\rm fid}, z) &= \frac{1}{2\pi^2}  \sqrt{\frac{\pi}{2}} \int_0^\infty {\rm d}k\;k^2 \,  P_{\rm m}[k_\ell(z),z]\, \frac{J_3(k s^{\rm fid})}{(k s^{\rm fid})^4}\;.\label{eq:pl2xil-Iell-72}
\end{align}

\subsection{Implementation and validation} \label{subsec:mag_bias_imp}

Subsequently,  within \cloe, we implemented the option to take into account the impact of spectroscopic magnification when the cosmological parameter inference is carried out using the spectroscopic galaxy clustering (GCsp) 2PCF probe. This option is specified within the \cobaya\texttt{config.yaml} file, where the \texttt{use\_magnification\_bias\_spectro} entry can  be set to either \texttt{True} or \texttt{False}. Should it be set to the former, the specific calculation of the two contributions $\xi_\ell^{\mu\mu}(s^{\rm fid};z)$ and $\xi_\ell^{\rm g\mu}(s^{\rm fid};z)$ are then carried out respectively in the functions \texttt{multipole\_correlation\_function\_mag\_mag()} and \texttt{multipole\_correlation\_function\_dens\_mag()}, according to Eqs.~\eqref{eq:pl2xil-magn-magn} and \eqref{eq:pl2xil-dens-magn}. These functions are invoked within the \texttt{multipole\_correlation\_function()} function of the \texttt{spectro.py} module, after the non-magnified multipole correlation functions are calculated. To speed up the computation, the integration in Eqs.~\eqref{eq:pl2xil-kell}, \eqref{eq:pl2xil-Iell-32}, and \eqref{eq:pl2xil-Iell-72} can be implemented using the \texttt{fftlog} and \texttt{hankel} transform algorithms. 

Thus, calculating the magnification bias contributions would require the additional input parameter $b^{\rm spectro}_{\rm magn}$, one for each bin. They are defined as \texttt{magnification\_bias\_spectro\_bin\_i} within \cloe, where \texttt{i} represents the spectroscopic bin index. This parameter can either be fixed or sampled when carrying out the inference. 

We validate our implementation against the external code COrrelation Function Full-sky Estimator \citep[\coffe;][]{Tansella:2018sld}, which calculates the galaxy 2PCF and its multipoles using linear perturbation theory. We adopted the fiducial cosmology specified in the second column of \cref{tab:mag_bias_priors}.

\begin{figure}[h!]
    \centering
    \includegraphics[width=\columnwidth]{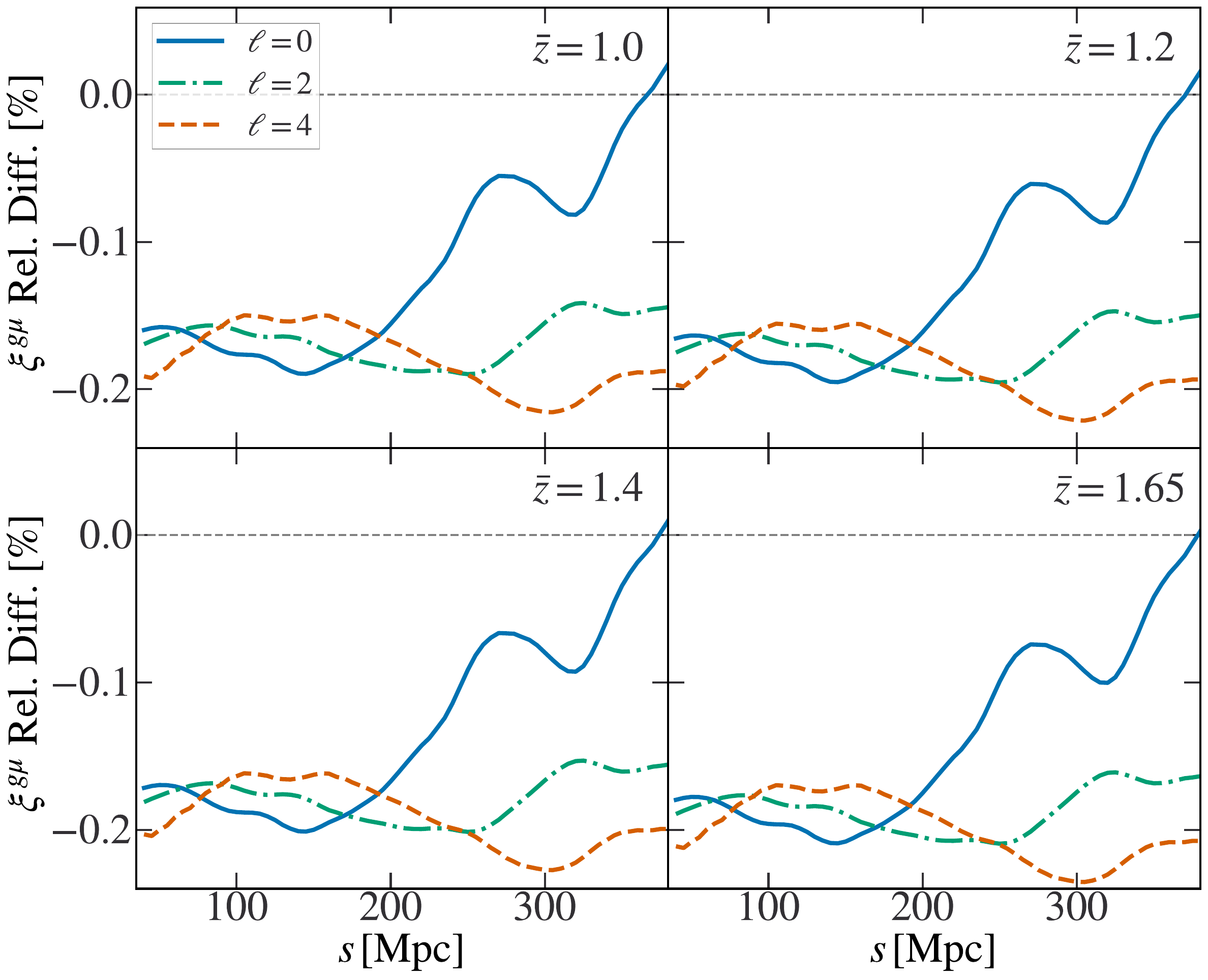}
    \caption{Relative percentage differences between the $\xi^{\rm g\mu}$ contribution as calculated by \cloe and \coffe, for the monopole (blue), quadrupole (green), and hexadecapole (orange) at the four mean redshifts. The grey dotted line denotes equality (zero per cent difference).}
    \label{fig:rel_diff_xi_g_mu}
\end{figure}

\begin{figure}[h!]
    \centering
    \includegraphics[width=\columnwidth]{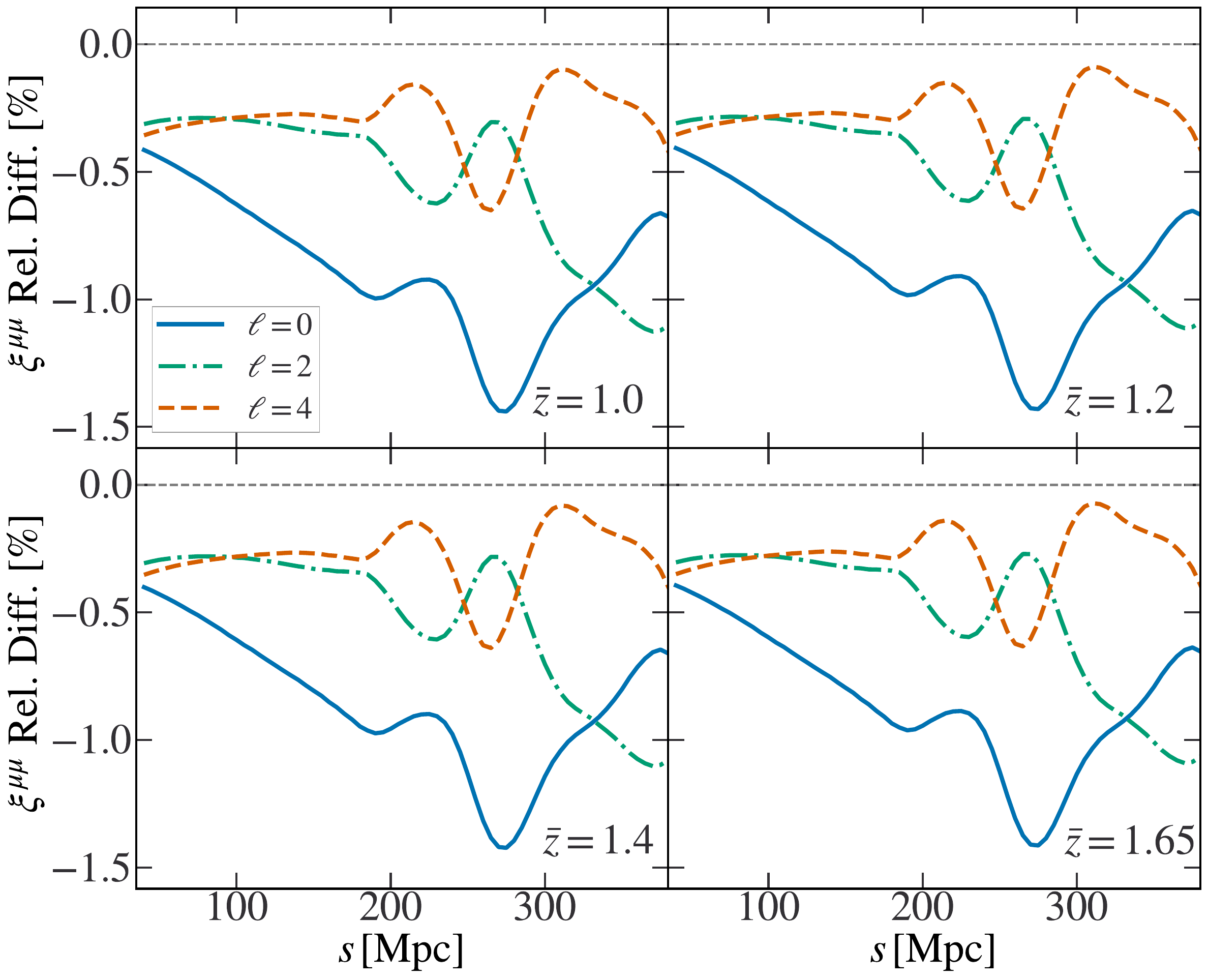}
    \caption{Relative percentage differences between the $\xi^{\mu\mu}$ contribution as calculated by \cloe and \coffe, for the monopole (blue), quadrupole (green), and hexadecapole (orange) at the four mean redshifts. The grey dotted line denotes equality (zero per cent difference).}
    \label{fig:rel_diff_xi_mu_mu}
\end{figure}

Assuming fiducial cosmological and nuisance values as detailed in the second column of \cref{tab:mag_bias_priors}, we calculate the density and magnification auto-correlation and cross-correlation functions, $\xi_\ell^{\mu\mu}$ and $\xi_\ell^{\rm g\mu}$, comparing our results between \cloe and \coffe, and plot their relative per cent differences in Figs.~\ref{fig:rel_diff_xi_g_mu} and \ref{fig:rel_diff_xi_mu_mu} respectively. For each redshift bin, we show the monopole, quadrupole, and hexadecapole for a separation range of $s \in [40,385]\, \text{Mpc}$ following \cite{EP-JelicCizmek}. We see that in the case of $\xi_\ell^{\rm g\mu}$, the relative difference is well within $0.2\%$ in all cases. For $\xi_\ell^{\mu\mu}$, it is less than $2\%$. As a sanity check, we also verify the galaxy density auto-correlation $\xi^{\rm{gg}}_{\rm{obs},\ell}$ against \coffe; the results are collected in Appendix~\ref{app:a}.

\subsection{Forecasts on cosmological constraints}
After validating the implementation of the magnification signal, we conduct a Bayesian likelihood analysis with \cloe to quantify the effect of lensing magnification on the resultant cosmological analysis. To this end, we used \cloe to generate synthetic data vectors in the form of 2PCF multipoles as described in Sects.~\ref{subsec:mag_bias_theory} and \ref{subsec:mag_bias_imp}.  
We assumed the fiducial values listed in Sect.~\ref{subsec:mag_bias_imp} for the cosmological and nuisance parameters. The latter includes the per-bin galaxy and magnification biases. The density and magnification auto-correlation and cross-correlation 2PCFs were incorporated into the data vector.

We then carried out nested sampling runs using \texttt{PolyChord} \citep{Polychord,Polychord2} to sample over the five cosmological parameters $\{\omega_{\rm b} \equiv \Omega_{\rm b} h^2$, $\omega_{\rm m} \equiv \Omega_{\rm m} h^2$, $n_{\rm s}$, $h$, $\sigma_8\}$ and the four galaxy bias parameters $\{b^{\rm spectro}_{\rm{gal},1},b^{\rm spectro}_{\rm{gal},2},b^{\rm spectro}_{\rm{gal},3},b^{\rm spectro}_{\rm{gal},4}\}$, one for each redshift bin, while keeping the local count slope parameter $s_{\rm{magn}}(z)$ fixed to $s_{{\rm{magn}},i}\in \{0.79, 0.87, 0.96, 0.98\}$ per redshift bin. \Cref{tab:mag_bias_priors} lists the prior ranges and distributions adopted for each parameter in the analysis. Additionally, we employ the theoretical Gaussian covariance matrix calculated by \coffe, which was produced at the fiducial cosmology with \textit{Euclid} DR3 sky area. It is also worth noting that only the linear matter power spectrum was considered and corrections to the Alcock--Paczynski~\citep[AP;][]{AlcPac1979} effect were ignored, following the setup of \cite{EP-JelicCizmek}. 

\begin{table} 
\caption{Prior ranges for the sampled cosmological and nuisance parameters.} 
  \centering
  \renewcommand{\arraystretch}{1.5}
    \begin{tabularx}{0.49\textwidth}{XXX}
         \hline
         \rowcolor{crisp}\textbf{Parameter} &\textbf{Fiducial}& \textbf{Prior} \\
         \hline
         \rowcolor{gray}\multicolumn{3}{c}{Cosmology} \\
         $\omega_{\rm m}$ & $0.143$ &$\mathcal{U}(0.133,0.153)$\\
         $\omega_{\rm b}$ & $0.022$ &$\mathcal{U}(0.018,0.026)$\\
         $h$ & $0.67$&$\mathcal{U}(0.37, 0.97)$\\
         $n_{\rm s}$ & $0.96$& $\mathcal{U}(0.923,0.997)$\\
         $\sigma_8$ & $0.83$ &$\mathcal{U}(0.65, 1.01)$\\
         \hline
        \rowcolor{gray}\multicolumn{3}{c}{Nuisance} \\
         $b^{\rm spectro}_{\rm{gal},1}$& $1.441$& $\mathcal{U}(1.24,1.64)$ \\
        $b^{\rm spectro}_{\rm{gal},2}$& $1.643$& $\mathcal{U}(1.44,1.84)$ \\
        $b^{\rm spectro}_{\rm{gal},3}$& $1.862$& $\mathcal{U}(1.55,2.15)$ \\
        $b^{\rm spectro}_{\rm{gal},4}$& $2.078$& $\mathcal{U}(1.57,2.57)$ \\
        \hline
  \end{tabularx}
  \tablefoot{$\mathcal{U}(\text{min},\text{max})$ denotes a uniform distribution with limits shown in the brackets. The per-bin local count slope parameters $s_{{\rm{magn}},i}$ were fixed in this analysis. Here $\omega_{\rm m}$ refers to the total matter density, combining both cold dark matter and baryons.}
  \label{tab:mag_bias_priors}
\end{table}

\begin{figure}[ht!]
    \centering
    \includegraphics[width=\linewidth]{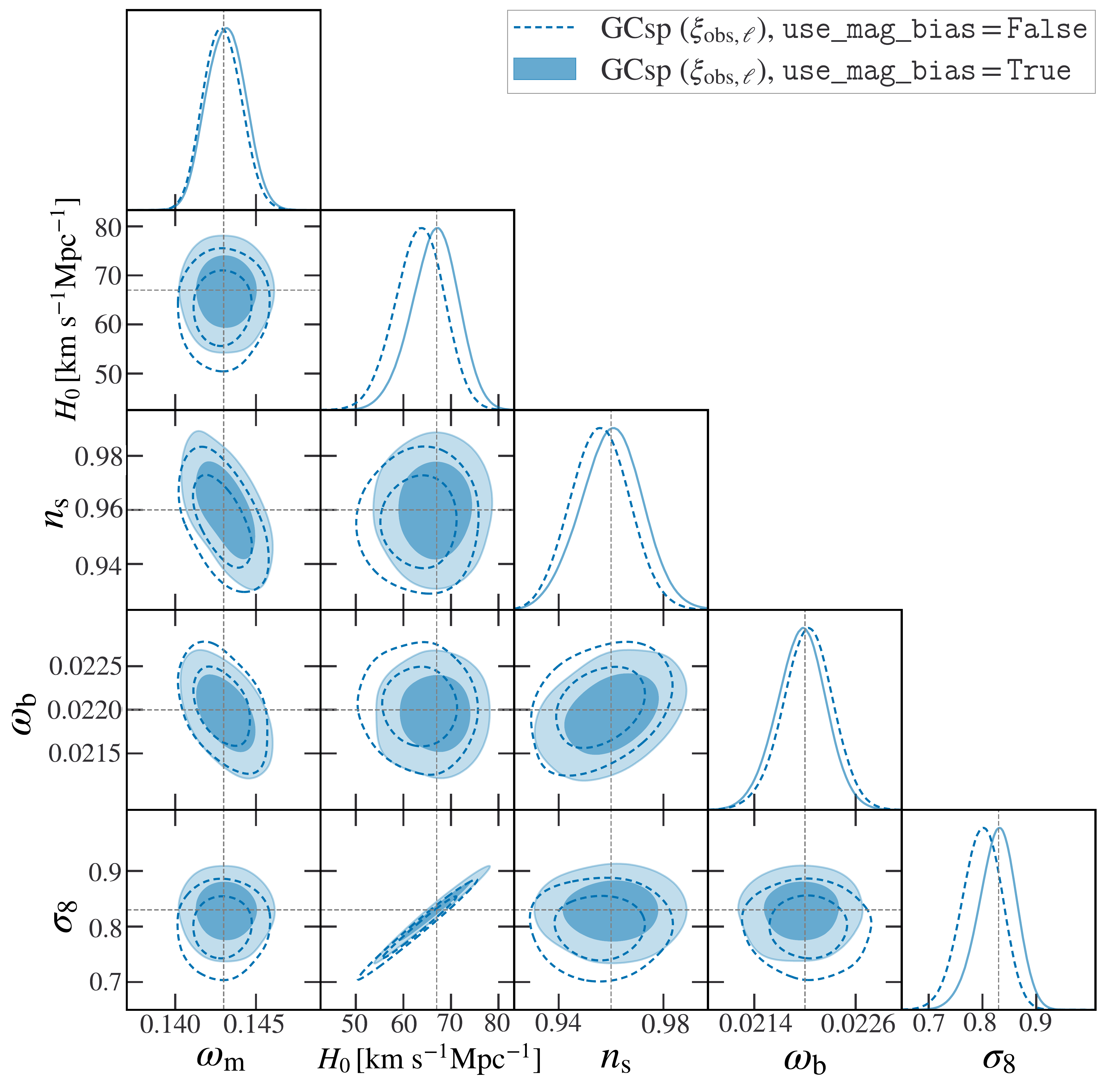}
    \caption{1D and 2D marginalised posteriors of the cosmological parameters when magnification bias is taken into account within the theoretical modelling of the multipole 2PCF $\xi_{\rm{obs},\ell}(s)$ in \cloe (solid contours, light blue) versus when it is not (dotted contours, dark blue). The fiducial values are denoted by the dotted grey lines.}
    \label{fig:mag_bias_contours}
\end{figure}

In Fig.~\ref{fig:mag_bias_contours}, we present the marginalised 2D posterior distributions of the five cosmological parameters, for both cases when magnification bias is and is not included within the calculation of the theory vector. Firstly, we see that we are able to recover the fiducial cosmology (denoted by the grey dotted lines) for all parameters when magnification bias is properly accounted for (blue solid contours). However, when this is not the case, there is a considerable shift in the contours (dotted compared to solid), most significantly for $\sigma_8$, $H_0$, and $n_{\rm s}$, where they are consistently underestimated. From Eq. \eqref{eq:2pcf_gg}, including the magnification contributions $\xi_\ell^{\mu\mu}(s^{\rm fid};z)$ and $\xi_\ell^{\rm g\mu}(s^{\rm fid};z)$ increases the amplitude of the overall correlation signal, hence the amplitude of clustering. This would then manifest as a higher value for $\sigma_8$ and a larger spectral tilt of the matter power spectrum, i.e. a larger $n_{\rm s}$. As for $H_0$, we see that since it features in both Eqs. \eqref{eq:pl2xil-magn-magn} and \eqref{eq:pl2xil-dens-magn}, including the magnification bias terms will naturally affect its recovered best-fit value. We also present the posteriors for the galaxy bias parameters $b_{\mathrm{gal},i}$ in Fig.~\ref{fig:mag_bias_nuis}, where we comment on the effect of magnification bias on these nuisance parameters.

To quantify the goodness-of-fit of the two cases (with and without magnification bias), we checked that the value of the log evidence, $\ln{Z}$, obtained by \texttt{PolyChord} is $\ln{Z}=-22.7$ for the case where magnification bias is included. Then, we have $\ln{Z}=-26.1$ for the case where it is not, demonstrating that the model with magnification bias taken into account is strongly favoured.

Additionally, we confirmed that even with the marginal deviations of the 2PCFs presented in Figs.~\ref{fig:rel_diff_xi_g_mu} and \ref{fig:rel_diff_xi_mu_mu}, the data vectors produced by \cloe still give mean values of the cosmological parameters that are consistent with those presented in \cite{EP-JelicCizmek}, well within $1\sigma$. Thus, this also acts as a verification of the accuracy of the $\xi^{g\mu}$ and $\xi^{\mu\mu}$ calculations detailed in the previous subsection, further rendering this exercise an important step towards validating \cloe against external verified codes.

\section{Beyond \LCDM : incorporating Weyl potential modifications into \cloe}
\label{sec:by_gamma}

Modified-gravity theories often introduce additional fields, which constitute extra degrees of freedom beyond those in GR, and typically break the equality between the Weyl potential $(\Phi+\Psi)/2$ and the Newtonian potential $\Psi$. To capture deviations from GR in a model-independent way, it is common to introduce two phenomenological functions, $\mu_{\rm{mg}}(k,z)$ and $\Sigma_{\rm{mg}}(k,z)$, that alter the Poisson equations according to \cite{Zhang:2007nk}, \cite{ Amendola:2007rr}, \cite{Pogosian:2010tj}, and \cite{2016A&A...594A..14P}:
\begin{align}
-k^2 \Psi &= \frac{4 \pi G}{c^2 (1+z)^2}  \mu_\mathrm{mg}(k, z)\left[\bar{\rho} \Delta+3\left(\bar{\rho}+\frac{\bar{p}}{c^2}\right) \sigma\right]\;, \\
-\frac{k^2}{2}(\Phi+\Psi) &= \frac{4 \pi G}{c^2(1+z)^2} \bigg\{\Sigma_\mathrm{mg}(k, z)\left[\bar{\rho} \nonumber  \Delta+3\left(\bar{\rho}+\frac{\bar{p}}{c^2}\right) \sigma\right] \\ 
& \qquad \qquad -\frac{3}{2} \mu_\mathrm{mg}(k, z)\left(\bar{\rho}+\frac{\bar{p}}{c^2}\right) \sigma\bigg\}\;, 
\end{align}
where $k$ represents the wavenumber, $G$ is Newton's gravitational constant, $\bar{\rho} = \bar{\rho}_{\rm m} + \bar{\rho}_{\rm r}$ is the background energy density, $\bar{p} = \bar{p}_{\rm m} + \bar{p}_{\rm r}$ is the background pressure, $\Delta$ is the comoving density contrast, and $\sigma$ is the anisotropic stress of the relativistic species.

Assuming that at late times, $\sigma$ and $\bar{\rho}_{\rm r}$ are negligible, we can rewrite these equations as
\begin{align}
-k^2\Psi & =\frac{4 \pi G}{c^2}\frac{\bar{\rho}_{\mathrm{m}}(z) \Delta_{\mathrm{m}}(k,z)}{(1+z)^2} \mu_{\rm{mg}}(k,z)\;, \label{eq:mu_poisson} \\
-\frac{k^2}{2}(\Phi+\Psi) & =\frac{4 \pi G}{c^2} \frac{\bar{\rho}_{\mathrm{m}}(z) \Delta_{\mathrm{m}}(k,z)}{(1+z)^2}  \Sigma_{\rm{mg}}(k,z)\;.\label{eq:sigma_poisson}
\end{align}

The evolution of the $\Phi$ and $\Psi$ potential can be constrained using the \Euclid\ primary probes. On the one hand, galaxy clustering will provide information on the distribution of galaxies. This traces the distribution of overdensities in the Universe and, consequently, it can provide information on the potential $\Psi$. On the other hand, cosmic shear provides information on the lensing potential $\psi$ (see Eq. \ref{eq:LensigPot}) by observing the impact of gravitational lensing deflections of light rays on the galaxy shapes.

Rather than expressing the cosmic shear power spectrum $C_{ij}^{\gamma\gamma}(\ell)$ directly in terms of the lensing potential, \cloe makes explicit use of Eq.~\eqref{eq:sigma_poisson}, modelling cosmic shear  (see Appendix~\ref{app:lensing} for more details)
as\begin{equation}\label{eq:cl_gamma_std}
  C_{ij}^{\gamma\gamma}(\ell) = c\int_0^\infty{{\rm d}z \; \frac{W_i^\gamma(z)\,W_j^\gamma(z)}{H(z)f_K^2(z)} \,P_{\rm m}[k_\ell(z),z]}\;, 
\end{equation}
where the matter power spectrum $P_{\rm m}$ enters. The window function $W_i^\gamma$ contains both the lensing efficiency term and the conversion factor between matter and lensing power spectra:
\begin{align}\label{eq:kern_gamma_std}
    W_i^\gamma = \,&  \frac{3 H_0^2\Omega_\mathrm{m,0}}{2 c^2}\, (1 + z)\, \Sigma_\mathrm{mg}(k,z)\,f_K\left[r(z)\right]\nonumber \\
   & \times \int_{z}^{z_\mathrm{max}}{\mathrm{d}z^{\prime}\; n_{i}^\mathrm{L}\left(z^{\prime}\right)\,\frac{f_K\left[r(z^{\prime}) - r(z)\right]}{f_K\left[r(z^{\prime})\right]}}\;,
\end{align}
where $n_i^{\mathrm{L}}$ is the galaxy density distribution in the $i$th tomographic bin.

While the inclusion of $\Sigma_\mathrm{mg}(z,k)$ in Eq.~\eqref{eq:kern_gamma_std} can in principle accommodate deviations from the standard \LCDM lensing prediction, the current structure of \cloe requires this function to be coded within the software itself, as no interface is currently available to retrieve such a function from a Boltzmann solver. On the other hand, the impact of a modification of gravity on the matter power spectrum is not accounted for in the same way in \cloe, and the software relies on retrieving the modified $P_{\rm m}$ from an Einstein--Boltzmann solver.

 Therefore, to obtain the theoretical predictions on \Euclid\ observables for a modified gravity model, we would need to modify two separate codes:
 \begin{itemize}
     \item a Boltzmann solver, where the modified $P_{\rm m}$ is computed, either through Eq. \eqref{eq:mu_poisson} or for some specific model.
     \item \cloe itself in order to include the $\Sigma_{\rm mg}(z)$ function corresponding to the chosen model or parametrization.
 \end{itemize}
 
Other than being cumbersome, with the need to modify different codes, this approach is also prone to errors, as we would need to pay particular attention to the consistency of the two modifications in order to obtain meaningful results.

For this reason, we decide to change this approach, at least in the context of modified gravity models: rather than parametrising $\mu_\mathrm{mg}$ and $\Sigma_\mathrm{mg}$ in two separate codes, we propose handling both within a single modified Boltzmann solver by constructing a quantity that simultaneously captures modifications to lensing and structure growth. Subsequently, we can propagate this quantity to \cloe and reformulate the definition of the angular power spectrum.  

\subsection{Theoretical description for implementation}

In order to do so, we implement a modification, noticing that Eq.~\eqref{eq:kern_gamma_std} can be seen as two separate contributions
\begin{equation}
    W_i^\gamma = \Gamma(z)\,f_K\left[r(z)\right]\int_{z}^{z_\mathrm{max}}{\mathrm{d}z^{\prime}\; n_{i}^\mathrm{L}(z^{\prime})\,\frac{f_K\left[r(z^{\prime}) - r(z)\right]}{f_K\left[r(z^{\prime})\right]}}\;,
\end{equation}
where $\Gamma(z)$ is the factor relating the Weyl potential $(\Phi+\Psi)/2$ to $\Psi$, from which we compute $P_{\rm m}$ (see Eq. \ref{eq:Gamma_def}), while the rest of the equation is the lensing efficiency.

It is possible to translate Eq.~\eqref{eq:sigma_poisson} into a relation between power spectra, allowing us to write
\begin{align}
    P_{\Upsilon\Upsilon}[k_\ell(z),z] = & \left[\frac{3 H_0^2\Omega_\mathrm{m,0}}{2 c^2}\,(1+z)\,\Sigma_{\rm{mg}}(k,z)\right]^2P_{\rm m}[k_\ell(z),z] \nonumber \\ = &\, \Gamma^2(z)\,P_{\rm m}[k_\ell(z),z]\;,
    \label{eq:Weyl_matter}
\end{align}
where we defined $P_{\Upsilon\Upsilon}$ as the Weyl power spectrum, given by the Weyl transfer function,
$\Upsilon  =k^2 (\Phi+\Psi)/2$.
For all purposes, the approach of \cloe can also be written by defining a new power spectrum, $\tilde{P}_{\rm dd}$, included in the $C_{ij}^{\gamma\gamma}(\ell)$, assuming that the conversion above can be used also at non-linear scales (see Sect.~\ref{sec:assumptionsMG} for more details) as
\begin{equation}
    \tilde{P}_{\rm dd}[k_\ell(z),z] = \Gamma^2(z)\,P_{\rm m}^\mathrm{NL}[k_\ell(z),z]\;.
\end{equation}
This leads to the definition of a new window function,
\begin{equation}\label{eq:kernel_gamma_mod}
    \tilde{W}_i^\gamma(z) = f_K\left[r(z)\right]\int_{z}^{z_\mathrm{max}}{\mathrm{d}z^{\prime} \;n_{i}^\mathrm{L}(z^{\prime})\,\frac{f_K\left[r(z^{\prime}) - r(z)\right]}{f_K\left[r(z^{\prime})\right]}}\;,
\end{equation}
which only depends on geometrical quantities. The angular power spectrum can be written as
\begin{equation}\label{eq:cl_gamma_mod}
  C_{ij}^{\gamma\gamma}(\ell) = c\int_{z_\mathrm{min}}^{z_\mathrm{max}}{{\rm d}z \;\frac{\tilde{W}_i^\gamma(z)\,\tilde{W}_j^\gamma(z)}{H(z)f_K^2(z)}\,\tilde{P}_\mathrm{dd}[k_\ell(z),z]}\;.
\end{equation}

We want to introduce this change of definition in \cloe, redefining the shear window function to contain only the lensing efficiency, while the deflection spectrum and its cross terms take the form
\begin{equation}\label{pdd}
\begin{aligned}
    \tilde{P}_\mathrm{dd}(k,z) &= \Gamma^2(z)\,P_{\rm m}^\mathrm{NL}(k,z)\;,\\
    \tilde{P}_\mathrm{dg}(k,z) &= \Gamma(z)\,b(z)\,P_{\rm m}^\mathrm{NL}(k,z)\;,\\
    \tilde{P}_\mathrm{dI}(k,z) &= \Gamma(z)f_{\rm IA}(z)\,P_{\rm m}^\mathrm{NL}(k,z)\;,\\
\end{aligned}
\end{equation}
where $b(z)$ is the linear galaxy bias, directly connecting perturbations of the galaxy field to the underlying matter density contrast,
\begin{equation}
\label{eq:linearbias}
    \delta_\mathrm{g}(k,z) = b(z)\,\delta_\mathrm{m}(k,z). \;
\end{equation}
The scale independence of the galaxy bias is known to only work well at linear scales and for simple cosmologies \citep{Desjacques_2018}. The scale dependence induced through massive neutrinos is discussed in Sect.~\ref{sec:nus}. Furthermore, $f_{\rm IA}(z)$ includes the terms responsible for intrinsic alignment,
\begin{equation}
    f_{\rm IA}(z)=-\mathcal{A}_{\rm IA}\mathcal{C}_{\rm IA}\frac{\Omega_{\rm m, 0}}{D(z)} 
    [(1 + z)/(1+z_{\rm p})]^{\eta_{\rm IA}}[\langle L \rangle(z)/L_{\star}(z)]^{\beta_{\rm IA}}\;,\label{eq: IA_function}
\end{equation}
where  $\langle L \rangle(z)$ is the redshift-dependent average luminosity and $L_{\star}(z)$ is the characteristic luminosity of source galaxies, obtained from the luminosity function.
We adopt the redshift-dependent non-linear alignment (zNLA) model for intrinsic alignments by setting $\beta_{\rm IA} = 0$. The parameters $\eta_{\rm IA}$ and $\mathcal{A}_{\rm IA}$ are treated as free parameters in the model, while $\mathcal{C}_{\rm IA}$ = 0.0134 and the pivot redshift $z_{\rm p} = 0$ are fixed in our analysis. See for example \cite{Bridle2007} and \cite{Blanchard-EP7} for a review of different intrinsic alignment models.

In order to use modified Boltzmann solvers, we want \cloe to compute the conversion factor from quantities that it can retrieve from them. Therefore, we have
\begin{equation}\label{eq:conversion}
    \Gamma^2(z) = \frac{P_{\Upsilon\Upsilon}(k,z)}{P_{\rm m}(k,z)}\;.
\end{equation}

\subsection{Assumptions and range of validity}\label{sec:assumptionsMG}

The approach we outlined allows us to take into account models that modify both lensing and the growth of structures, such as modified-gravity models. While remaining quite general in its derivation and allowing the inclusion of a more extended set of theories with respect to the standard recipe, it still relies on assumptions. Therefore, it cannot account for all effects that we would expect when working with modified cosmological models.

A first limitation can be seen in Eq.~\eqref{eq:conversion}, where the conversion factor $\Gamma(z)$ is assumed to be scale-independent. Indeed, the ratio between the two power spectra could, in general, exhibit a scale dependence, which needs to be accounted for in the conversion factor. Such a dependence could be easily accounted for, as the conversion factor, computed directly from power spectra retrieved from the Boltzmann solver, is now applied directly to the power spectra, and therefore can take a scale dependence within the structure of \cloe. However, such an effect could imply that other modifications need to be included in the recipe used by \cloe, such as opening the possibility for a scale-dependent growth factor $D(k,z)$ when modelling intrinsic alignment effects (\citealt{EP-CLOE1}).

It is important to stress that the modelling of systematic effects should be put under scrutiny when dealing with extended theories. Effects such as galaxy bias and intrinsic alignments are directly related to gravitational interactions and modifications of these, such as those encompassed by our approach, might require a change in the modelling of these effects \citep{Reischke:2021ndq}.

Another drawback of this method is that $\Gamma (z)$ is computed from linear power spectra. This assumes that the relation between the two potentials does not change when going to non-linear scales. However, we know that viable modifications of the Poisson equation need to be screened at very small scales, where standard predictions need to be recovered to account for the very precise measurements in the local Universe. Such screening mechanisms require $\Gamma(z)$ to reach its GR value for sufficiently small scales.\footnote{Our implementation method thus covers the class of theories that need no screening or are not naturally screened (we do not consider theories where the fifth force is screened without being captured by $\Sigma_{\rm mg}$).}

Therefore, in terms of power spectra, we can summarise the limitation as follows: even though \cloe can in principle use a scale-dependent conversion factor (provided the linked Boltzmann solver already accounts for MG deviations in both $P_{\Upsilon\Upsilon}$ and $P_{\rm m}$) and propagate it through Eqs. \eqref{pdd}, two caveat remain. First, screening at small scales forces the non-linear ratio $P_{\Upsilon\Upsilon}(k,z)/{P_{\rm m}(k,z)}$ to recover its GR value for $k\gg k_{\rm scr}$ (with $k_{\rm scr}$ the screening wavenumber), which requires model-specific matching in the construction of deflection spectra, beyond simply extrapolating the linear calibration via the conversion factor. Second, non-linear mode coupling can modify the relation between the scalar potentials itself, so applying a linearly calibrated $\Gamma$ (even with scale dependency) to $P_{\rm m}^{\rm NL}$ may be inconsistent unless other parts of the modelling are also generalized (e.g. accounting for scale dependent growth factor in intrinsic alignment modelling within \cloe). Therefore, in practice, we confine inference to scales where this mapping is accurate and rely on specific non-linear modelling with a case by case treatment of screening at small scales, as discussed in \cite{Euclid:2025tpw}. 

\begin{figure*}
    \centering
    \includegraphics[width=\textwidth]{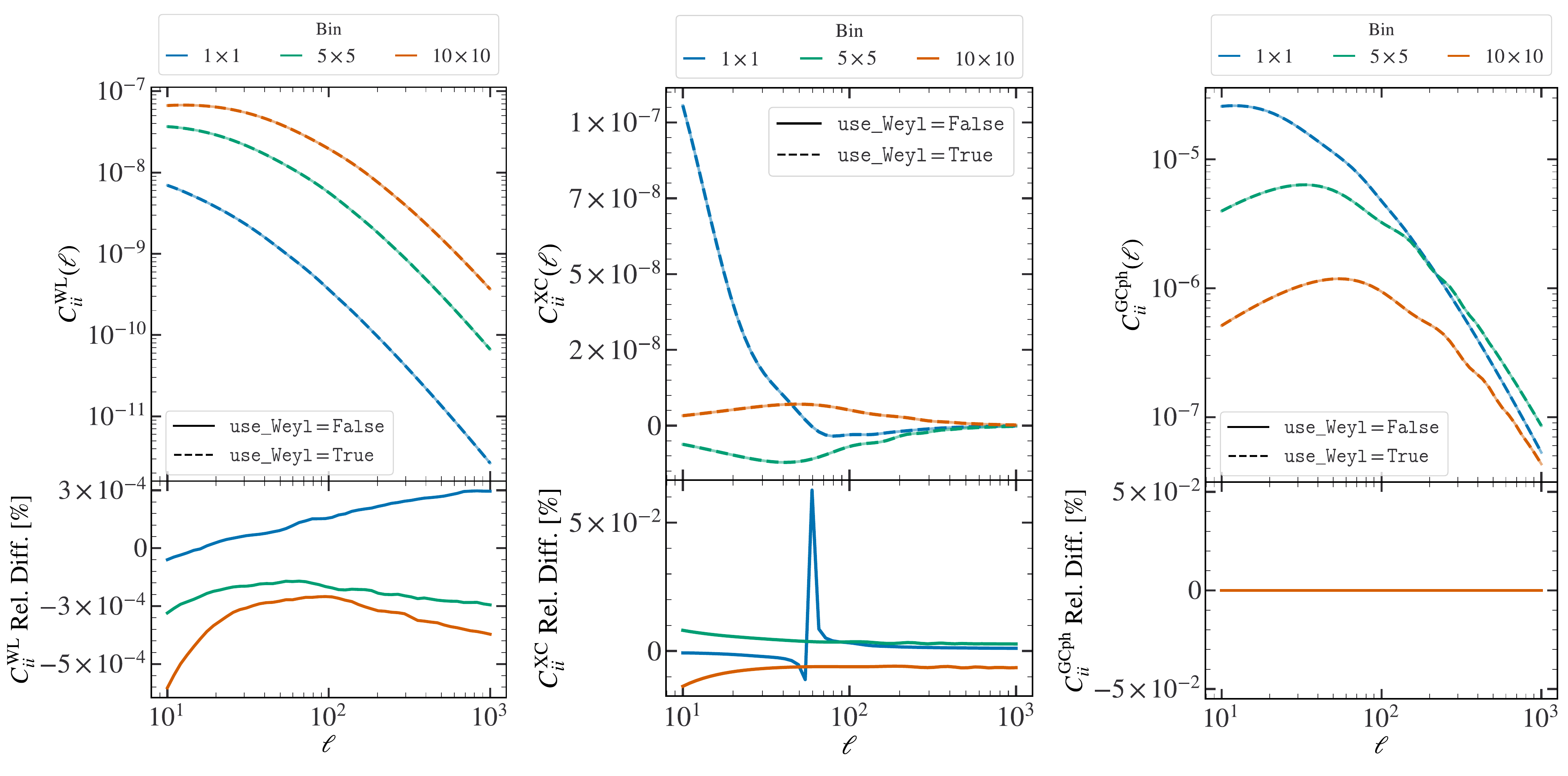}
        \caption{Angular power spectra of WL (left), XC (middle), and GCph (right) at fiducial values of \LCDM parameters of \cref{tab:w0waCDM} across different redshift bins (upper panels), and the relative percentage differences between the cases where \texttt{use\_Weyl} flag is set to \texttt{True} and \texttt{False} (lower panels). The spike in the XC lower panel occurs at the $\ell$ where the $C_\ell$ crosses zero, causing a numerical divergence rather than a physical feature. }
    \label{fig:fiducial}
\end{figure*}

\subsection{Implementation and validation}
\label{sec:weyl_test}
\begin{figure}[ht!]
\centering
\includegraphics[width=\linewidth]{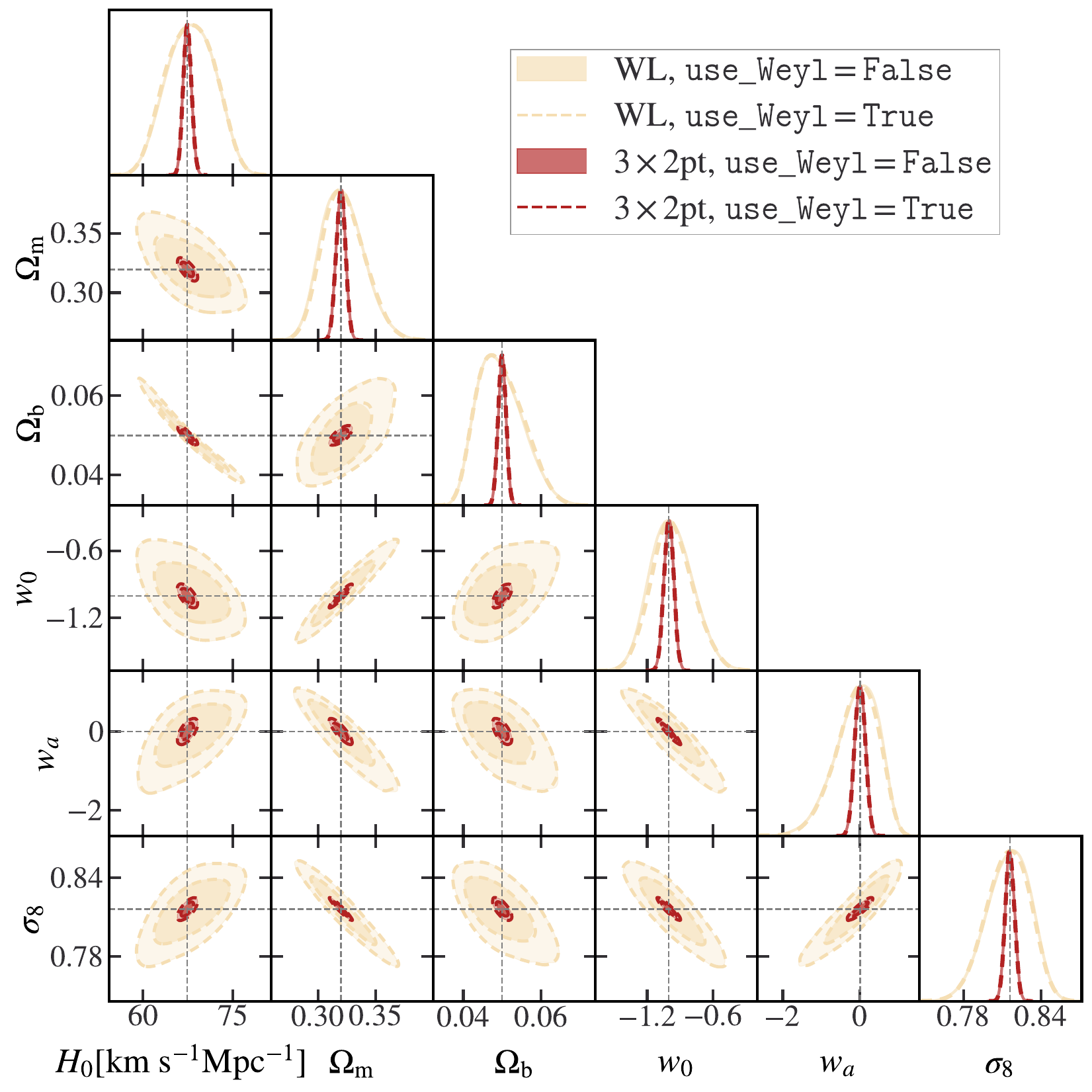}
\caption{Comparison of the 1D and 2D marginalised posterior distributions of a subset of cosmological parameters in \wowaCDM model for the \threetimestwo and WL analyses. Dashed lines and contours correspond to results with the \texttt{use\_Weyl} flag enabled, while solid lines and contours show results with the  \texttt{use\_Weyl} flag disabled.}
\label{fig:contoues_comp}
\end{figure}

To implement this approach within \cloe, we introduce the \texttt{Weyl\_matter\_ratio\_def} function inside the \texttt{cosmology.py} file. This function takes redshift and wavenumber as inputs and computes the conversion factor, as defined in Eq.~\eqref{eq:conversion}, through the division of the linear Weyl and matter power spectra, both of which are obtained from the relevant Boltzmann code. In addition, the \texttt{use\_Weyl} flag has been added to the \texttt{EuclidLikelihood.yaml} file, which can be set to either \texttt{True} or \texttt{False}. Setting it to \texttt{True}, modifies the deflection spectrum and its cross terms according to Eqs. \eqref{pdd}, and uses a new window function as per Eq.~\eqref{eq:kernel_gamma_mod}. As a result, the angular power spectra are calculated in the form of Eq.~\eqref{eq:cl_gamma_mod}. All these modifications are implemented in the \texttt{photo.py} file. 

The implementation of the Weyl conversion factor has been tested and verified for the \LCDM model where $\Sigma_{\rm mg}(k,z) = 1$ and $\Gamma^2 (z)$ factor is given by
\begin{equation}
   \Gamma^2(z) = \frac{P_{\Upsilon\Upsilon}(k,z)}{P_{\rm m}(k,z)} =  \left [ \frac{3 H_0^2\Omega_\mathrm{m,0}}{2 c^2}(1+z)\right ]^2\;.
   \label{Gamma-LCDM}
\end{equation}
Consequently, in the \LCDM model, Eqs.~\eqref{eq:cl_gamma_mod} and \eqref{eq:cl_gamma_std} are equivalent. To validate this equivalence, we obtained angular power spectra for the weak lensing (WL), galaxy-galaxy lensing (XC) and photometric galaxy clustering probes (GCph) from \cloe using the fiducial cosmological parameters listed in \cref{tab:w0waCDM}. In Fig.~\ref{fig:fiducial}, we show the comparison of the angular power spectra between the cases where the \texttt{use\_Weyl} flag is set to \texttt{True} and \texttt{False}. 
We verify our implementation in the \wowaCDM limit by observing that the relative differences are well below the percentage level for WL and XC probes. For GCph, the relative difference is exactly zero, as the modifications discussed earlier, are irrelevant for this probe and do not modify the galaxy-galaxy power spectrum. 

To ensure a more robust validation of the implementation's consistency, we performed forecasts for WL and \threetimestwo probes within the $w_0w_a$CDM model, to obtain consistent posterior distributions of parameters. \Cref{tab:w0waCDM} shows the fiducial values and prior ranges for the cosmological and nuisance parameters used in our forecast. In this section, we fix the neutrino parameters to $\sum m_\nu = 60 ~$meV$\,c^{-2}$ and $N_{\rm {eff}} = 3.046$. For the scale cuts, we set $\ell_{\rm min} = 10$ for all the probes, $\ell_{\rm max} = 5000$ for the WL probe, and $\ell_{\rm max} = 3000$ for the XC and GCph probes.

Furthermore, since our analysis involves a higher dimensional parameter space compared to the previous section, we use \nautilus,\footnote{\url{https://github.com/johannesulf/nautilus}} a boosted importance nested sampler (INS) algorithm, to perform parameter sampling more efficiently. Unlike traditional methods that calculate integrals over nested shells, \nautilus employs deep learning to construct optimised sampling boundaries \citep{10.1093/mnras/stad2441}. Our setup includes 4000 live points, 16 neural networks for the estimator, 512 likelihood evaluations per step, and a pool number of 50 processes for parallelisation of likelihood calls and sampler calculations. 

\begin{figure*}[ht!]  
    \centering
    \includegraphics[width=\textwidth]{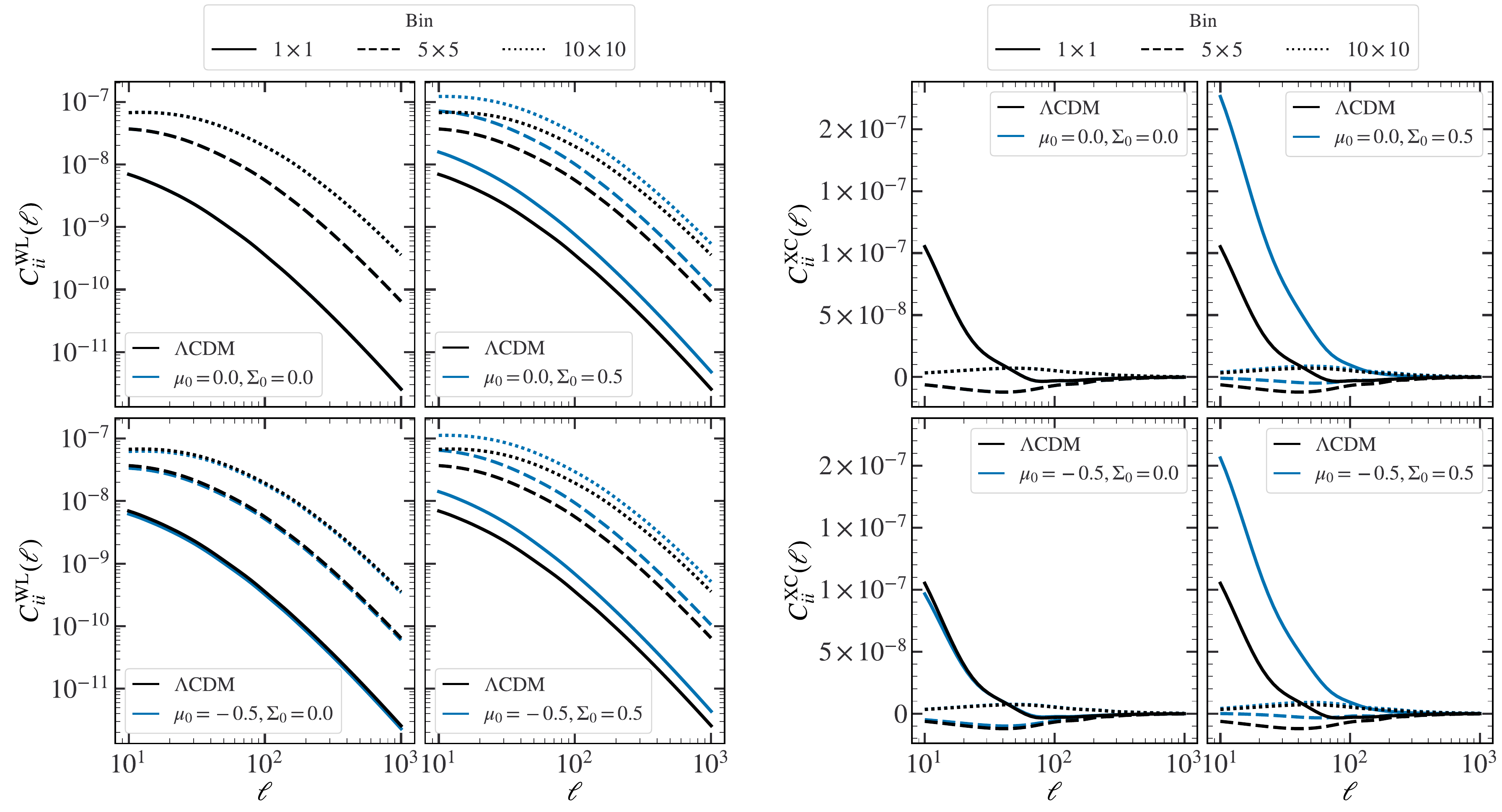}
    \caption{
    Angular power spectra of WL (left) and XC (right), for different values of parameters, $\mu_0$ and $\Sigma_0$, alongside the \LCDM predictions, across different redshift bins. }
    \label{fig:C_WL_XC}
\end{figure*}

\begin{table*}[h!] 
\caption{Fiducial values and prior ranges of the sampled cosmological and nuisance parameters in the forecasts used in Sects.~\ref{sec:by_gamma} and \ref{sec:nus}.} 
\centering
\begin{tabularx}{\textwidth}{lXXX}
\hline
\rowcolor{crisp}\multicolumn{2}{l}{\textbf{Parameters}} & \textbf{Fiducial value} & \textbf{Prior} \\
\hline
\rowcolor{gray}\multicolumn{4}{c}{\LCDM} \\
The Hubble constant & $H_0 [\mathrm{km}\,\mathrm{s}^{-1}\,\mathrm{Mpc}^{-1}]$ & $\phantom{0}67.37$ & $\mathcal U(55, 91)$ \\
Present-day physical baryon density & $\omega_\mathrm{b}$ &  $\phantom{0}0.0227$ & $\mathcal N(0.0227, 0.00038)$\\
Present-day physical cold dark matter density & $\omega_\mathrm{c}$ & $\phantom{0}0.1219$ & $\mathcal U(0.01, 0.37)$\\
Slope of primordial curvature power spectrum  & \ns & $\phantom{0}0.966$ & $\mathcal U(0.87, 1.07)$ \\
Amplitude of scalar perturbations  & $\ln(10^{10}\,\As)$ & $\phantom{0}3.04$ & $\mathcal U(1.6, 3.9)$ \\
\hline
\rowcolor{gray}\multicolumn{4}{c}{Chevalier--Linder--Polarski dark energy} \\
Time-independent component  & \multirow{1}{*}{$w_0$}& \multirow{1}{*}{$-1$} & \multirow{1}{*}{$\mathcal U(-3.0, 0)$} \\
Time-dependent component  & \multirow{1}{*}{$w_a$} & \multirow{1}{*}{$\phantom{0}0$} &  \multirow{1}{*}{$\mathcal U(-3.0, 3.0)$}\\
\hline
\rowcolor{gray}\multicolumn{4}{c}{Baryonic feedback model} \\
Baryonic feedback efficiency factor & \multirow{2}{*}{$\log_{10}(T_{\rm AGN} [\mathrm{K}])$} & \multirow{2}{*}{$7.75$} & \multirow{2}{*}{$\mathcal N(7.75, 0.17825)$} \\
of \texttt{HMCode2020} \\
\hline
\rowcolor{gray}\multicolumn{4}{c}{Neutrino parameters} \\
Cosmological neutrino mass &$\sum m_\nu\: [\mathrm{meV}\,c^{-2}]$ & $60$ &$\mathcal{U}(59, 750)$\\
Additional number of massless relics &$\Delta N_\mathrm{eff}$ & $0$& $\mathcal{U}(0, 1.954)$ \\
\hline
\rowcolor{gray}\multicolumn{4}{c}{\threetimestwo nuisance parameters} \\
Per-bin shear multiplicative bias & $m_{i=1\ldots13}$ & $0.0$ & $\mathcal N(0.0, 0.0005)$\\
Amplitude of intrinsic alignments & \multirow{1}{*}{\aIA} &  \multirow{1}{*}{$0.16$} & \multirow{1}{*}{$\mathcal U(-2, 2)$}\vspace{0.1cm}\\
Power-law slope evolution & \multirow{2}{*}{\etaIA} &  \multirow{2}{*}{$1.66$} & \multirow{2}{*}{$\mathcal U(0.0, 3.0)$}\\
of intrinsic alignment redshift\vspace{0.1cm}\\
Galaxy bias coefficients & \multirow{2}{*}{$b_{\mathrm{gal}, i=0\ldots3}$} & $\{1.33291,$ $-0.72414,$ & \multirow{2}{*}{$\mathcal U(-3, 3)$}\\
for the cubic polynomial&& $\phantom{\{}1.01830,$ $-0.14913\}$\vspace{0.2cm}\\

Magnification bias coefficients & \multirow{2}{*}{$b_{\mathrm{mag}, i=0\ldots3}$} & $\{-1.50685,$ $1.35034,$  & \multirow{2}{*}{$\mathcal U(-3, 3)$}\\
for the cubic polynomial&&$\phantom{\{-}0.08321,$ $0.04279\}$\vspace{0.2cm}\\
Per-bin mean redshift shift  & $\Delta z^{\rm L}_{i=1\ldots13}$ & $\{-0.025749,$ $0.022716,$
  & $\mathcal N\left[z_i^{\rm fid}, 0.002\,(1+z_i^{\rm fid})\right]$\\
 &&\!$\phantom{\{}-\!0.026032,$ $0.012594,$\\
 &&$\phantom{\{-}0.019285,$ $0.008326,$\\
 &&$\phantom{\{-}0.038207,$ $0.002732,$\\
 &&$\phantom{\{-}0.034066,$ $0.049479,$\\
 &&$\phantom{\{-}0.066490,$ $0.000815,$\\
 &&$\phantom{\{-}0.049070\}$\\
 \hspace{0.01cm}\\
\hline
\end{tabularx}
\tablefoot{For the forecast in Sect.~\ref{sec:by_gamma} we fix the neutrino parameters. $\mathcal{N}(\mu,\sigma)$ denotes a Gaussian distribution with mean $\mu$ and standard deviation $\sigma$.}
\label{tab:w0waCDM}
\end{table*}

Figure~\ref{fig:contoues_comp} shows the 2D posterior distribution for a subset of cosmological parameters, comparing the two cases where the \texttt{use\_Weyl} flag is set to \texttt{True} and \texttt{False}, for both WL and \threetimestwo analyses. The matching posteriors indicate that the modification introduced by \texttt{use\_Weyl} flag in \wowaCDM model does not affect the constraints on the cosmological parameters as expected. In this way, we were able to validate our implementation.

\subsection{Angular power spectra in MG theories}
The approach outlined in this section allows us to incorporate extended cosmological models in \cloe when their effect is accounted for in modified Boltzmann solvers such as \texttt{MGCAMB}\footnote{\url{https://github.com/sfu-cosmo/MGCAMB}} \citep{Wang:2023tjj} or \texttt{MGCLASS}\footnote{\url{https://gitlab.com/zizgitlab/mgclass--ii}} \citep{Sakr:2021ylx}.
For our quantitative tests, we followed the approach of \cite{DES:2018ufa} and \cite{Euclid:2025tpw}, adopting a late-time, scale-independent parametrization of $\mu_\mathrm{mg}$ and $\Sigma_\mathrm{mg}$, such that
\begin{equation}
    \mu_{\rm{mg}}=1+\mu_0 \frac{\Omega_{\mathrm{de}}(z)}{\Omega_{\mathrm{de}, 0}} \;,\quad \Sigma_{\rm{mg}}=1+\Sigma_0 \frac{\Omega_{\mathrm{de}}(z)}{\Omega_{\mathrm{de}, 0}}\;.
    \label{DES}
\end{equation}
Here, $\Omega_\mathrm{de,0}$ denotes the dark energy density parameter today, and the constants $\mu_0$ and $\Sigma_0$ determine the magnitude of the modifications to GR. Setting $\mu_0 = \Sigma_0 = 0$ restores the standard \LCDM model.

\begin{figure}[h!]
\centering
\includegraphics[width=\columnwidth]{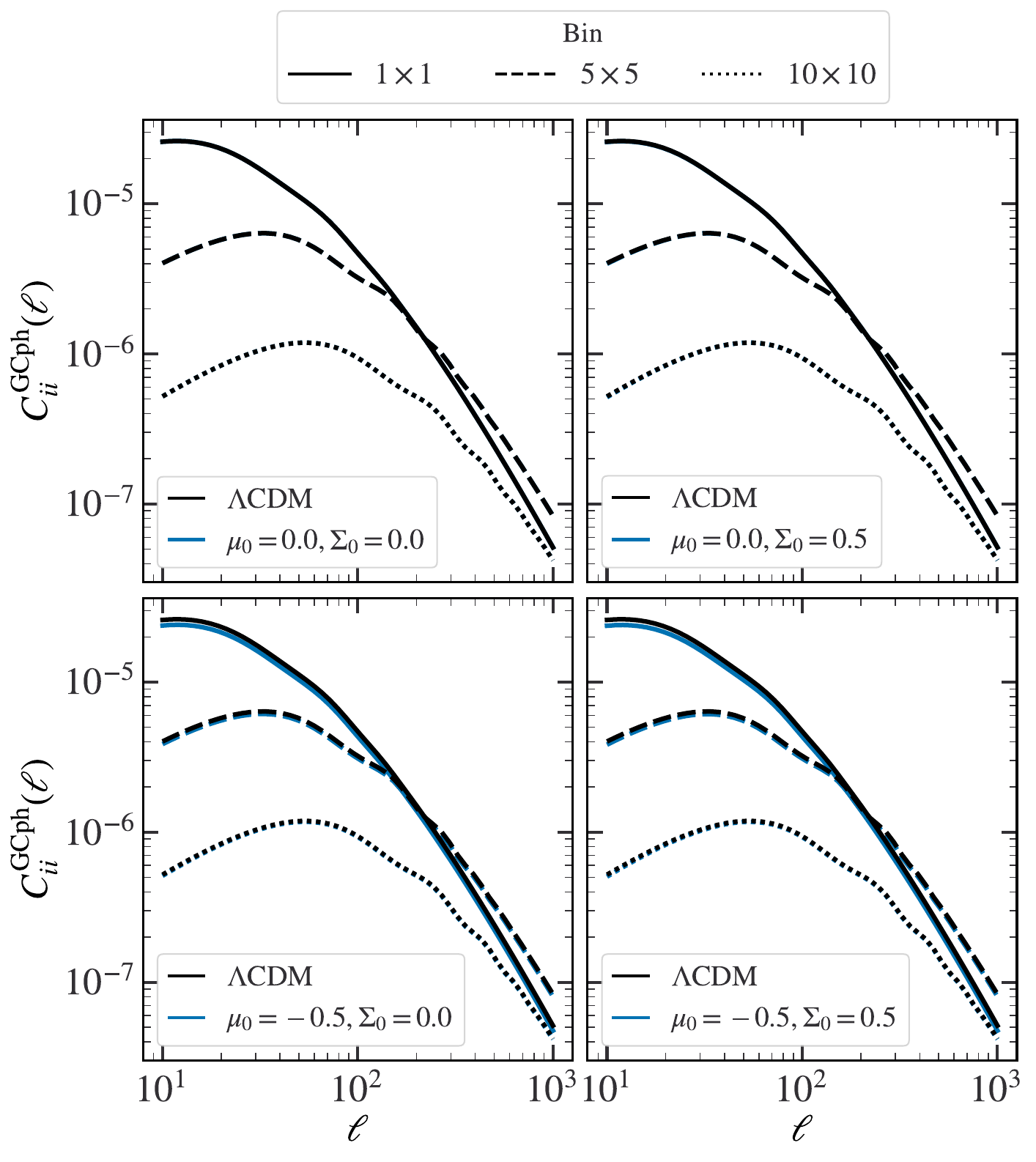}
\caption{Angular power spectrum of photometric galaxy clustering (GCph) for different values of parameters $\mu_0$ and $\Sigma_0$.}
\label{fig:C_GCph}
\end{figure}

To investigate how these two parameters affect the angular power spectra in modified gravity theories, we integrated \texttt{MGCAMB}  within \cloe.\footnote{\texttt{MGCAMB} was validated to match the \texttt{MGCLASS} code used in \cite{Euclid:2025tpw}.} We considered two fiducial sets of values for $\mu_0$ and $\Sigma_0$: $\{\mu_0,\Sigma_0\} = \{0,0\}$ and $\{\mu_0,\Sigma_0\} = \{-0.5,0.5\},$ which are taken from \cite{Euclid:2025tpw}, and are referred to as PMG-1 and PMG-2, respectively. Subsequently, with the \texttt{use\_Weyl} flag set to \texttt{True}, we compute the angular power spectra for WL, XC, and GCph on grid values generated from these two sets. The results are presented in Figs.~\ref{fig:C_WL_XC} and \ref{fig:C_GCph} alongside the predictions from the \LCDM model. The changes observed in WL and XC power spectra (Fig.~\ref{fig:C_WL_XC}) are primarily driven by $\Sigma_{\rm mg}$, which directly modifies the lensing signal by altering the interaction of relativistic particles with the gravitational potential of matter fields. Additionally, there is a secondary effect on WL and XC power spectra through $\mu_{\rm mg}$, which governs the growth of matter overdensities. On the other hand, on sub-horizon scales, the galaxy clustering power spectrum is solely sensitive to and influenced by $\mu_{\rm mg}$ and remains unaffected by changes in $\Sigma_{\rm mg}$.

Consequently, WL, which scales roughly as $\Gamma^2 P_{\rm m}$, and XC, which is proportional  $\Gamma P_{\rm m}$, exhibit partially degenerate amplitude responses to $\Sigma_0$ and $\mu_0$. Incorporating GCph provides an independent constraint on the growth amplitude in $P_{\rm m}$ via $\mu_0$, and therefore can help break the $\Sigma_{\rm mg}-\mu_{\rm mg}$ degeneracy.

\section{Consistent implementation of photometric observables with massive neutrinos} \label{sec:nus}

Over the last two decades, the continuous improvement of the precision and accuracy of cosmological observations, especially with the current generation of cosmic microwave background (CMB) and LSS experiments, has opened a window to constrain neutrino properties, such as the number of relativistic particles or the mass of neutrinos. In this regard, and despite the great progress in the precision of $\beta$-decay experiments, cosmology provides the most stringent constraints to date on the absolute neutrino mass scale. However, even with the combination of most of the current probes, such as CMB, baryon acoustic oscillations, supernovae, and LSS clustering measurements, only the tightening of the upper bound has been possible. The latest release from the DESI Collaboration yielded an upper limit of $\sum m_\nu < 0.071 \,\mathrm{eV}c^{-2}$ at the $95\%$ confidence level (CL) when combined with CMB measurements. This showed that there might be a possibility of indirectly constraining the neutrino mass hierarchy using cosmological data \citep{desicollaboration2024desi2024viicosmological, elbers2025constraintsneutrinophysicsdesi}. One of the primary science goals of \Euclid is to further improve the cosmological constraints on the neutrino mass \citep{Laureijs11}, possibly delivering evidence of a non-zero value. This has to be accomplished while confirming the robustness of such a discovery against the variations of the number of neutrinos or the modelling of dark energy. These two extra degrees of freedom further degrade the confidence found by other previous cosmological probes.  With the combination of different probes, it is possible to break these parameter correlations and \Euclid will play a vital role with its highly complementary probes. 
\subsection{Theoretical description}
\label{sec:sac_my_nus}
The massive neutrinos effect on the LSS can be divided into three phenomenological effects, which can be essentially quantified through the fraction of massive neutrinos to total matter,
\begin{equation}
    f_\nu = \frac{\Omega_\nu}{\Omega_\mathrm{m}},\:
\end{equation}
where we use the fraction of massive neutrinos to the total energy density budget $\Omega_\nu$. We identify three main effects of massive neutrinos on the matter power spectrum:
\begin{itemize}
    \item[1.] Due to the high relic velocity of massive neutrinos, the clustering of neutrinos is suppressed on scales smaller than the neutrino-free-streaming scale $k_\mathrm{fs}$. This leads to the  matter perturbations on scales smaller than the free-streaming scale losing the contributions from neutrinos \citep{Lesgourgues_Mangano_Miele_Pastor_2013}, expressed as 
    \begin{align}
        \delta_\mathrm{m} &= f_\nu\,\delta_\nu+(1-f_\nu)\,\delta_\mathrm{cb}&\\
        &\simeq (1-f_\nu)\,\delta_\mathrm{cb}&\text{for}\,k\gg k_\mathrm{fs}\:,
    \end{align}
    where we used the matter density contrast of CDM+baryons (cb) $\delta_\mathrm{cb}$. This effect on its own suppresses the matter power on the smallest scales by a factor of $(1-2\,f_\nu)$.
    \item[2.] The minimum scale, $k_\mathrm{min}$, for which neutrinos were unable to cluster at some time during the evolution of the Universe, is given by the scale of the non-relativistic transition.
    For reasonable neutrino masses, we find that the non-relativistic transition happens during matter domination. For these scales, the massive neutrinos contribute to the Hubble drag but not to the matter perturbations. This leads to a reduction of the growth, slowing down structure formation by an additional factor \citep{Lesgourgues_Mangano_Miele_Pastor_2013}
    \begin{align}
             P_{\rm m}(k) \sim (1-6\,f_\nu)\,P^{f_\nu=0}_{\rm m}(k)&&\text{for}\,k\gg k_\mathrm{min}\;.   
    \end{align}
    This effect leads to a smooth step-like suppression of the total matter power spectrum for scales smaller than $k_\mathrm{min}$.
    \item[3.] Neutrinos contribute to the energy density of ultra-relativistic constituents at matter--radiation equality. When fixing the amount of total matter today, the wavenumber of equality is shifted. 
\end{itemize}
While the second and third effects impact the total matter power spectrum, an additional contribution from the first effect remains to be noted.
It concerns the definition of the galaxy bias from Eq.~\eqref{eq:linearbias}. On small scales, neutrinos would not enter the galaxy perturbations, since they have never clustered in haloes. They do, however, enter the total matter perturbations. This leads to an intrinsic scale dependence of this bias, even on linear scales. When measuring the total neutrino mass using galaxy clustering probes, this effect changes the response of the observable to the neutrino mass and could thus bias the parameter inference and the reported uncertainty. It was shown, for example in \cite{Villaescusa_Navarro_2014}, that the bias defined using the cb field is scale independent. We can thus define
\begin{equation}
    \delta_\mathrm{g} = \hat{b}\,\delta_\mathrm{cb}\;,
\end{equation}
and use the scale-independent linear galaxy bias $\hat{b}$. Effectively, this changes the power spectrum of galaxies to no longer be affected by the first effect but still, the other two, leading to the following final adopted phenomenological scaling \citep[\kpnu hereafter]{EP-Archidiacono}, expressed as
\begin{equation}
\label{eq:nuscaling}
\begin{aligned}
P_{\rm m}(k) &\sim (1-8\,f_\nu)\,P^{f_\nu=0}_{\rm m}(k)\;,&\\
P_\mathrm{gg}(k) &\sim \hat{b}^2\,(1-6\,f_\nu)\,P^{f_\nu=0}_\mathrm{cb}(k)&\text{for}\,k\gg k_\mathrm{min}\:,
\end{aligned}
\end{equation}
where $P_\mathrm{cb}$ is the cb auto power spectrum.

The galaxy clustering probe has an additional contribution coming from RSD. They also trace the underlying matter distribution and can be used to measure the growth rate. It was demonstrated in simulations by \cite{Villaescusa_Navarro_2018} that the underlying density field as well as the growth, which enters this computation, are better described by the cb ones. We compute the effective growth directly from the cb power spectrum, 
\begin{equation}
    f_\mathrm{cb}^\mathrm{eff}(k,z) = -(1+z)^2\, \frac{\mathrm{d}\sqrt{P_\mathrm{cb}(k,z)/P_\mathrm{cb}(k,0)}}{\mathrm{d}z}\;.
\end{equation}

These considerations are the same for both \Euclid catalogues, using spectroscopic and photometric redshifts, for which the different recipes are outlined in \cite{EP-CLOE1}. The difference lies in the computation of the non-linear corrections. As detailed before, the power spectrum of the spectroscopic probe uses perturbation theory to compute its non-linear corrections. There are methods, which are particularly designed to predict these corrections in cosmologies with massive neutrinos, based on computing directly the corrections on $\delta_{\rm cb}$ \citep{Noriega_2022}.

The cosmic shear and photometric galaxy clustering probes, on the other hand, cover much smaller scales, for which the perturbation theory approach breaks down. In this case, we have to run $N$-body simulations and create fast and reliable functions to extract the non-linear power spectra. The most fundamental treatment of massive neutrinos in these $N$-body simulations would be to add them as separate massive particles to the simulation itself. Example implementations include \texttt{MassiveNuS} \citep{massivenus}, or the suite of \texttt{Quijote} simulations \citep{Quijote_sims}. However, due to the high thermal velocity and small masses in comparison to the dark matter particles, the numerical treatment could quickly become costly. In the following, we explain in further detail how we adjusted the model of the photometric probes to account for the neutrino-induced scale-dependent bias.

While typical emulators and (semi-analytical) fitting functions are built to compute the non-linear corrections of the total matter power spectrum, an open question about the computation of the non-linear cb power spectrum remains to be answered. We followed the prescription presented in \kpnu. It was shown that at first order, the non-linear cb power spectrum can be computed from the non-linear matter power spectrum by removing the linear power spectrum of massive neutrinos \citep{EP-Adamek}. This is expressed as
\begin{equation}\label{eq:pknlcb}
    P^\mathrm{NL}_{\rm m}(k) \approx f_\mathrm{cb}^2\,P^\mathrm{NL}_\mathrm{cb}(k)+ 2\,f_\mathrm{cb}\,f_\nu\,P_{\mathrm{c} \nu}(k)+
    f_\nu^2\,P_{\nu}(k)\;.
\end{equation}
Here, we denote $P_{\nu}$ as the neutrino auto power spectrum and $P_\mathrm{c \nu}$ as the cross-correlation power spectrum of cb and neutrinos. $f_\mathrm{cb}=1-f_\nu)
$ is the fraction of CDM and baryons of the total matter density. This approximation would be exact if it connected either the linear total matter power spectra or the non-linear ones. Nevertheless, it works well following Eq.~\eqref{eq:pknlcb} for two reasons:
\begin{itemize}
    \item[1.] We can use the linear neutrino power spectrum since typically the neutrino free-streaming scale is larger than the non-linear scale; thus, neutrino perturbations can be treated as linear on all scales.
    \item[2.] The neutrino perturbations are strongly suppressed on scales smaller than the free-streaming scale. Thus, the neutrino and cb cross-correlation is strongly suppressed on non-linear scales and we can stick to the linear power spectrum on larger scales. 
\end{itemize}
As the \Euclid \threetimestwo probe also includes galaxy--galaxy lensing, the same considerations must be made. To compute the galaxy--galaxy lensing angular power spectrum we can calculate
\begin{equation}
    \cl{\ell}[ij][\gamma\mathrm{G}] = c\int_0^\infty \mathrm{d}z \;\frac{W_i^\mathrm{\gamma}(z)\,W_j^\mathrm{G}(z)}{H(z)\,r^2(z)}\,P_{\rm dg}(k(\ell,z),z) \;,
\end{equation}
for the galaxy--cosmic shear cross-correlation. The power spectrum that appears here is the galaxy--displacement and would be computed using Eq.~\eqref{pdd}. The equation still applies even without modifications to the Weyl potential. In models without modified gravity, the displacement field can be related to the total matter field through the scale-independent function $\Gamma$ defined in Sect.~\ref{sec:by_gamma}. Applying the same logic as for the galaxy auto-correlation here, we encounter the correlator of cb and total matter as
\begin{align}
    \langle\delta_\mathrm{d}(\boldsymbol{k})\,\delta_\mathrm{g}(\boldsymbol{k}')\rangle=&
    \Gamma\,b(k)\,P_{\rm m}^\mathrm{NL}(k)\,(2\,\pi)^3\,\delta_\mathrm{D}^{(3)}(\boldsymbol{k}+\boldsymbol{k}')\;, \\
    =&\Gamma\, \hat{b}\,\langle\delta_\mathrm{m}(\boldsymbol{k})\,\delta_\mathrm{cb}(\boldsymbol{k}')\rangle\;.
\end{align}
Here, we  introduce the displacement density contrast, $\delta_\mathrm{d}$, and the 3D Dirac delta function, $\delta_\mathrm{D}^{(3)}$. We compute this correlator as the geometric mean of the cb power spectrum and the total matter power spectrum as an approximation,
\begin{equation}
\label{eq:correlator_cbtot}
    \langle\delta_\mathrm{m}(\boldsymbol{k})\,\delta_\mathrm{cb}(\boldsymbol{k}')\rangle \!=\! \sqrt{P^\mathrm{NL}_\mathrm{m}(k)\,P^\mathrm{NL}_\mathrm{cb}(k)}\,\,(2\,\pi)^3\,\delta_\mathrm{D}^{(3)}(\boldsymbol{k}+\boldsymbol{k}')+\mathcal{O}(f_\nu^2)\;.
\end{equation}
This approximation and its validity are further explained in Appendix~\ref{app:b}.
For the full \threetimestwo analysis, there is an additional contribution to the lensing signal: the intrinsic alignment (IA). There are models to estimate this effect by relating it to nearby galaxies influencing each other's orientation through their tidal fields.
In the (extended) non-linear alignment models, (e)NLA, the intrinsic alignments are related to the local density contrast at the time of galaxy formation. The galaxies then align with the tidal field. This is encompassed by a linear bias relationship, where the free bias is a phenomenological function $\delta_\mathrm{IA} = A_\mathrm{IA}\,\delta_\mathrm{m}$. The definition of $A_\mathrm{IA}$ can be found in \cite{EP-CLOE1}. It is a function with two free parameters $\mathcal{A}_\mathrm{IA}$ and $\eta_\mathrm{IA}$. Like in the case of the galaxy--cosmic shear cross-correlation, we encounter for the galaxy--intrinsic alignment cross-correlation the correlator of total matter and cb. Using the same approximation, Eq.~\eqref{eq:correlator_cbtot}, the cross-correlation power spectrum of IA and galaxies becomes
\begin{align}
    P_\mathrm{g,IA} = b\,A_\mathrm{IA}\,P_{\rm m}=\hat{b}\,A_\mathrm{IA}\,\sqrt{P^\mathrm{NL}_\mathrm{m}\,P^\mathrm{NL}_\mathrm{cb}} \;.
\end{align}

\subsection{Implementation and validation}
Following this reasoning, we added the flag \texttt{GC\_use\_cold\_matter\_tracer}. If set to \texttt{True}, this implements changes in the code to compute the galaxy power spectrum connecting the scale-independent galaxy bias and the cb power spectrum. This also unifies the galaxy bias from the photometric and spectroscopic surveys, because the latter uses the cb power spectrum in the context of its EFTofLSS approach, see \cite{EP-CLOE1} or Euclid Collaboration: Crocce et al.\ (in prep). 

We implemented Eq.~\eqref{eq:pknlcb} in a general way to compute the non-linear correction. The non-linear power spectrum could either be computed directly inside the Einstein--Boltzmann solver (EBS) or computed from the linear spectra using a boost computed from an external code.  This is the first type of boost that could be obtained, for example from \texttt{EuclidEmulator2}\xspace \citep{Knabenhans-EP9} or \texttt{pyhmcode}\xspace \citep{Tr_ster_2022}. As discussed above, in both cases, either directly from the EBS or through boosts, we should apply the neutrino approximation afterwards. This was explicitly checked in \kpnu.

After obtaining the non-linear power spectrum, we could choose to also add the effect of baryonic feedback. Again this could be achieved within the EBS, for example with the recipe of \texttt{HMCode2020}\xspace \citep{Mead:2020vgs}, or as a boost from a separate code, for example, \texttt{BCemu}\xspace \citep{Giri:2021qin}. The library presented in \cite{10.1093/mnras/stz3199} shows that the effect of baryonic feedback only depends weakly on the neutrino mass. This follows our physical intuition since neutrinos are decoupled from the baryonic field and should only affect the cb field at second order through its gravitational back reaction. Because of this, we apply Eq.~\eqref{eq:pknlcb} also after applying the effect of baryonic feedback. Like this, it is also self-consistent between computing the boost or getting the power spectrum from the EBS with the boost already applied.

For additional modifications of the power spectrum, for example, through the effect of modified theories of gravity, it has to be checked how neutrinos are affected by this and if the boosts should be applied before or after applying Eq.~\eqref{eq:pknlcb}.

For example, in the case of the phenomenological \gammag-Linder model, we assume that the growth modification applies uniformly to all power spectra. Specifically, this means that $P_{\rm cb \nu}$ and $P_\nu$ are rescaled by the same factor as the total matter power spectrum. Consequently, to maintain consistency, the scaling boost must be applied to $P_{\rm cb}^{\rm NL}$ after Eq.~\eqref{eq:pknlcb}.

This is implemented using a new class in the \texttt{non\_linear} module: \texttt{NonlinearNeutrinoApprox}. A key member function of this class is \texttt{wrap\_linear\_neutrino\_approx}, which allows users to input a total matter power spectrum and any number of boosts, provided as callable functions. By multiplying all boosts with the input matter power spectrum, the resulting non-linear matter power spectrum is computed. The input power spectrum can be either linear or non-linear. If a linear spectrum is provided, at least one non-linear boost must be included. Additional boosts, such as those accounting for baryonic feedback or modified theories of gravity, can also be applied. If no boosts are used, the input power spectrum must already be non-linear. The function retrieves the neutrino auto power spectrum and the neutrino--cb cross power spectrum from the cosmology module to compute Eq.~\eqref{eq:pknlcb}. The workings of the function are detailed in Sect. \ref{algo:wrap_neutrinos}. This approach enables the flexible application of boosts to the matter power spectrum, either before or after computing the non-linear cb power spectrum.

\SetAlgoNlRelativeSize{-1} 
\SetAlgoNoLine 
\SetAlgoVlined 

\begin{algorithm}
\caption{ Method to compute the non-linear baryon+CDM power spectrum with linear neutrino approximation}
\label{algo:wrap_neutrinos}
\KwIn{Total matter power spectrum $P_{\rm{m}}$, boost functions ,$B_1, B_2, \ldots, B_n$, wavenumber ,$k$, redshift, $z$},
\KwOut{non-linear baryon+CDM power spectrum $P_{\text{cb}}^{\text{NL}}(z, k),$}

$f_{\text{cb}} \gets \frac{\Omega_\text{b} + \Omega_{\text{CDM}}}{\Omega_{\text{m}}}$,\\
$f_{\nu} \gets 1 - f_{\text{cb}}$,

$P_{\rm{m}}^\mathrm{NL} \gets P_{\text{m}}(z, k)$,\\
\ForEach{boost function: $B_i$ in $B_1, B_2, \ldots, B_n$}{
    $P_{\text{m}}^\mathrm{NL} \gets P_{\text{m}}^\mathrm{NL} \times B_i(z, k),$
}

$P_{\text{cb}\times\nu} \gets P_{\mathrm{c}\nu}(z, k),$ \\
$P_{\nu\nu} \gets P_{\nu}(z, k),$ \\

$P^{cc} \gets \frac{P_{\rm{m}}^\mathrm{NL} - 2 \, f_{\text{cb}} \, f_{\nu} \, P_{\text{cb}\times\nu} - f_{\nu}^2 \, P_{\nu\nu}}{f_{\text{cb}}^2}$,

\Return{$P^{\text{cc}}$}
\end{algorithm}
After calling \texttt{wrap\_linear\_neutrino\_approx}, the class becomes a 'callable' and passing it a redshift and a wavenumber gives us the computed power spectrum. Within the code, we used this new class whenever we had to compute the non-linear cb power spectrum. When boosts are modelled to directly rescale the cb power spectrum,  they can either be multiplied directly after calling the \texttt{NonlinearNeutrinoApprox} class or applied to the linear neutrino auto power spectrum and neutrino--cb cross power spectrum and additionally passed to \texttt{wrap\_linear\_neutrino\_approx}. Both methods are equivalent, but in the latter case, the output of linear power spectra is consistent with the modelling choice.

\begin{figure}
    \centering
    \includegraphics[width=\linewidth]{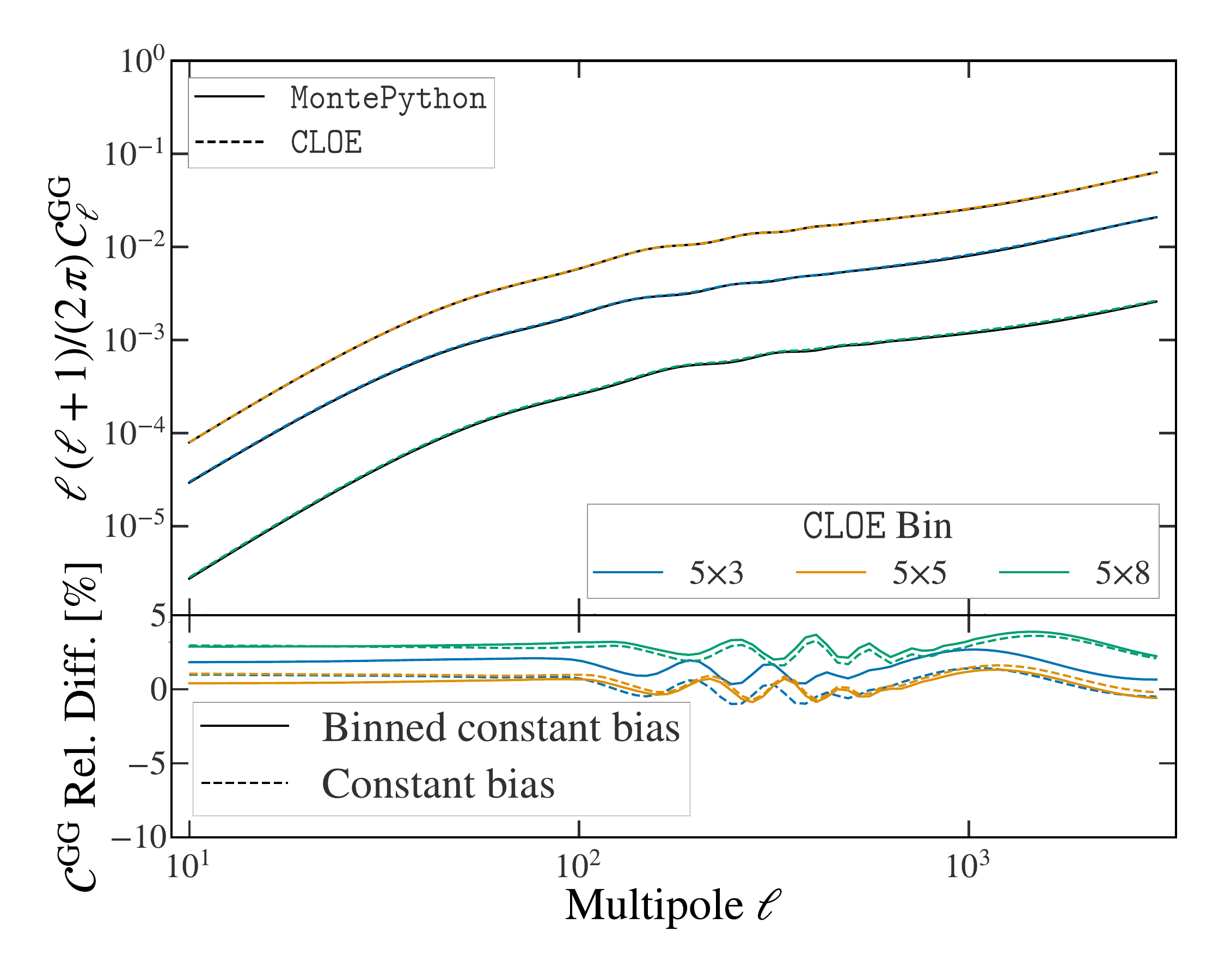}
    \caption{ \texttt{Top}: Normalised angular power spectra for the GCph probe calculated using \cloe (dashed lines) and \montepython (solid lines). Different colours indicate various redshift bin combinations. \texttt{Bottom}: The ratio of \cloe to \montepython results for two different galaxy bias models available in \cloe, distinguished by their respective linestyles. The colours correspond to the same redshift bins as in the top panel.}
    \label{fig:mpvcloe_fid}
\end{figure}

Within the code, all necessary places have been adjusted according to our discussion in Sect.~\ref{sec:sac_my_nus}. Finally, to validate our implementation against the previous work presented in \kpnu, we decided to perform a comparison on the level of the angular power spectrum. The \kpnu used a different likelihood and sampler code, \montepython\citep{Audren:2012wb, Brinckmann:2018cvx}. To start with, we compared the output at the fiducial cosmology of \cloe with \montepython in Fig.~\ref{fig:mpvcloe_fid}.

Both codes offer a good comparison against each other, with relative differences on the level of $5\%$. One important difference between them is the EBS that the code is based on. While \cloe uses \camb, \montepython internally calls \class\citep{Diego_Blas_2011}. Both codes are known to have small differences already for the same cosmological input due to differences in the treatment of certain species as well as in their respective approximation schemes. This is most noticeable in the treatment of neutrinos. To get both codes to match on a sub-percentage level, the cosmological parameter input has to be fine-tuned and the precision of both codes has to be greatly enhanced\,(see \kpnu). Without this, the fiducial power spectra already disagree at $1\%$ level. For an MCMC, however, the choice of precision parameters has little impact on the final result since any numerical noise will get averaged over during sampling. For this comparison, and the forecast in the next section, we compute the non-linear matter power spectrum using the \texttt{HMCode2020} recipe presented in \cite{Mead:2020vgs}.

Furthermore, the \montepython implementation has a slightly different implementation of the galaxy bias than within \cloe; for this reason, we additionally added the prescription of \montepython as a possible flag within \cloe. In this implementation, the galaxy bias is modelled as one discontinuous step-like function that is constant within each bin. We compare this model with another bias model within \cloe where the bias is modelled as a constant for each individual bin. The small difference is that, in regions outside of the bin edges of a given redshift bin, the bias function takes a different value. This makes a difference as, due to finite detector resolution, galaxies of a given bin can have measured redshifts that fall into a different bin. The \montepython implementation models the galaxy bias such that galaxies where the bin was misidentified should also have the galaxy bias belonging to that bin, while the default \cloe implementation assigns to these galaxies the bias belonging to their ``True'' bin. As we can see from Fig.~\ref{fig:mpvcloe_fid}, this effect is small in comparison to the differences between \montepython and \cloe.\\
There are small differences in the numerical computations between both codes. For example, the cut-off up to which redshift the window functions are computed. While \montepython goes up to a redshift of 2.5, \cloe computes them for larger redshifts until 4. At these high values of $z$, these window functions only give small contributions, but this leads to a systematic underprediction of \montepython with respect to \cloe. Keeping these differences in mind, we consider the agreement of the fiducial spectra as sufficient, and we can continue with our validation.

After we found good agreement within the fiducial angular power spectra, we wanted to validate the effect of our new flag. The suppression of the power spectrum caused by massive neutrinos is relatively small compared to the differences between the \cloe and \montepython codes. To better isolate and amplify the neutrino effect, we can examine the ratios of the angular power spectrum while varying the neutrino mass. This approach highlights the neutrino-induced suppression without equally amplifying numerical discrepancies. However, some differences between the codes persist, primarily due to the distinct approximation schemes for neutrinos implemented in \class and \camb.

\begin{figure}
    \centering
    \includegraphics[width=\linewidth]{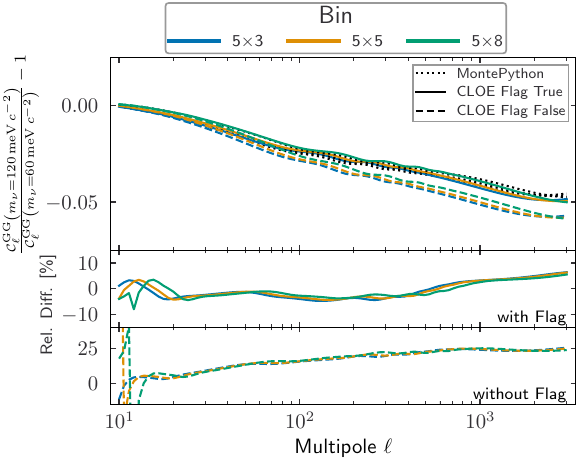}
    \caption{Response of the photometric galaxy clustering angular power spectrum to a doubling of the neutrino mass.
    \texttt{Top:} Response for \cloe with and without the \texttt{GC\_use\_cold\_matter\_tracer} shown in solid and dashed lines, respectively. The different colours correspond to different correlations of bins. We compare the responses to \montepython with the cb prescription shown in dotted black lines.
    \texttt{Middle:} Relative difference to \cloe with the neutrino flag to the \montepython response.
    \texttt{Lower:} Same as the middle panel, but for the default \cloe implementation.
    }
    \label{fig:mpvcloe_neutrinoresponse}
\end{figure}

This is further explained in \kpnu. The amplification of these effects is much smaller than for the main neutrino suppression. We call the ratio of angular power spectra the response.From our discussions in Sect.~\ref{sec:sac_my_nus}, we know that the rescaling should reduce to the response as described in Eq.~\eqref{eq:nuscaling}. In Fig.~\ref{fig:mpvcloe_neutrinoresponse}, we show how the response changes when switching between our new option and the default of \cloe. We can see that with the flag, the response of \cloe matches the response of \montepython.  Notwithstanding these differences, we considered the validation to be sufficient since we can reproduce the fiducial angular power spectrum as well as qualitatively reproduce the response of the power spectrum from \kpnu.

\subsection{Comparison to previous forecasts for the photometric probe}

In this section, we describe how we performed a forecast for the neutrino parameters. A similar forecast was already done in \kpnu and, thus, we regularly compare the choices taken here with those reported in that publication. Following a validated implementation, we can present the forecast on the additional neutrino parameters. This forecast is based on the one presented in Sect.~\ref{sec:weyl_test} and \cite{EP-CLOE3}. It is the third of the four companion papers of this work and to stick to the nomenclature of these other papers, we refer to it as Paper 3 hereafter.

One important difference is the choice of the scale cut for the WL probe. In Paper 3, the scale cut was performed in multipole space with a hard cut at $\ell^\mathrm{WL}_\mathrm{max}=3000$. Because the WL probe measures deep into the non-linear regime, we set a more conservative cut at $\ell^\mathrm{WL}_\mathrm{max}=1500$. For the smallest scales ($k\gtrsim1\,h\,\mathrm{Mpc}^{-1}$), the \texttt{HMCode2020} recipe starts deviating from simulations with massive neutrinos at the $2\%$ level. With our choice for the scale cut, we also match the scale cuts in \cite{Blanchard-EP7} and \kpnu.

Furthermore, while the \kpnu followed the prescriptions for the observables described in \cite{Blanchard-EP7}, Paper 3 introduced additional systematic effects. These comprise magnification biases for the photometric probes, shifts in the lens and source galaxy distribution, and a binned multiplicative bias to the shear signal. For an overview of these parameters, see \cref{tab:w0waCDM}. We forego opening up this study to these additional systematic effects. It was shown that these are prior-dominated and thus do not strongly degrade the measured sensitivities. With this choice, we match the methodology of \kpnu: among the nuisance parameters, we only varied the IA parameters and the galaxy biases. Additionally, all other cosmological parameters listed in \cref{tab:w0waCDM} are varied.

It was shown in \kpnu that the constraints on the cosmological neutrino mass from \Euclid's \threetimestwo probe strongly degrade when additionally opening up the additional number of ultra-relativistic relics \dnnu. Since we want to compare our results to the forecast presented in \kpnu, we additionally open up this parameter.

The baseline model for the \Euclid observables described in Paper 3 additionally differs from \kpnu through an additional contribution of RSD to the photometric galaxy clustering probe. It was shown in \cite{Tanidis-TBD} that this does not influence the measured sensitivity much, but is important to mitigate biasing the parameter inference. We also decided to keep a prior on $\omega_b$ from big bang nucleosynthesis that was present in Paper 3 but not in \kpnu. This prior is important since it breaks a strong degeneracy between $\omega_\mathrm{b}$ and $H_0$ that would otherwise lead to the exploration of unphysical cosmologies. We additionally consider baryonic feedback effects within the \texttt{HMCode2020} recipe through a one-parameter model varying $T_\mathrm{AGN}$, an effective heating parameter. For the modelling of the galaxy bias, we also stick to the model of Paper 3, which models it as a polynomial of third order.

\begin{figure}[!htpb]
    \centering
    \includegraphics[width=\linewidth]{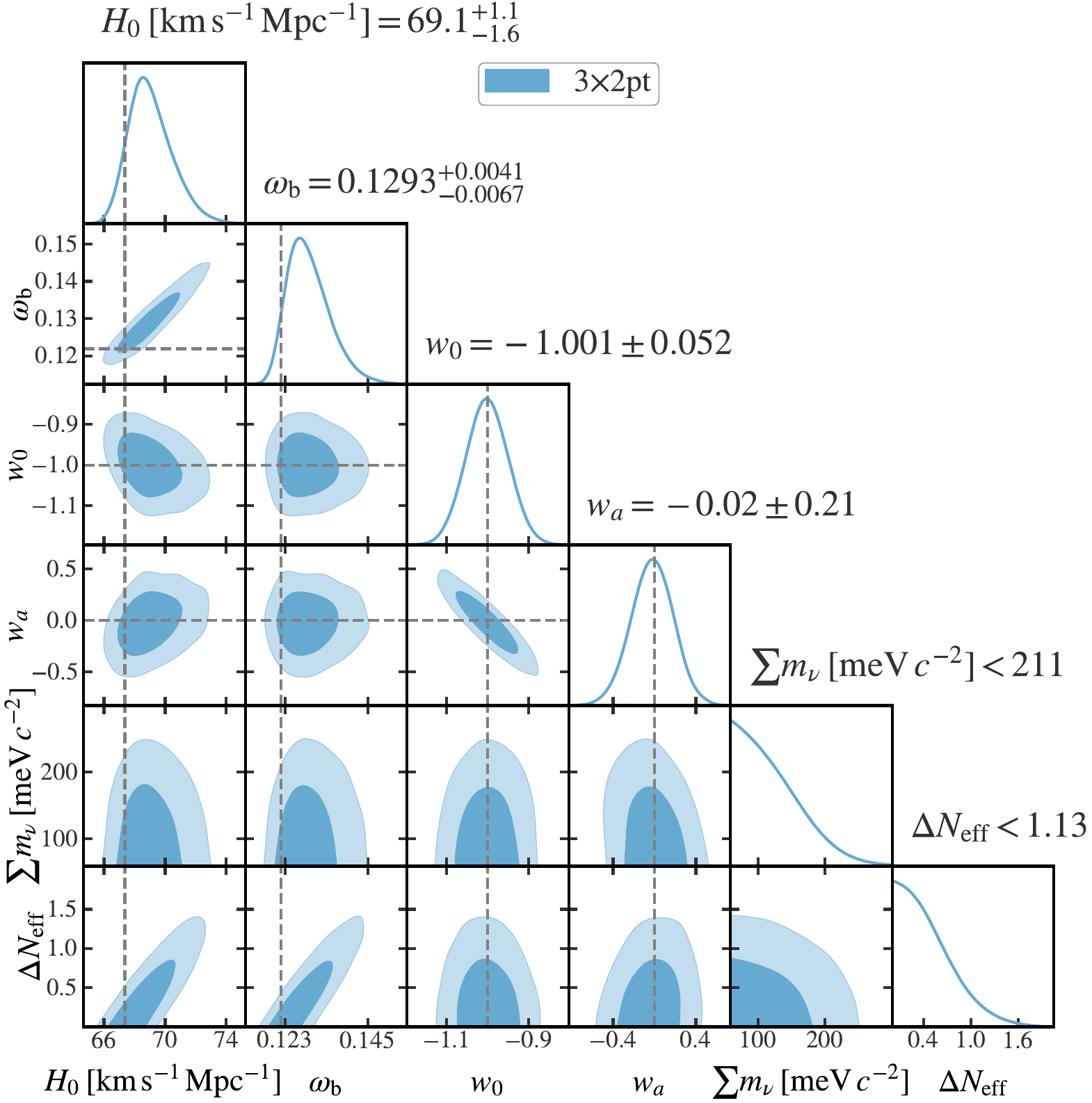}
    \caption{Forecast for the sensitivity of \Euclid's \threetimestwo probe to the neutrino parameters. We present the 1D and 2D marginalised posteriors for some selected cosmological parameters. The dashed lines represent the fiducial cosmology. The quoted uncertainties are the two-sided $68\%$ CL limits for all parameters, except for $\sum m_\nu$ and $\Delta N_\mathrm{eff}$, where we quote one-sided $95\%$ CL upper limits.}
    \label{fig:P_w0waMNT_mini}
\end{figure}

The forecast is performed using the \nautilus sampler as described in Sect.~\ref{sec:by_gamma}. The results of our forecast are presented in Fig.~\ref{fig:P_w0waMNT_mini}. For both $\sum m_\nu$ and \dnnu, the posteriors are compatible with the prior edge at a $68\%$ CL. In cases like this, the standard deviation is not a good measure since it will underpredict the actual uncertainty. We thus give the $95\%$ CL upper limit for the neutrino parameters.

The parameter inference could reliably recover the fiducial cosmology. Interestingly enough, by looking at the one-dimensional (1D) marginalised posteriors, we see a shift in $H_0$ and $\omega_\mathrm{c}$.  In both cases, we can understand this as a projection effect. Both of these parameters are strongly correlated with \dnnu, which hits the lower prior bound at 0. Due to this, the correlation direction towards lower values for $H_0$ and $\omega_\mathrm{b}$ is cut off and the mass of the posterior is shifted towards higher values. If we look at the two-dimensional (2D) marginalised contour of these parameters with \dnnu, it becomes clear, that the fiducial value is at the maximum of the posterior of \dnnu, signalling that this shift in the parameters only appears after the marginalisation over \dnnu. The same projection effect can be seen in Fig. 13 of \kpnu, and the direction of the bias matches the one presented here, notwithstanding the different set of sampled parameters.

\begin{figure}[!htpb]
    \centering
    \includegraphics[width=\linewidth]{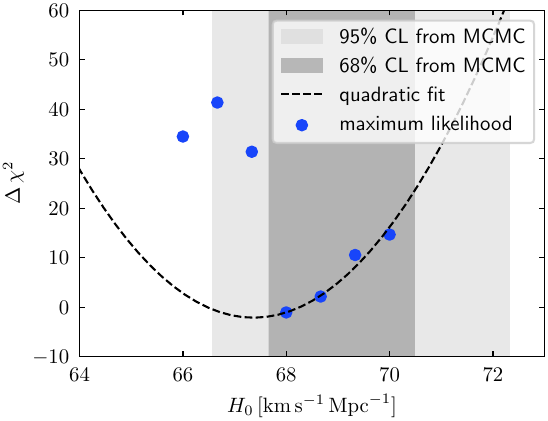}
    \caption{Comparison of the confidence levels obtained from our MCMC to a profile likelihood method. The results are presented for $H_0$. The shaded regions correspond to the marginalised 68 and 95\% confidence levels. The blue dots are the values of the $\chi^2$ found by a minimiser varying all other free parameters but $H_0$. To estimate the best-fit and uncertainty, we fit a parabola to the minimum. This is plotted with the black dashed line.}
    \label{fig:profiler_H0}
\end{figure}

To further study this aspect, we performed a profile likelihood analysis for $H_0$. This method maximises the likelihood for a given value of the profiled parameter. The resulting shape of the likelihood curve can be used to extract the best fit and uncertainty. This method can mitigate potential projection and prior-volume effects. This is presented in \cref{fig:profiler_H0}. The profiler has found a best fit value of $H_0=(67.4\pm0.6)\,\mathrm{km}\,\mathrm{s}^{-1}\,\mathrm{Mpc}^{-1}$. This is in excellent agreement with the fiducial value of $H_0$ and has a comparable uncertainty. A lower uncertainty is expected since typical minimisers have problems finding the best fit in a high-dimensional parameter space. 
The result of the profiler also unveils strong non-Gaussianity in the $H_0$ parameter. This can again be explained by \dnnu hitting the theoretical prior bound and thus not being flexible enough to absorb the low values of $H_0$ through the geometric degeneracy. To study the exact shape of the likelihood, a parametric extension of the cosmological model, which allows for a negative contribution of ultrarelativistic species to the radiation, would be preferred. Thus, we omitted these points from the fit.

Our reported uncertainty on $\sum m_\nu$ is very comparable to the results of \kpnu. Our uncertainty is $18.8\%$ smaller. This can largely be explained through the prior on $\omega_\mathrm{b}$, as well as the modelling of the galaxy bias. The \threetimestwo probe finds a strong parameter correlation between $H_0$ and $\Omega_\mathrm{b}$ (this can be for example seen in Fig. 10 of \kpnu), and a weaker correlation of $H_0$ and $\sum m_\nu$. Adding a tight prior from measurements of light element abundances from Big Bang nucleosynthesis\,(BBN) to these partially breaks this degeneracy, thus improving the sensitivity to the neutrino mass. Furthermore, the galaxy biases are correlated with the neutrino mass as they both change the amplitude of the matter power spectrum. In our case, this correlation is also rather weak though due to our conservative scale cuts also cutting away large parts of the signal of massive neutrinos. Still, since our bias model has less freedom than the bias model in \kpnu (four coefficients for a polynomial for us and 10 free bias parameters in the binned model of \kpnu) this additionally tightens our constraints on $\sum m_\nu$. Our forecast uncertainties on \dnnu are less comparable, but also here we understand where the discrepancy comes from. Our reported uncertainty is $33.7\%$ smaller. This comes clearly from the prior on $\omega_\mathrm{b}$ owing to the fact that, together with $H_0$ and $n_\mathrm{s}$, this is the main correlation direction of \dnnu for the \threetimestwo probe. Again by directly breaking the correlation, and indirectly through breaking the $H_0$--$\Omega_\mathrm{b}$ degeneracy, this drastically improves the uncertainty on \dnnu, and showcases the vast compatibility of the \threetimestwo and BBN.

\section{Conclusions and future outlook}\label{sec:conc}
In this paper, we demonstrated that the \cloe code can be successfully modified to test cosmological model extensions and incorporate novel relativistic effects within one of the key \Euclid observables. Specifically, we focus on the inclusion of the magnification bias term in the spectroscopic 2PCF, a relativistic effect that has the potential to improve our understanding of the LSS and the underlying cosmological parameters. By integrating this term, we find that our results are consistent with those obtained by \citet{EP-JelicCizmek} within 2\% for the correlation functions, further validating our approach. Moreover, we assess the impact of neglecting the magnification bias assuming a DR3 setup, revealing that doing so can introduce significant biases in key cosmological parameters, particularly a $0.4\sigma$ deviation in the Hubble parameter, $H_0$, and a $0.6\sigma$ deviation in the clustering parameter, $\sigma_8$, even within the standard \LCDM framework. This highlights the importance of including such relativistic effects in precision cosmology.

In addition, we developed a novel strategy that bypasses the need to redefine \Euclid's primary observables in terms of the Weyl potential, a complication that is often encountered when linking any photometric software that computes theoretical predictions with modified Boltzmann solvers, such as \cloe. By circumventing this requirement, we can directly connect \cloe with these solvers without altering the foundational structure of the observables themselves. This allows for a more seamless integration of modified gravity models and other extensions to the \LCDM paradigm. We  thoroughly assessed this new implementation in the \LCDM regime by sampling the posterior distribution of the parameters of interest in the \LCDM regime and found that it behaves as expected in recovering the known \LCDM results. Furthermore, we have demonstrated that this framework can be used to produce the corresponding photometric observables in the $\mu$-$\Sigma$ modified gravity regime by activating this new functionality and linking \cloe with \texttt{MGCAMB}, a modified version of CAMB that includes additional modified gravity cosmological parameters, using the forecast predictions shown in \citet{Frusciante23}.

In addition, we incorporated the neutrino parametrisation outlined in \kpnu into \cloe. This addition allows for a more accurate treatment of neutrinos in the cosmological model and we validated this extension by comparing it with a similar implementation in the \montepython software. The validation ensures that \cloe is fully compatible with the latest neutrino modelling used in \kpnu, providing a robust tool for future cosmological analyses that require precise treatment of neutrinos in preparation for achieving the scientific requirements of the \Euclid missions, as highlighted in \citet{EuclidSkyOverview}. As an interesting further study, we could consider combining the latter two features in the modelling of the photometric probes; however this would necessitate an accurate prescription for massive neutrinos in modified gravity theories, which we shall leave for future work.

Looking forward, we identified several promising directions for future development of \cloe to be ready to fully exploit the unprecedented statistical power of \Euclid. For instance, evaluating the likelihood function requires the output of an Einstein--Boltzmann solver together with some recipe to model non-linear scales, which is time-consuming (especially for \LCDM extensions). An accurate modelling of the non-linear matter power spectrum is crucial in order to extract precise and unbiased constraints for different cosmological models \citep{EP-Bose}. We recommend the use of emulators to speed up the computation of observables at nonlinear scales, using tools such as \texttt{EuclidEmulator2}, \texttt{CosmoPower} \citep{SpurioMancini:2021ppk}, \texttt{CosmicNet} \citep{Gunther:2022pto}, \texttt{CONNECT} \citep{Nygaard:2022wri}, and \texttt{Effort} \citep{Bonici:2025ltp}, or baryonic feedback emulators such as \texttt{BCEmu}. Modified gravity emulators like \texttt{ReACT} \citep{Bose_2020}, \texttt{Forge} \citep{Arnold_2022}, and \texttt{e-Mantis} \citep{S_ez_Casares_2023} are also able to capture the effects of beyond-\LCDM physics in a fast and accurate manner, extending the modelling possibilities beyond those already studied within this paper. Implementing these emulators within \cloe is relatively straightforward and would allow for a massive reduction in computational costs when testing beyond-\LCDM extensions. 

Even with accelerated theoretical predictions, an exploration of the parameter space can still be very demanding with classical inference techniques like nested sampling. This is mainly due to the large number of nuisance parameters to be marginalised over, as well as the complex parameter degeneracies that are usually introduced by extended models. In light of these difficulties, it will be very important to consider more efficient and scalable Bayesian inference methods, which includes techniques that have been recently developed in the framework of simulation-based inference \citep[SBI;][]{FrancoAbellan:2024tbj} and Hamiltonian Monte Carlo \citep{Piras:2024dml}. An additional advantage of SBI methods is that they do not need an explicit evaluation of the likelihood; rather, they simply draw samples from it via a stochastic simulator, which is constructed by means of computing theoretical predictions of the observables. This enables the modelling of systematic effects that would be computationally prohibitive or analytically intractable with standard likelihood-based methods \citep{vonWietersheim-Kramsta:2024cks}.

In conclusion, the modifications and improvements presented in this paper expand the capabilities of \cloe, extending its tests to a broader range of cosmological models and incorporating important relativistic effects into the spectroscopic probe. These advancements pave the way for more accurate and efficient cosmological analyses using incoming \Euclid Data Release 1 results. These data will contribute to the ongoing effort to better understand the nature of dark energy, dark matter, and the fundamental forces that govern the Universe.

\begin{acknowledgements} 
\AckEC  A portion of the MCMC forecasts were performed on UNIGE's High Performance Computing (HPC) service, Baobab (\url{https://doc.eresearch.unige.ch/hpc/start}) and on the Snellius Computing Cluster at SURFsara (\url{https://servicedesk.surf.nl/wiki/display/WIKI/Snellius}). Part of this work was carried out using the Feynman cluster of the Institut de recherche sur les lois fondamentales de l'Univers (Irfu) at CEA Paris-Saclay, France. ANZ and MK acknowledge funding from the Swiss National Science Foundation. GCH acknowledges support through the ESA research fellowship programme. ZS acknowledges funding from DFG project 456622116 and support from IRAP lab. EMT acknowledges support through funding from the European Research Council (ERC) under the European Union’s HORIZON-ERC-2022 (grant agreement no. 101076865). SP acknowledges support through the \textit{Conceptión Arenal Programme} of the Universidad de Cantabria and funding from the project UC-LIME (PID2022-140670NA-I00), financed by MCIN/AEI/ 10.13039/501100011033/FEDER, UE. BB is supported by a UK Research and Innovation Stephen Hawking Fellowship (EP/W005654/2).
\end{acknowledgements}

\bibliographystyle{aa}
\bibliography{biblio, Euclid}

\begin{appendix}

\section{Validation of the spectroscopic galaxy clustering two-point correlation function}\label{app:a}

\begin{figure}[h!]
    \centering
\includegraphics[width=0.85\linewidth]{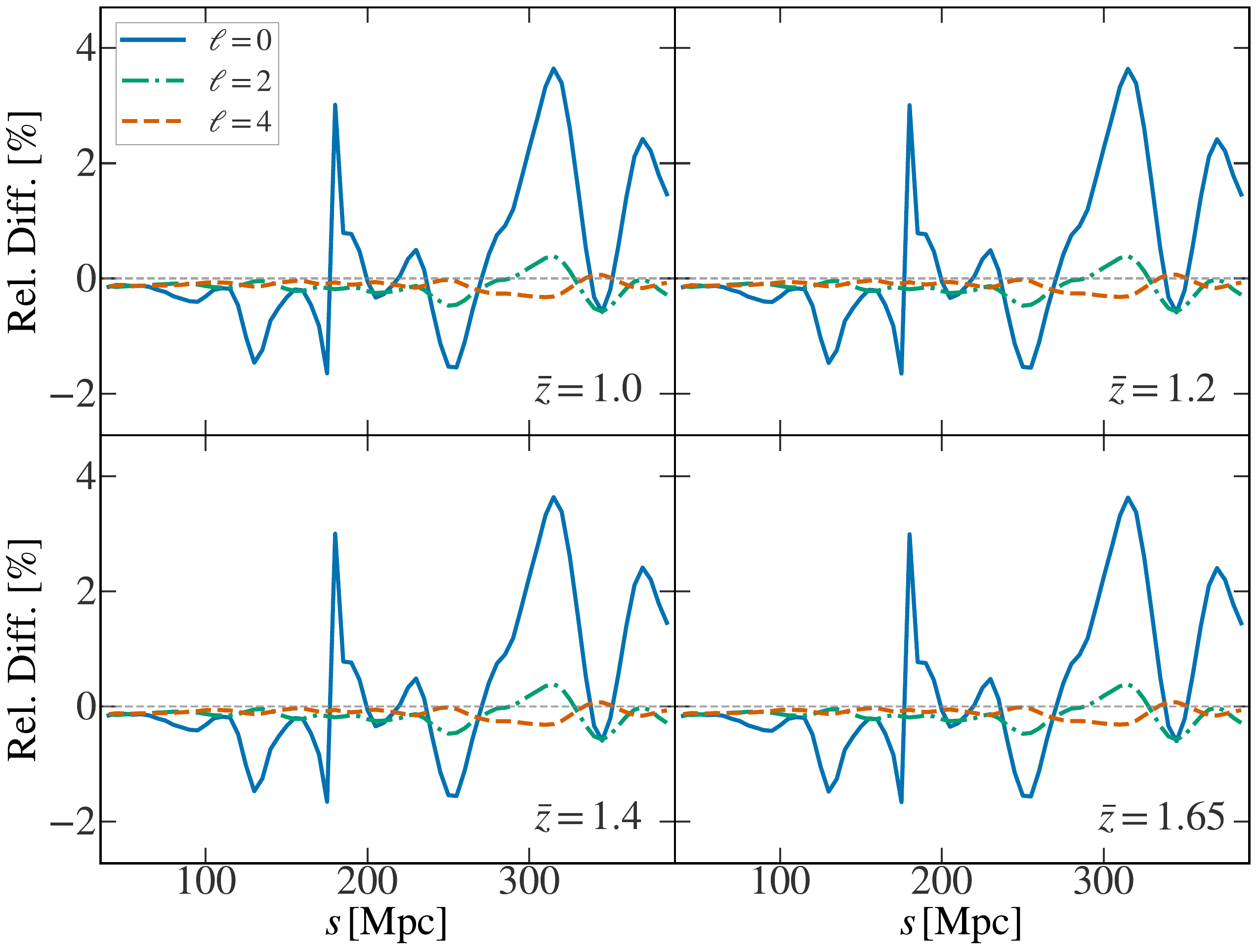}
    \caption{Relative percentage difference between the $\xi^{gg}$ correlation functions as calculated by \cloe and \coffe, for the monopole (blue), quadrupole (green), and hexadecapole (orange). The grey dotted line denotes equality (zero percentage difference).}
    \label{fig:gcsp_xi_gg}
\end{figure}

We present in Fig.~\ref{fig:gcsp_xi_gg} the percentage relative differences between the galaxy density-density auto-correlation function $\xi^{\rm gg}_{{\rm obs},\ell}\,(s;z)$ computed by \coffe versus \cloe. The percent-level discrepancy is mainly due to the difference in Boltzmann codes employed to calculate the matter power spectrum: \cloe uses \camb, while \coffe uses \texttt{CLASS}. We see that for all multipoles at all redshifts, the difference is within $5\%$. The monopole has consistently larger errors and oscillations than the quadrupole and hexadecapole.

\section{Effect of magnification bias on nuisance parameters}\label{app:b}

\begin{figure}[h!]
    \centering
\includegraphics[width=0.85\linewidth]{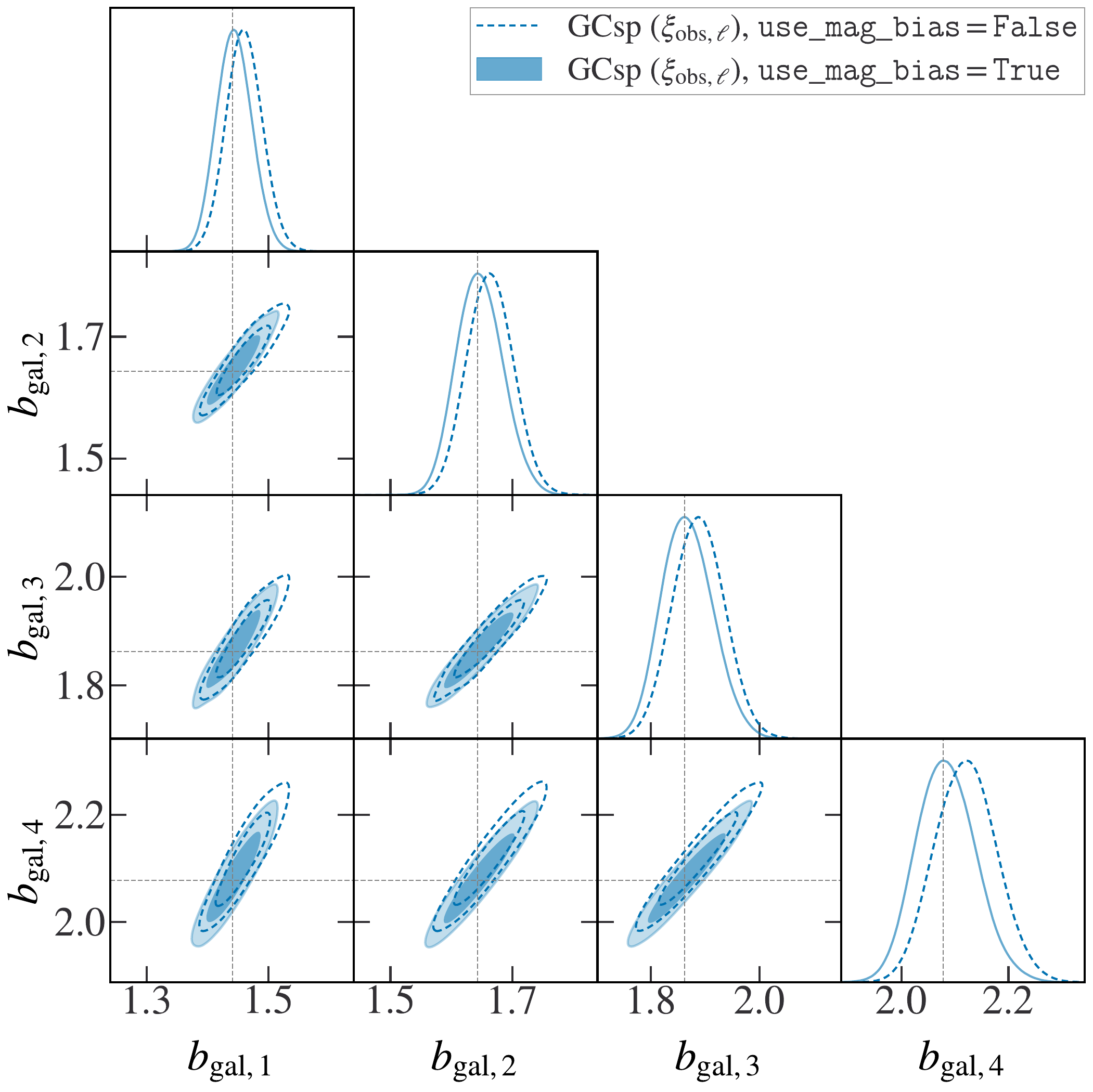}
    \caption{1D and 2D marginalised posteriors of the galaxy bias parameters, $b_{\mathrm{gal},i}$, when magnification bias is taken into account within the theoretical modelling of the multipole 2PCF $\xi_{\rm{obs},\ell}(s)$ in \cloe (solid contours, light blue) versus when it is not (dotted contours, dark blue). The fiducial values are denoted by the dotted grey lines.}
    \label{fig:mag_bias_nuis}
\end{figure}

In Fig.~\ref{fig:mag_bias_nuis}, we present the 1D and 2D posterior distributions of the nuisance parameters, the galaxy bias for each spectroscopic redshift bin, $b_{\mathrm{gal},i}$. We see that not appropriately accounting for the magnification bias gives rise to a shift in their best-fit values from the fiducial, with the magnitude of this shift generally increasing with redshift. This can be understood by looking at Eq. \eqref{eq:pl2xil-dens-magn}: since the fiducial value of $b_{\mathrm{gal},i}$ increases with redshift bin, this increases the contribution of the density-magnification cross-correlation term, in turn leading to greater modelling inaccuracies at higher redshifts.

\section{The shear power spectrum in the Limber approximation}
\label{app:lensing}
Following the approach outlined in \cite{Kilbinger:2017lvu}, we derive the
shear power spectrum by first expressing the lensing potential $\psi$ at a position $(\theta,\varphi)$ in the sky. In the Born approximation, this is the projection of the 3D Weyl potential ${\Psi}_W = (\Phi+ \Psi)/2$ \citep{Kaiser_1998} expressed as
\begin{equation}
   \psi(\theta, \varphi) = \frac{2}{c^2}
   \int_{0}^{\infty}
   \mathrm{d}z \frac{{\Psi}_W\,q[r(z)]}{f_K[r(z)]}
   \frac{c}{H(z)}\;,
   \label{eq:LensigPot}
\end{equation}
where the lensing efficiency $q$ is defined as
\begin{equation}
    q[r(z)]=
  \int_{z}^{z_{\mathrm{max}}}
  \mathrm{d}z'\;n(z')\,\frac{f_K[r(z')-r(z)]}{f_K[r(z')]} \;.
\end{equation}
The spherical harmonic power spectrum of the lensing potential can then be written as
\begin{align}
C_{ij}^{\psi}(\ell)
&=
\frac{8}{\pi\,c^{4}}
\int_{0}^{\infty}\!\mathrm{d}z\; \nonumber
\frac{c}{H(z)}\;\frac{q_i[r(z)]}{f_K[r(z)]}\\
&\times\int_{0}^{\infty} \mathrm{d}z'\; 
\frac{c}{H(z')}\;\frac{q_j[r(z')]}{f_K[r(z')]} \\
&\times \int_0^\infty\mathrm{d}k\;k^{2}\,
j_{\ell}\bigl[k\,f_K[r(z)]\bigr]\;
j_{\ell}\bigl[k\,f_K[r(z')]\bigr]\;
P_{{\Psi}_W} \bigl(k,\,z,\,z'\bigr) \;,\nonumber
\end{align}
where the survey is divided into tomographic bins with
\begin{equation}
    q_i[r(z)]
  \;=\;
  \int_{z}^{z_{\mathrm{max}}}
  \mathrm{d}z'\;n_i(z')\,\frac{f_K[r(z')-r(z)]}{f_K[r(z')]} \;.
\end{equation}
To facilitate a joint analysis with galaxy position power spectra, it is convenient to write $C_{i j}^\psi(\ell)$ in terms of the matter power spectrum $P_{\rm m}(k, z)$, rather than the Weyl power spectrum $P_{{\Psi}_W}(k, z)$. These spectra are defined by 
\begin{equation}
\begin{aligned}
\left\langle\hat{\Psi}_W(\boldsymbol{k}, z) \hat{\Psi}_W^*\left(\boldsymbol{k}^{\prime} ; z^{\prime}\right)\right\rangle&=(2 \pi)^3 \delta_{\mathrm{D}}\left(\boldsymbol{k}-\boldsymbol{k}^{\prime}\right) P_{{\Psi}_W}\left(k , z, z^{\prime}\right)\;,\\
\left\langle\hat{\delta}(\boldsymbol{k} ; z) \hat{\delta}^*\left(\boldsymbol{k}^{\prime} ; z^{\prime}\right)\right\rangle&=(2 \pi)^3 \delta_{\mathrm{D}}\left(\boldsymbol{k}-\boldsymbol{k}^{\prime}\right) P_{\mathrm{m}}\left(k, z, z^{\prime}\right)\;.
\end{aligned}\label{eq:spectra}
\end{equation}
Using Eqs.~\eqref{eq:sigma_poisson} and \eqref{eq:spectra}, we can relate the two power spectra\footnote{Note that \camb's definition of the Weyl transfer function as $T_\mathrm{Weyl}  =k^2 (\Phi+\Psi)/2$ automatically incorporates an extra $k^4$ factor in the Weyl power spectrum. That is why no explicit $k^4$ appears in Eq. \eqref{eq:Weyl_matter}.} via

\begin{align}
    P_{{\Psi}_W} (k,z,z') &=   \Gamma(k,z) \Gamma(k,z') \frac{P_\mathrm{m}(k,z,z')}{k^4}\label{eq:Gamma_def}
\end{align}
with $\Gamma$ being defined as the Weyl-matter conversion factor,
\begin{equation}
\Gamma(k,z)=\frac{4 \pi G}{c^2} \frac{\bar{\rho}_{\rm m}(z)}{(1+z)^2}  \Sigma_{\rm{mg}}(k,z)\;.
\end{equation}
Using $\bar\rho_\mathrm{m} = \bar\rho_\mathrm{m,0} (1+z)^3$ and $  4\pi G \bar\rho_\mathrm{m,0} = 3 H_0^2\Omega_\mathrm{m,0} /2$, the conversion factor can be rewritten as
\begin{equation}
\Gamma(k,z)=\frac{3}{2}\frac{H_0^2}{c^2}\Omega_\mathrm{m,0}(1+z)\Sigma_{\rm{mg}}(z,k)\;.
\end{equation}
Given this, and under the assumption that $\Gamma$ is scale-independent, the lensing potential power spectrum becomes
\begin{align}
C_{ij}^{\psi}(\ell)
\;&=\;
\frac{8}{\pi\,}
\int_{0}^{\infty}\!\mathrm{d}z\; \nonumber
\frac{c}{H(z)}\;\frac{q_i[r(z)]}{f_K[r(z)]}\Gamma(z)\\
&\times\int_{0}^{\infty}\!\mathrm{d}z'\; 
\frac{c}{H(z')}\;\frac{q_j[r(z')]}{f_K[r(z')]}\Gamma(z') \\
&\times\int_{0}^{\infty} \frac{\mathrm{d}k}{k^{2}}\,
j_{\ell}\bigl[k\,f_K[r(z)]\bigr]\;
j_{\ell}\bigl[k\,f_K[r(z')]\bigr]\;
P_{\rm m}\bigl(k,\,z,\,z'\bigr)\;. \nonumber
\end{align}
Adopting the flat sky approximation, the shear power spectrum is related to the lensing potential power spectrum by
$C_{ij}^{\gamma\gamma}(\ell) = \ell^4 C_{ij}^{\psi}(\ell)/4$, which leads to
\begin{align}
C_{ij}^{\gamma\gamma}(\ell)
&=
\frac{2}{\pi\,}\ell^4
\int_{0}^{\infty}\!\mathrm{d}z\; \nonumber
\frac{c}{H(z)}\;\frac{q_i[r(z)]}{f_K[r(z)]}\Gamma(z)\\
&\times\int_{0}^{\infty}\!\mathrm{d}z'\; 
\frac{c}{H(z')}\;\frac{q_j[r(z')]}{f_K[r(z')]}\Gamma(z') \\
&\times\int_{0}^{\infty} \frac{\mathrm{d}k}{k^{2}}\,
j_{\ell}\bigl[k\,f_K[r(z)]\bigr]\;
j_{\ell}\bigl[k\,f_K[r(z')]\bigr]\;
P_{\rm m}\bigl(k,\,z,\,z'\bigr)\;. \nonumber
\end{align}
Finally, by applying first-order Limber approximation \citep{Kilbinger:2017lvu}, the shear power spectrum is simplified to 
\begin{align}
C_{ij}^{\gamma\gamma}(\ell)
\simeq
c
\int_{0}^{\infty}\mathrm{d}z\;
\frac{W^\gamma_i(z) W^\gamma_j(z) }{H(z)f_K^2[r(z)]} P_{\rm m}\bigl[k_\ell(z),z]\;,
\end{align}
where we introduce the lensing window function as
\begin{equation}
W^\gamma_i(z) =\Gamma(z)f_K[r(z)] q_i[r(z)]\;. 
\end{equation}

\section{Approximating the cb--total matter correlator}
\label{app:d}
As described in Sect.~\ref{sec:nus}, when dealing with the cross-correlation of WL and GC probes, the correlator of cb with total matter shows up. We approximate this through a geometric mean of the respective auto power spectra, as described in Eq.~\eqref{eq:correlator_cbtot}
\begin{equation}
    \langle\delta_\mathrm{m}(\boldsymbol{k})\,\delta_\mathrm{cb}(\boldsymbol{k}')\rangle =\! \sqrt{P^\mathrm{NL}_\mathrm{m}(k)\,P^\mathrm{NL}_\mathrm{cb}(k)}\,(2\,\pi)^3\,\delta_\mathrm{D}^{(3)}(\boldsymbol{k}+\boldsymbol{k}')+\mathcal{O}(f_\nu^2)\;.
\end{equation}
Here we want to go into further detail regarding this approximation. For solely linear spectra, we can find the exact value of this correlator
\begin{align}
    \langle\delta_\mathrm{m}(\boldsymbol{k})\,\delta_\mathrm{cb}(\boldsymbol{k}')\rangle &=  \langle \left[f_\mathrm{cb}\,\delta_\mathrm{cb}(\boldsymbol{k})+f_\nu\,\delta_\nu(\boldsymbol{k})\right]\,\delta_\mathrm{cb}(\boldsymbol{k}')\rangle \label{eq:lienarcorrelator} \\
    &= \left[f_\mathrm{cb}P_\mathrm{cb}^\mathrm{L}(k)+f_\nu\,P^\mathrm{L}_{\mathrm{c}\nu}(k)\right]\,(2\,\pi)^3\,\delta_\mathrm{D}^{(3)}(\boldsymbol{k}+\boldsymbol{k}')\;. \nonumber
\end{align}
Neglecting terms of order $f_\nu^2$, the geometric mean of the linear spectra becomes
\begin{align}
    \sqrt{\pcb\,\pmm} =& \sqrt{\pcb\,\left(f_\mathrm{cb}^2\,\pcb^\mathrm{L}+2\,f_\nu\,f_\mathrm{cb}\,P_{\mathrm{c}\nu}+f_\nu^2\,P_\nu\right)} \label{eq:geomean}\\
    =&\sqrt{\pcb\,\left[\left(1-2f_\nu\right)\,\pcb^\mathrm{L}+2\,f_\nu\,P^\mathrm{L}_{\mathrm{c}\nu}\right]}+\mathcal{O}(f_\nu^2)\phantom{\;.}\nonumber\\
    =&\pcb\sqrt{1+2\,f_\nu\left(\frac{P_{\mathrm{c}\nu}}{\pcb}-1\right)}+\mathcal{O}(f_\nu^2)\phantom{\;.}\nonumber\\
    =&\pcb\left[1+f_\nu\left(\frac{P_{\mathrm{c}\nu}}{\pcb}-1\right)\right]+\mathcal{O}(f_\nu^2)\phantom{\;.}\nonumber\\
    =&f_\mathrm{cb}\,\pcb+f_\nu\,P_{\mathrm{c}\nu}+\mathcal{O}(f_\nu^2)\;.
\end{align}
For linear power spectra, Eq.~\eqref{eq:correlator_cbtot} is valid. However, computing the non-linear correlator remains an open question. Since the nonlinear cb power spectrum is only approximate, replacing both power spectra in the geometric mean with the nonlinear spectra (right-hand side of Eq.~\ref{eq:correlator_cbtot}) is functionally equivalent to replacing only the cb power spectrum in Eq.~\eqref{eq:lienarcorrelator}. We chose the former approach, but note that this has to be checked against simulations.

\end{appendix}

\end{document}